\documentclass[twoside,12pt]{article}
\usepackage{epsfig}
\def\FBS{{\em Few-Body Syst.} }
\def\Journal#1#2#3#4{{#1} {#2} (#4) #3 }

\def\NP{{\em Nucl. Phys.}}
\def\NPA{{\em Nucl. Phys.} A}

\def\PRO{{\em Prog. Theor. Phys.}}
\def\NPB{{\em Nucl. Phys.} B}

\def\PLB{{\em Phys. Lett.} B}

\def\PRL{\em Phys. Rev. Lett.}
\def\PREV{\em Phys. Rev.}
\def\PREP{\em Phys. Rep.}
\def\PRA{{\em Phys. Rev.} A}
\def\PRD{{\em Phys. Rev.} D}
\def\PRE{{\em Phys. Rev.} D}
\def\PRC{{\em Phys. Rev.} C}
\def\PRB{{\em Phys. Rev.} B}

\def\ZPA{{\em Z. Phys.} A}
\def\ANNP{\em Ann. Phys. (N.Y.)}
\def\RMP{{\em Rev. Mod. Phys.}}
\def\CHEM{{\em J. Chem. Phys.}}

\def\r{\bf r}

\def\p{\bf p}
\def\P{\bf P}

\def\x{\bf x}
\def\y{\bf y}
\def\z{\bf z}
\def\ss{\mbox{\boldmath $\sigma$}}
\def\xx{\mbox{\boldmath $x$}}
\def\veta{\mbox{\boldmath $\eta$}}
\def\vpi{\mbox{\boldmath $\pi$}}
\newcommand{\be}{\begin{equation}}
\newcommand{\ee}{\end{equation}}
\newcommand{\bea}{\begin{eqnarray}}
\newcommand{\eea}{\end{eqnarray}}

\newcommand{\vlowk}{V_{{\rm low}\,k}} 
\begin{document}

\title{ \vspace{1cm} Modern Ab Initio
 Approaches and Applications in Few-Nucleon Physics with $A \ge 4$}
\author{Winfried\ Leidemann and Giuseppina\ Orlandini\\
\\
Dipartimento di Fisica, Universit\`a di Trento, I-38123 Trento, Italy\\
INFN, Gruppo Collegato di Trento, I-38123 Trento, Italy}
\maketitle

\begin{abstract}
We present an overview of the evolution of ab initio methods for few-nucleon systems with $A \ge 4$, tracing
the progress made that today allows precision calculations for these systems.
First a succinct description of the diverse approaches is given. In order to identify analogies and differences
the methods are grouped according to different formulations of the quantum mechanical many-body problem. Various significant
applications from the past and present are described. We discuss the results with emphasis on the developments following the original
implementations of the approaches. In particular we highlight benchmark results which represent
important milestones towards setting an ever growing standard for theoretical calculations.
This is relevant for meaningful comparisons with experimental data. Such comparisons may reveal whether
a specific force model is appropriate for the description of nuclear dynamics.
\end{abstract}

\section{Introduction}\label{sec:INTRO}

The last two decades have witnessed decisive progress in nuclear physics by ab initio calculations. 
In the previous years ab initio calculations were restricted mainly to applications in two- and 
three-nucleon problems while in the recent past these techniques have been extended to considerably
larger nuclei, which, in some cases, consist of even more than ten nucleons. Here we
include such larger systems in our definition of few-nucleon systems. 
Progress has been made in various directions. Methods which were already well established in the field have been further
improved. As examples we mention Faddeev calculations, which have been extended from three- to four-nucleon 
systems~\cite{GlK93a,CiC99} and which have solved the problem of inclusion of the Coulomb interaction in momentum space~\cite{DeF05}, 
and Green's function Monte Carlo (GFMC) calculations, which have reached a degree of
sophistication which allows inclusion of local realistic nuclear forces up to a nucleon number of $A=12$~\cite{Pi08}. 
In addition, novel approaches such as the no-core shell model 
(NCSM)~\cite{NaV00} and the Lorentz integral transform (LIT)~\cite{EfL94} methods have
been introduced. Last, but not least, growing 
computational resources and improved numerical algorithms have allowed calculations which
were previously unthinkable.

At the same time enormous progress has also been made in the description of the dynamical input of an ab initio 
calculation, i.e. the nuclear interaction.
In the 1990s modern realistic phenomenological and meson-theoretical nucleon-nucleon (NN) potential models,
which provide  high precision descriptions of both NN bound state and the wealth of NN scattering state
observables, have been developed.
However, the failure of these potentials to account for the binding energy and other observables 
in the three-body sector has demonstrated the necessity of including three- and in principle even more-nucleon forces. 
Three-body forces of different forms have been proposed and their parameters have been obtained by
fitting to data. At the same time a new line of research dedicated
to the derivation of the nuclear interaction from chiral perturbation theory was opened up. 
This new ansatz
has led to the construction of competitive modern realistic potentials that contain chiral many-nucleon forces  
consistent with the two-body force~\cite{EpM11,EpH09,MaE11}.

The complications in ab initio calculations, generated on the one hand  by the strongly repulsive
nature of all the modern realistic potentials of the 1990s and on the other hand by the three-body forces,
have stimulated an interesting discussion and led to a corresponding line of research.  
Thus the last few years saw considerable evolution in obtaining phase-shift equivalent NN potential models,
which often lead to a great simplification in ab initio calculations. In fact one may apply unitary transformations to the 
two-body Hamiltonian, and to other operators at the same time, without effecting any change in the two-nucleon observables. 
This leads to an infinite number of equally valid NN potentials some of which could have characteristics that are simpler
to treat in numerical calculations. For example
one can aim at  softer potentials at short range. By applying a unitary transformation at the three-body level one could even
trade a three-nucleon force with a  two-body potential that has a different off-shell behavior.
Examples for this procedure are the potentials obtained by the renormalization group transformations like the so-called 
$\vlowk$~\cite{BoK03}, 
and the free or in medium similarity  renormalization group (SRG) potentials~\cite{BoF10}, as well as the force obtained 
by the unitary correlation operator method (UCOM)~\cite{RoN10} and the J-matrix inverse scattering potential (JISP)~\cite{ShW07}. 

Though it is true that all the various potentials  mentioned above are phase-shift equivalent,
they may not be identical with the
original potential models, since, in general, the off-shell structure can be different. This may lead
to different results for observables in the $A>2$ sector, such as
e.g. the triton binding energy. It is clear that even applying a specific unitary transformation at the three-body level, one may have
different results for $A>3$ observables and one may regard the transformed Hamiltonian as a new potential model. Since there are an infinite
number of different unitary transformations one has an infinite number of different potentials. Only the comparison to experiment of 
precise ab initio results for $A>3$ observables can discriminate among these potentials.
In this work we will not discuss nuclear forces, although in various places
we do mention different models for NN potentials 
and three-nucleon forces (for a recent review of the latter see~\cite{KaE12}).

The present work aims at giving 
an overview of the, by now, numerous ab initio methods
which are applied to obtain ab initio solutions of nuclear many-body problems for bound and
scattering states, as well as for observables of hadronic and perturbation-induced reactions.
Since in the last years the term {\it ab initio} has been rather loosely used,
we first would like to clarify what we consider to be an {\it ab initio method} and an 
{\it ab initio result}. For this purpose
let us consider an $A$ nucleon system that is described by a well-defined 
microscopic Hamiltonian $H$ with $A$ nucleon degrees of freedom and where the internal relative motion
is treated correctly. 
If a method enables one to obtain the observable under consideration  by
solving the relevant quantum mechanical many-body equations, without any uncontrolled approximation, 
we consider it to be an {\it ab initio method}. Controlled approximations, however, are allowed.
In fact a controlled approximation, e.g. a limited number of channels in a Faddeev calculation, can be  
increasingly improved up to the point that convergence is reached for the observable. Such 
a converged result we denote as a {\it precise ab initio} result. The comparison of {\it precise ab initio} results 
with nuclear data then allows an indisputable answer as to whether or not the chosen Hamiltonian
appropriately describes the nuclear dynamics. Any uncontrolled
approximation in the calculation would not lead to such a clear-cut conclusion.
Quite naturally, {precise ab initio} results obtained with different {ab initio methods} but 
with the same Hamiltonian as input, have to agree and are often referred to as  {\it benchmark results}. 

The aim of a quantum mechanical treatment of a particle system is the determination of observables.
Often this aim can be achieved by rather different formulations.
The organization of our work is guided by this diversity. 
In the present few-body case all formulations start from the same point, namely from 
a Hamiltonian that has nucleons as degrees of freedom and that contains a well-defined interaction model. 
While pursuing a specific formulation one may still have a choice between different methods, some of which can be
innate to the strategy under consideration, or others just adapted from elsewhere. 
Therefore the first part of the work is devoted to such different formulations. The diversity of methods  
are described in the corresponding subsections, which contain, in addition, some illustrative examples from the literature.
Selected results are presented and discussed in a following section. 

Here a comment is in order on the issue of whether a given method is more suited to bound- and/or continuum-state
problems. In principle all the various methods presented here can be applied to both cases. The only exceptions are the LIT 
and the complex scaling methods, which do not treat bound-state problems, but which are dedicated to the calculation 
of observables in the continuum. On the other hand it is true that some methods are more often used, either for bound-state or for
continuum-state problems. For instance, variational techniques are applied more frequently to bound states, but, as
discussed in Section~\ref{sec:VAR}, continuum-state calculations can also
be carried out via the Kohn variational principle. A similar observation can be made for the GFMC technique 
(see Section~\ref{sec:GFMC}). 

The methods we are about to present in this work are quite general and not restricted to a specific field.
However, we limit our discussion mainly to examples from nuclear physics. In order to make the wide range of applications
more explicit we list here topics and related review articles, where ab initio techniques are receiving a growing 
interest from the few- and many-body communities: universality in few-body systems and 
Efimov physics~\cite{BrH06}, few-body ultracold atomic and molecular systems~\cite{Bl12}, quantum degenerate Fermi 
gases~\cite{GiP08}, and last but not least quantum dots~\cite{ReM02}. For the latter case there is also a recent benchmark 
calculation with diffusion Monte Carlo and coupled cluster techniques~\cite{PeH11}), where in addition a brief overview
on the state-of-the-art of the physics of quantum dots is given in the introduction.

In more detail, our work is organized as follows. In Section~\ref{sec:FB}
 we characterize the so-called few-body problem very briefly.
In Section~\ref{sec:FY} we start with 
short introductions to the Faddeev and Faddeev-Yakubovsky formalisms. A reformulation of the 
Faddeev approach and the Alt-Grassberger-Sandhas (AGS) method are discussed in Section~\ref{sec:AGS}. An introduction to the variational 
formulation of the few-body 
problem is given in Section~\ref{sec:VAR}, followed by various subsections describing specific variational 
methods. Another formulation, 
largely used both in few-and many-body physics, especially for the construction of effective interactions, 
is based on similarity transformations. Thus, in Section~\ref{sec:SIM}, the similarity  transformation 
formulation is outlined and few-body methods that make use of it are described. Formulations that are suitable
for using Monte Carlo techniques are discussed in Section~\ref{sec:MC}.
After the discussion of the various ab initio methods, we show in Section~\ref{sec:RECRES} a selection of results that in our 
view represent some advancement of the field. There it will be seen that we place a certain emphasis on benchmark calculations. 
In Section~\ref{sec:4_hist} we report on the early stage of ab initio calculations for the $A=4$ nuclear system, 
because there
the fundamentals for many later developments have been laid and we think that this pioneering work deserves not to be forgotten. 
The more modern stage of $A=4$ calculations initiated in the late 1980s is discussed in Section~\ref{sec:4_mod}. 
Results for nuclear systems with $A>4$ are described 
in Sections~\ref{sec:AG4_SPECTRA} and ~\ref{sec:AG4_FURTHER}. Finally  summary and outlook are given in Section~\ref{sec:SUM}. 

\bigskip 
\section{The Few-Body Problem}\label{sec:FB}

The dynamics of a system of $A$ non-relativistic nucleons of equal mass $m$
is governed by the translation invariant nuclear Hamiltonian $H$, 
which consists of kinetic energy $T$ and potential $V$:
\bea
\nonumber
H &=& T + V \\
  &=& \sum_{i=1}^A {\frac {{\p}^2_i} {2m} } - \frac {{\P}_{\rm \rm CM}^2} {2Am} + \sum_{i<j}^A V_{ij} 
+ \sum_{i<j<k}^A V_{ijk} \,.\label{Hint}
\eea
In the equation above ${\p}_i$ is the momentum of the $i$-th nucleon, $\P_{\rm CM}$ is the center of mass (CM) momentum,  
while $V_{ij}$ and $V_{ijk}$ denote the NN  
potential $V_{\rm NN}$, and the three-nucleon potential $V_{\rm NNN}$, respectively.
We will refer to the latter 
as {\it three-nucleon force} (3NF). Since the nucleons are effective degrees of freedom,   
the nuclear Hamiltonian $H$  should in principle contain potentials involving more than three nucleons.
However, at present such many-nucleon 
forces seem to play a negligible or at least a minor role in the dynamics of $A$-nucleon systems (see e.g.~\cite{EpM11}).

The basic dynamical equation to be solved ab initio is the Schr\"odinger equation,
\be
(H-E) |\Psi\rangle = 0 \,,
\ee
where $E$ and $|\Psi\rangle$ denote one of the eigenenergies and eigenfunctions of $H$, respectively.
The spectrum of $H$, represented by the infinite set of eigenenergies 
$E$, is discrete below, and continuous above, 
the first break-up threshold ($E_{\rm th})$ of the $A$-nucleon system. 

In order to solve the Schr\"odinger equation one has to supply proper
boundary conditions. For example, in coordinate space calculations one requires the regularity condition at
$r_{ij}=0$, where $r_{ij}$ is the relative distance between two nucleons. The asymptotic
boundary condition depends on the energy of the system. For $E < E_{\rm th}$ the wave function represents a bound state
and thus is described by a square integrable (localized) function.
For $E \ge E_{\rm th}$ the asymptotic boundary condition for the corresponding scattering state remains quite simple 
in the case of $A=2$. 
On the contrary the asymptotic boundary conditions pose
a serious problem if both $E \ge E_{\rm th}$ and $A>2$. How one solves that problem is discussed in the following
subsection. In the $A=2$ case one may start not from the Schr\"odinger equation but from its integral form,
the Lippmann-Schwinger (LS) equation 
\be
|\Psi^{(\pm)}\rangle = |\Phi\rangle + {\frac {1}{E \pm i\epsilon - T }} V |\Psi^{(\pm)}\rangle \,,
\ee
where 
$|\Phi\rangle$ is the free solution describing a plane wave, i.e. satisfying 
$(T -E) |\Phi\rangle =0$.
The nice feature of the LS equation  is that it already contains the asymptotic boundary
condition, i.e. an outgoing and an incoming spherical scattered wave for $|\Psi^{(+)}\rangle$ and $|\Psi^{(-)}\rangle$,
respectively~\cite{GoW04}.

This simple example shows that it can be helpful to reformulate the original quantum mechanical many-body problem so that it 
becomes possible to tackle it  with a different technique.
It is this possibility that generates the richness of methods in many-body theory.

\subsection{The Faddeev and Faddeev-Yakubovsky (FY) Formulation}\label{sec:FY}

Today it is well known that a direct application of LS-type equations to a 
scattering problem does not lead to a unique solution for a system with more than two particles. 
However, how such a unique solution can be obtained was not clear for quite some time.
In his seminal work Faddeev showed how the problem can be  treated \cite{Fa61}. Considering the
three-body system he derived a set of equations, {\it the Faddeev equations}. We do not illustrate 
Faddeev's original derivation,  but discuss here a few selected points of the  didactic derivation 
of the Faddeev equations given by Gl\"ockle \cite{Gl83}. The important result of Faddeev is
that a specific three-body scattering state should obey not just one LS equation,
but two additional equations at the same time. Rewriting the Hamiltonian as
\begin{equation}
 H= T+ \sum_{i=1}^3 V_i\,,
\end{equation}
where $V_1=V_{23},V_2=V_{13},V_3=V_{12}$, the three equations for the $(2+1) \rightarrow (2+1)$ reaction that ensure the 
correct boundary conditions read as follows 
\bea
\label{Psi_1}
|\Psi_\alpha^{(+)}\rangle &=& |\Phi_\alpha\rangle + G_\alpha(E)  V^\alpha |\Psi_\alpha^{(+)}\rangle \,,\\
\label{Psi_2}
|\Psi_\alpha^{(+)}\rangle &=& \,\,\,\,\,\,\,\,\,\,\,\,\,\,\,\,\,\, G_\beta(E)  V^\beta |\Psi_\alpha^{(+)}\rangle \,,\\
\label{Psi_3}
|\Psi_\alpha^{(+)}\rangle &=& \,\,\,\,\,\,\,\,\,\,\,\,\,\,\,\,\,\,G_\gamma(E)  V^\gamma |\Psi_\alpha^{(+)}\rangle \,,
\eea
where $\alpha\ne \beta\ne \gamma$  can take the values 1, 2, and 3 , $E$ is the total energy of the three-nucleon (3N) system and
 $|\Psi_\alpha^{(+)}\rangle$ is the outgoing  scattering wave in channel $\alpha$ (i.e. where asymptotically 
one has particle $\alpha$ and the pair formed by the other two particles). This function satisfies
\begin{equation}
 H |\Psi^{(+)}_\alpha\rangle= E |\Psi^{(+)}_\alpha\rangle\,,
\end{equation}
while $|\Phi_\alpha\rangle$ is eigenfunction of the channel Hamiltonian $H_\alpha\equiv T + V_\alpha$, i.e. it fulfills
\begin{equation}
 H_\alpha |\Phi_\alpha\rangle = E |\Phi_\alpha\rangle \,.
\end{equation}
The potential term $ V^\alpha $ is defined as $\,V^\alpha \equiv H - H_\alpha$ and $G_\alpha(z)$ 
is the resolvent operator for channel $\alpha$ 
\be
G_\alpha(z) = {\frac {1}{z-H_\alpha} } \,\,. 
\ee
The inhomogeneous LS equation (\ref{Psi_1}) alone does not define the boundary conditions
uniquely. This is the case only with the two   
additional homogeneous equations (\ref{Psi_2}) and (\ref{Psi_3}). Also for the (1+1+1) 
scattering one arrives at three analogous equations.  
However, all these sets of equations do not directly allow practical solutions.
In fact for any equation set the same state has to fulfill three different equations. In order to
come to a more feasible task the three separate equations should be transformed into a coupled equation system.
The proper strategy is to decompose the wave function in three parts, the so-called Faddeev components
$|\Psi_{\alpha,\mu}\rangle$:
\be
|\Psi_\alpha^{(+)}\rangle = \sum_{\mu=1}^3 |\psi_{\alpha,\mu}\rangle \,,
\ee
where $|\psi_{\alpha,\mu}\rangle\equiv G_0(E)V_\mu|\Psi_\alpha^{(+)}\rangle$ with $G_0$ representing the free propagator. 
Using this decomposition one arrives at 
the following coupled equation system
\bea
\label{Psi_a1}
|\Psi_{\alpha,\alpha}\rangle &=& |\Phi_\alpha\rangle + G_\alpha(E)  V^\alpha (|\Psi_{\alpha,\beta}\rangle 
                                                                            + |\Psi_{\alpha,\gamma}\rangle) \,,\\
\label{Psi_a2}
|\Psi_{\alpha,\beta}\rangle &=& \,\,\,\,\,\,\,\,\,\,\,\,\,\,\,\,\,\, G_\beta(E)  V^\beta (|\Psi_{\alpha,\gamma}\rangle 
                                                                            + |\Psi_{\alpha,\alpha}\rangle) \,,\\
\label{Psi_a3}
|\Psi_{\alpha,\gamma}\rangle &=& \,\,\,\,\,\,\,\,\,\,\,\,\,\,\,\,\,\,G_\gamma(E) V^\gamma (|\Psi_{\alpha,\alpha}\rangle 
+ |\Psi_{\alpha,\beta}\rangle) \,.
\eea
These are then the equations that are usually called Faddeev equations. For bound states
one has a modified set of Faddeev equations, where the inhomogeneous term in Eq.~(\ref{Psi_a1}) is dropped.

The Faddeev equations in the form discussed above are naturally suitable for a solution in momentum space. In fact most applications
for the three-nucleon (3N) system have been carried out in this representation (for a review see~\cite{GlW96}). 
Also Faddeev calculations for the four-nucleon (4N)
bound state have been made in momentum space as discussed below. Coordinate space Faddeev calculations 
have been performed for both the 3N and 4N cases.  A particularly interesting
example of using the Faddeev formalism in coordinate space is in low-energy (3+1) scattering.
Therefore we devote the rest of this section to the Faddeev formalism in coordinate space.
First we consider the 3N system and assume that the nucleons interact via a short-range two-body potential $V_{ij}$.

Denoting with ${\r}_i$ the position of nucleon $i$ one has the following sets of Jacobi coordinates
\be
{\x}_{ij} = {\r}_j - {\r}_i \,, \,\,\,\,\,\,\, {\y}_{ij,k} = {\r}_k - {\frac {{\r}_i + {\r}_j}{2} } \,.
\ee
There are six different permutations and hence six different sets. However,   ${\x}_{ij}$
and ${\x}_{ji}$ describe the same physical situation, thus there are only three independent sets. 
In the following step the wave function is decomposed into the aforementioned three Faddeev components $\psi_i$: 
\be
\langle {\x} {\y} \, | \Psi \rangle = \psi_1({\x}_{23},{\y}_{23,1}) + \psi_2({\x}_{31},{\y}_{31,2}) + \psi_3({\x}_{12},{\y}_{12,3}) \,,
\ee
where we have suppressed spin and isospin degrees of freedom.
Then, one requires that the  Faddeev components satisfy the following set of coupled equations,
\bea
\label{Fad_Amp1}
(T - E + V_{23}) \psi_1({\x}_{23},{\y}_{23,1}) &=& -V_{23}\, (\psi_2({\x}_{31},{\y}_{31,2}) + \psi_3({\x}_{12},{\y}_{12,3})) \,, \\ 
\label{Fad_Amp2}
(T - E + V_{13}) \psi_2({\x}_{31},{\y}_{31,2}) &=& -V_{13}\, (\psi_1({\x}_{23},{\y}_{23,1}) + \psi_3({\x}_{12},{\y}_{12,3})) \,, \\
(T - E + V_{12}) \psi_3({\x}_{12},{\y}_{12,3}) &=& -V_{12}\, (\psi_1({\x}_{23},{\y}_{23,1}) + \psi_2({\x}_{31},{\y}_{31,2}))  \,,
\label{Fad_Amp3}
\eea
which are also known as Faddeev-Noyes equations. 
Adding up the three Eqs.~(\ref{Fad_Amp1}-\ref{Fad_Amp3}) one easily verifies that the Schr\"odinger equation,
\be
(T - E + V_{23} + V_{13} + V_{12}) \langle {\x} {\y} \, | \Psi\rangle = 0 \,,
\ee
is fulfilled.
For the  case of identical particles (the nucleons) considered here, there is a nice feature of 
the Faddeev formalism, namely that the various $\psi_i$
 can be obtained from each other by particle permutations.
In fact one has
\be 
\psi_2({\x}_{31},{\y}_{31,2}) = P_{12} P_{23} \psi_1({\x}_{23},{\y}_{23,1}) \,,\,\,\,\,\,
\psi_3({\x}_{12},{\y}_{12,3}) = P_{23} P_{12} \psi_1({\x}_{23},{\y}_{23,1}) \,,
\ee
and therefore it is sufficient to solve a single equation  
\be
\label{Fad_last}
(T - E + V_{23}) \psi_1({\x}_{23},{\y}_{23,1}) = -V_{23}\, (P_{12} P_{23} + P_{23} P_{12}) \, \psi_1({\x}_{23},{\y}_{23,1}) \,.
\ee

Let us consider a 3N state with total angular momentum $J$ and parity $\pi$.
To solve Eq.~(\ref{Fad_last}) for this state $|J^\pi\rangle$ the Faddeev component $\psi_i$, from here on
simply denoted by $\psi$, is expanded into {\it channels} 
\be
\label{Fad_Expsn}
\psi({\x},{\y}) = \sum_{\{\lambda \}}{\frac {\phi^{\{\lambda \}}(x,y)}{xy} } f^{\{\lambda \}}({\hat x},{\hat y}) \,,
\ee
where for simplicity the subscripts of variables ${\x}$ and ${\y}$ and the dependence of the functions
$\psi$ and $f^{\{\lambda \}}$  on spin and isospin variables are dropped. Here the different {\it channels}
refer to different sets $\{\lambda \}$ of quantum numbers.
As to boundary conditions, one has 
$\phi^{\{\lambda \} }(x\,$=0,$\,y)\,$=$\,\phi^{\{\lambda \}}(x,\,y\,$=\,$0\,)$=$\,0$ % \,.
for bound and scattering states. 
Moreover, in the bound-state case, one has the asymptotic boundary condition
$\phi^{\{\lambda \}}(x\,$=$\,X_{\rm max},\,y)\,$=$\,\phi^{\{\lambda \}}(x,\,y\,$=$\,Y_{\rm max})\,$=$\,0$, 
where the values for $X_{\rm max}$ and $Y_{\rm max}$ have to be chosen properly.

The correct configuration space asymptotic boundary conditions for the 3N scattering problem were discussed
by Merkuriev, Gignoux, and Laverne \cite{MeG76}. 
It should be obvious that the solution of the coordinate space Faddeev equation poses a greater challenge
at energies above rather than below the three-body break-up 
threshold. In fact below the threshold one has to consider only the (2+1) channel. Then, for the asymptotic
solution one may divide the sets ${\{\lambda \}}$ in those which have the proper quantum numbers to allow for
a N-deuteron state and those where this is not possible. The $\phi^{\{\lambda \}}$ of the latter have to vanish 
asymptotically, whereas the $\phi^{\{\lambda \}}$ of the
former can be factorized asymptotically into a radial deuteron wave function in variable $x$ times a free solution in variable $y$.  
Normally the set $\{\lambda \}$ of quantum numbers includes $L,S,J$, where the channel spin $S$, 
with possible values 1/2 and 3/2, results from 
the coupling of deuteron spin $j_d=1$ and single nucleon spin 1/2. Then $S$ is coupled to $L$,
the quantum number for the orbital 
angular momentum associated with variable $y$, to total spin $J$. For any state $|J^\pi\rangle$ one has either two ($J=1/2$) or 
three ($J>1/2$) possible combinations of $S$ and $L$. The asymptotic free solution in $y$ consists of regular and irregular parts, 
which are described by Bessel and Neumann functions, respectively. Their relative admixture is governed by the usual scattering 
parameters, namely by phase shifts and mixing parameters.
For $J=1/2$ one has two phase shifts and one mixing parameter, while for $J>1/2$ there are
three phase shifts and three mixing parameters.

The inclusion of a 3NF is rather unproblematic in the Faddeev formalism, since it is a short-range force 
(for details see \cite{Gl83}). More complicated however is the problem of taking into account
the long-range Coulomb force although how this is accomplished was shown by Merkuriev \cite{Me80}.

Distinct from the NN problem, where not more than two partial waves are coupled, 
in the case of the 3N problem one has in principle an infinite number of coupled channels for any 
$|J^\pi\rangle$ state. However, by restricting 
the interaction to specific NN partial waves one obtains a finite number of channels. For example, if one restricts
the interaction only to $^1s_0$ and $^3s_1$-$^3d_1$ NN partial waves, one obtains five coupled channels in a 3N 
ground-state calculation. The number of channels is increased to 34
if one includes the interaction of all NN partial waves with total angular momentum $j \le 4$. 

The generalization of the Faddeev formalism to more than three particles was carried out by Yakubovsky \cite{Ya67}. 
Here we turn directly to the 4N problem in configuration space. 
In case of four nucleons there are two different classes of sets of Jacobi coordinates, namely a $K$ and a $H$ class with
\bea
\label{K_set}
{\x}_{ij}^K &=& {\x}_{ij} \,,\,\,\,\,\, {\y}_{ij,k}^K = {\y}_{ij,k} \,,\,\,\,\,\,\ 
z_{ijk,l}^K = {\r}_l - {\frac {{\r}_i + {\r}_j + {\r}_k}{3} } \,,\\
\label{H_set}
{\x}_{ij}^H &=& {\x}_{ij} \,,\,\,\,\,\, {\y}_{kl}^H = {\r}_l - {\r}_k \,, \,\,\,\,\,\, 
z_{ij,kl}^H = {\frac {{\r}_k + {\r}_l}{2} } - {\frac {{\r}_i + {\r}_j}{2} } \,.
\eea
Since there are 24 permutations one has 48 different sets. Again the cases with  ${\x}_{ij}$
and ${\x}_{ji}$ describe the same physical situation, and in addition one has to consider that for the
$H$ sets  ${\y}_{kl}$ and ${\y}_{lk}$ are also equivalent. Thus one remains with 18 (12$K$+6$H$) physically
non-equivalent Jacobi coordinate sets.

Similar to the three-body case one splits the wave function $\Psi$ into the FY components $\psi_{ij}$. By means
of the FY components one obtains a set of coupled equations, here 18 equations, which added together
lead to the four-body Schr\"odinger equation as detailed below. 

The FY components are defined by 
\be
\label{Psi_4}
\langle {\x} {\y} {\z} \,|\Psi \rangle = \sum_{i<j} \psi_{ij}({\x}_{ij},{\y},{\z}\,) \,,
\ee
 where 
\be
\psi_{ij}({\x}_{ij},{\y},{\z}\,) \equiv \psi_{ij}^K({\x}_{ij},{\y}^K_{ij,k},{\z}^K_{ijk,l})
                          + \psi_{ij}^K({\x}_{ij},{\y}^K_{ij,l},{\z}^K_{ijl,k})
                          + \psi_{ij}^H({\x}_{ij},{\y}^H_{kl},{\z}^H_{ij,kl}) \,,
\ee
with $k<l$.
Again, if one requires the six coupled equations (one for each $ij$ pair),
\be
\label{FY_Amp}
(T - E) \psi_{ij}({\x}_{ij},{\y},{\z}\,) = -V_{ij}\, \sum_{k<l} \psi_{kl}({\x}_{ij},{\y},{\z}\,) \,,
\ee
to be satisfied, one easily verifies that the resulting $\Psi$ of Eq.~(\ref{Psi_4}) fulfills the 4N Schr\"odinger equation,
\be
(T - E + \sum_{i<j} V_{ij}) \langle {\x} {\y} {\z} \, |\Psi \rangle = 0\,.
\ee
Each of the six equations is further split into three sets of coupled equations. 
The first set is given by 
\bea
\label{FY_set1}
\nonumber
 (T \!\!\!&-&\!\!\! E + V_{ij})\,\psi_{ij}^K({\x}_{ij},{\y}^K_{ij,k},{\z}^K_{ijk,l}) = -V_{ij} \,
\bigg[\psi_{ik}^K({\x}_{ik},{\y}^K_{ik,j},{\z}^K_{ikj,l}) + \psi_{ik}^K({\x}_{ik},{\y}^K_{ik,l},{\z}^K_{ikl,j}) \\
&+&\!\!\! \psi_{jk}^K({\x}_{jk},{\y}^K_{jk,i},{\z}^K_{jki,l})
+ \psi_{jk}^K({\x}_{jk},{\y}^K_{jk,l},{\z}^K_{jkl,i}) 
+ \psi_{ik}^H({\x}_{ik},{\y}^H_{lj},{\z}^H_{ik,lj}) + \psi_{jk}^H({\x}_{jk},{\y}^H_{il},{\z}^H_{jk,il})\bigg] \,.
\eea
For the second set one has the same six relations (\ref{FY_set1}) but with $k \leftrightarrow l$, while the third
set reads  
\be
(T - E + V_{ij}) \psi_{ij}^H({\x}_{ij},{\y}^H_{kl},{\z}^H_{ij,kl}) = -V_{ij} \,
\bigg[\psi_{kl}^K({\x}_{kl},{\y}^K_{kl,i},{\z}^K_{kli,j}) + \psi_{kl}^K({\x}_{kl},{\y}^K_{kl,j},{\z}^K_{klj,i})
+ \psi_{kl}^H({\x}_{kl},{\y}^H_{ij},{\z}^H_{kl,ij}) \bigg] \,.
\ee
If one of the  components $\psi_{\lambda,\mu}^K({\x}_{\lambda,\mu},{\y}^K_{\lambda\mu,\nu},{\z}^K_{\lambda\mu\nu,\sigma})$
should appear in one of the right hand sides of the three sets of equations above with $\lambda> \mu$, it 
has to be understood as being identical to that obtained  by exchanging $\lambda$ and $\mu$.

Similar to the Faddeev case with three nucleons it is not necessary to solve the full set of coupled equations
for the case of identical nucleons.
In fact using permutations one obtains from one $\psi^K$ and one $\psi^H$ component the other eleven $\psi^K$ and 
five $\psi^H$ components, respectively. Thus one remains with a system of two coupled equations 
(for further details see \cite{MeY83,MeY84,CiC98}).  

In order to solve the remaining two coupled equations, and in analogy to Eq.~(\ref{Fad_Expsn}),
one makes the following expansion of the FY components in channels ${\{\lambda \}}$:
\be
\psi^{H/K}({\x},{\y},{\z}\,) = \sum_{\{\lambda \}} {\frac {\phi_{H/K}^{\{\lambda \}}(x,y,z)}{xyz} } F^{\{\lambda \}}_{H/K}(\hat x,\hat y,\hat z) \,,
\ee
where, again, for simplicity the subscripts of variables $x$, $y$ and $z$ and the dependence of the functions
$\psi^{H/K}$ and $F^{\{\lambda \}}_{H/K}$ on spin and isospin variables are dropped. 
As for the 3N case one has the regularity condition for bound and scattering states
\be
\phi_{H/K}^{\{\lambda \}}(x=0,y,z) = \phi_{H/K}^{\{\lambda \}}(x,y=0,z) = \phi_{H/K}^{\{\lambda \}}(x,y,z=0) = 0 \,.
\ee
For bound states the asymptotic boundary condition reads
\be
\phi_{H/K}^{\{\lambda \}}(x=X_{\rm max},y,z) = \phi_{H/K}^{\{\lambda \}}(x,y=Y^{H/K}_{\rm max},z) = 
\phi_{H/K}^{\{\lambda \}}(x,y,z=Z^{H/K}_{\rm max}) = 0 
\ee
with proper values for $X_{\rm max}$, $Y^{H/K}_{\rm max}$, and $Z^{H/K}_{\rm max}$.
For the asymptotic boundary condition the situation is more complicated. In case of (3+1) scattering below 
the first inelastic threshold one has for $\phi_H^{\{\lambda \}}$ the same boundary condition as for the bound-state
case, and for $\phi_K^{\{\lambda \}}$ one has to divide the sets ${\{\lambda \}}$ in those where a factorization into a 3N bound-state 
Faddeev component 
with total angular momentum $j_{\bf xy}=1/2$ and a free solution in variable $z$
is possible, while the other sets have to vanish asymptotically. 
As for the 3N case the free solution consists of regular and irregular parts combined with proper factors containing 
scattering parameters. The channel 
spin $S^K$, which results from the coupling of $j_{\bf xy}$ and single nucleon spin 1/2, can be equal to 0 or 1 
and couples with the orbital angular momentum quantum number $L$ associated with variable $z$ to total spin $J$. 
For a given state $|J^\pi\rangle$ one has either one phase shift ($J=0$) or two phase shifts and one mixing parameter ($J>0$).
For the boundary conditions in the presence
of an open (2+2) channel or further break-up channels we refer to Ref.~\cite{MeY84}.

As was the case for the 3N problem
a 3NF and in particular the Coulomb force can also be built into the FY $r$-space formalism for
the 4N problem (see \cite{La03,LaC04,La09}). 
However, unlike the 3N case, a restriction of the interaction to specific NN partial waves does not lead to
a reduction of the number of coupled channels. In an actual calculation it is of course impossible not to make
a truncation, which then needs to be carefully checked for convergence. 
In order to better examine the circumstances we need to introduce a coupling scheme for the
angular momenta of $K$ and $H$ class coordinates. They are given by, 
$\{([(l^K_{\bf x} s^K_{\bf x}) j^K_{\bf x} 1/2] S^K l^K_{\bf y}) j^K_{\bf xy} (l^K_{\bf z} 1/2) i^K_{\bf z}\}J$ 
and $\{[(l^H_{\bf x} s^H_{\bf x}) j^H_{\bf x} l^H_{\bf z}]  S^H (l^H_{\bf y} s^H_{\bf y}) j^H_{\bf y}\}J$,
where $l^{H/K}_{\bf u}$ (${\bf u} \equiv {\bf x}$, ${\bf y}$, ${\bf z})$ is the angular momentum quantum number
associated with variable ${\bf u}$, while $s^{H/K}_{\bf u}$ denotes the total spin of the particle pair with
relative coordinate ${\bf u}$. 

\begin{table}
\begin{center}
\caption{{\label{channels}} Number of channels for given values of maximal inter-cluster orbital
angular momentum $l^K_{\bf z}=l^H_{\bf z}$ and maximal two-body angular momentum $j^{K/H}_{\bf x} = j_{\rm max}$ (from Ref.~\cite{KaG92a}).}
\begin{tabular}{c||c|c|c}
\hline \hline Partition 
 & $j_{\rm max}=1 $  & $j_{\rm max}=2$ & $j_{\rm max}=3$  \\
\hline
 $l^K_{{\bf z},\rm max}=l^H_{{\bf z},\rm max}=0$& & & \\
 $K$ & 10  & 18  &  26  \\ 
 $H$ & 10  & 18  &  26   \\
total & 20  & 36  &  52   \\
\hline
 $l^K_{{\bf z},\rm max}=l^H_{{\bf z},\rm max}=1$ &  & &  \\
 $K$ & 34  & 66  &  98  \\ 
 $H$ & 26  & 58  &  90   \\
total & 60  & 124  &  188   \\
\hline
 $l^K_{{\bf z},\rm max}=l^H_{{\bf z},\rm max}=2$ &  & &   \\
 $K$ & 62  & 130  &  202  \\ 
 $H$ & 34  & 98 &  170   \\
total & 96  & 228  &  372   \\
\hline
 $l^K_{{\bf z},\rm max}=l^H_{{\bf z},\rm max}=3$ &  & &  \\
 $K$ & 90  & 198  &  322  \\ 
 $H$ & 34  & 122  &  242   \\
total & 124  & 320  &  564   \\
\hline
 $l^K_{{\bf z},\rm max}=l^H_{{\bf z},\rm max}=4$ &    &     &       \\
  $K$ & 118  & 266  & 446  \\ 
 $H$ & 34  & 130  &  290   \\
total & 152  & 396  &  736   \\
\hline \hline
\end{tabular}
\end{center}
\end{table}
As has been observed in Ref.~\cite{KaG92a} one can limit the number of
states in a useful manner. For fixed $l^K_{\bf z}$ and fixed $J$ and total isospin $T$ the number
of $K$ states is strictly finite once the NN interaction is assumed to act only up to $j^K_{\bf x}=j_{\rm max}$.
In the same way the number of $H$ states is strictly finite for fixed $l^H_{\bf z}$ and $j^K_{\bf x} \le j_{\rm max}$.
The number of channels for various choices of $j_{\rm max}$ and maximal values $l^{K/H}_{{\bf z},\rm max}$ 
is illustrated in Table~\ref{channels} for $J=0$ and $T=0$. It is evident that these numbers
grow quite quickly with increasing $j_{\rm max}$ and $l^{K/H}_{{\bf z},\rm max}$.

In momentum space the calculation proceeds along
similar lines as in coordinate space, namely by the determination of FY components. 
The principal difference lies in the fact that the integral equations
automatically take care of the asymptotic boundary conditions.  
On the other hand, in momentum space calculations, one has to work with $t$-matrices which contain singularities 
and which have very complex structures (see Ref.~\cite{Gl83}).
Results for the $^4$He binding energy via rigorous FY momentum space calculations have first been obtained
by Kamada and Gl\"ockle~\cite{GlK93a,KaG92a,KaG92b,GlK93b} in the early 1990s. In Table~\ref{KaG} we display a 
few of these results for what were realistic NN potential models at that time.
Though the number of channels is infinite
even if one restricts the interaction to the NN partial waves $^3s_1$-$^3d_1$ and $^1s_0$, one sees that for this 
case 15$K$ and $9H$ channels are sufficient to obtain a convergent result.
Considering the full interaction, however,
many more channels are needed as can be implicitly deduced from the last column in Table~\ref{KaG},
\begin{table}
\begin{center}
\caption{{\label{KaG}} $^4$He binding energy in MeV for various NN potentials (no Coulomb force) from FY 
calculations~\cite{GlK93a,KaG92b}. Results of columns 2$\,$-$\,$4 include
interaction only in NN channels $^3s_1$-$^3d_1$ and $^1s_0$, whereas for those
of column 5 the full NN potential is taken setting $l^{K/H}_{{\bf z},\rm max}=1$ and 
$j^{K/H}_{{\bf x},\rm max}=3$ (see also Table~\ref{channels}). }
\begin{tabular}{c||c|c|c|c}
\hline \hline NN Potential & \multicolumn{3}{c|} {only $^3s_1$-$^3d_1$ and $^1s_0$ interaction} &
  full potential \\   
Number of Channels &5$K$+5$H$  & 15$K$+9$H$  & 27$K$+9$H$  & 98$K$+90H \\
\hline
 Nijmegen~\cite{NaR78} & 24.16  & 24.55  & 24.53 & 25.03 \\
 Paris~\cite{LaL80}    & 23.20  & 23.63  & 23.60 & 24.26 \\
 AV14~\cite{WiS84}     & 23.36  & 23.77  & 23.75 & 24.62\\
\hline \hline
\end{tabular}
\end{center}
\end{table}

In more recent FY bound-state calculations the precision has been further increased and the truncation scheme  
changed somewhat. Ref.~\cite{NoK02} employed modern realistic charge dependent NN potential models
(plus additional 3NFs) and the following channel truncation scheme: $j^{K/H}_{\bf u} \le 6$ 
and $l^{K/H}_{\bf u} \le 8$, as well as $\sum_{\bf u} l^K_{\bf u} \le 14$ and $\sum_{\bf u} l^H_{\bf u} \le 14$.
The most sophisticated calculation included couplings to $T=1$ and $T=2$ isospin channels and
required a total of 4200$K$ and 2000$H$ channels.

For the scattering problem truncation schemes different from that of Table~\ref{channels} have
been used. In order to describe $n$-$^3$H and $p$-$^3$He scattering below the three-body break-up threshold
with a precision of about 1\%  it is necessary to include all channels with 
$j_{\bf x}\le4$, $j_{\bf y}\le4$, and $j_{\bf z}\le3$~\cite{ViD11} in a $j$-$j$ coupling scheme
(e.g. $j$-$j$ coupling for $K$ sets: 
$\{[(l^K_{\bf x} s^K_{\bf x}) j^K_{\bf x} (l^K_{\bf y} 1/2) j^K_{\bf y}] j^K_{\bf xy} (l^K_{\bf z} 1/2) j^K_{\bf z}\}J$).

As pointed out above the Yakubovsky scheme has been formulated for any number of particles, but 
this scheme has not yet been used in calculations for $A>4$ systems. A first more formal application for $A=6$ 
has been given in \cite{GlW11}, where also some hints for a numerical implementation are provided.

\subsection{The Alt-Grassberger-Sandhas (AGS) Formulation}\label{sec:AGS}

Starting from Faddeev and FY theory for treating the many-body scattering problem
Alt, Grassberger and Sandhas introduced a different set of equations, which today are known as {\it the AGS equations}
\cite{AlG67}.
However, apart from just working with a different set of equations, the main motivation 
of practitioners of the AGS formalism  
was to seek possible further reductions of the problem. In fact the assumption
of separable forms for the NN $t$-matrix in the resulting AGS subsystem amplitudes leads to
one-variable integral equations which are much simpler to solve than the Faddeev or FY equations.

We describe the AGS formalism by considering the 3N scattering problem for the (2+1) partition as introduced
in the first part of Section~\ref{sec:FY}.
The central point of the AGS formalism is the calculation of a transition amplitude defined in analogy to the two-body case
\be
\langle \Phi_\beta | U_{\beta \alpha} |\Phi_\alpha \rangle \equiv \langle \Phi_\beta | V^\beta | \Psi_\alpha^{(+)} \rangle \,,
\ee
where $\Phi_\alpha$, $V^\beta$, and $\Psi_\alpha^{(+)}$ are defined as  in Section~\ref{sec:FY}.
Considering first the case $U_{\alpha\alpha}$ one has
\be
\label{type_A1}
U_{\alpha \alpha} |\Phi_\alpha \rangle \,\,=\,\, V^\alpha | \Psi_\alpha^{(+)} \rangle 
\,\,=\,\, (V_\beta + V_\gamma)| \Psi_\alpha^{(+)} \rangle \,.
\ee
Using Eqs.~(\ref{Psi_2}) and ~(\ref{Psi_3}) for $|\Psi_\alpha^{(+)}\rangle$ one finds
\be
\label{type_A2}
U_{\alpha \alpha} |\Phi_\alpha \rangle \,\,=\,\, V_\beta G_\beta  V^\beta| \Psi_\alpha^{(+)} \rangle +
                    V_\gamma G_\gamma  V^\gamma| \Psi_\alpha^{(+)} \rangle 
\,\,=\,\, V_\beta G_\beta  U_{\beta\alpha} |\Phi_\alpha \rangle +
                    V_\gamma G_\gamma  U_{\gamma\alpha} |\Phi_\alpha \rangle \,.
\ee
In a similar way one obtains for $U_{\beta \alpha}$ with $\beta \ne \alpha$:
\be
\label{type_B}
U_{\beta \alpha} |\Phi_\alpha \rangle = V_\alpha |\Phi_\alpha \rangle + V_\alpha G_\alpha  U_{\alpha\alpha} |\Phi_\alpha \rangle +
                    V_\gamma G_\gamma  U_{\gamma\alpha} |\Phi_\alpha \rangle \,.
\ee
In Eqs.~(\ref{type_A1}-\ref{type_B}) $\alpha$, $\beta$, and $\gamma$ are all different from each other and can take the 
values  1, 2, and 3. One should note that for a given entrance channel $\alpha$ is fixed
resulting in one equation of type (\ref{type_A2}) and two equations of type 
(\ref{type_B}). Thus, as in the Faddeev case, one has  a set of three coupled equations
for $U_{\alpha \alpha}$, $U_{\beta \alpha}$, and $U_{\gamma \alpha}$. As for the Faddeev components, one may benefit from the fact that the nucleons 
(neglecting the electromagnetic interaction) can be considered identical particles,
and reduce the AGS equation set to a single equation. Note that the operators $U_{\beta\alpha}$ act in the full 
Hilbert space and carry information of the three-body break-up channel as well. 

In order to bring the AGS equation set to its standard form one uses
\be
V_\alpha |\Phi_\alpha \rangle = G_0^{-1} |\Phi_\alpha \rangle \,,\,\,\,\,\,\,\,
V_\alpha G_\alpha = t_\alpha G_0 \,,
\ee
where $t_\alpha$ is the NN $t$-matrix, however, embedded in the 3N-space. Combining  Eqs.~(\ref{type_A2}) 
and (\ref{type_B}) into a single equation one finds
\be
\label{3N_AGS0}
U_{\beta\alpha}|\Phi_\alpha \rangle = G_0^{-1} \overline{\delta}_{\alpha\beta}|\Phi_\alpha \rangle + 
\sum_\gamma\overline{\delta}_{\beta\gamma} t_\gamma G_0 U_{\gamma\alpha}|\Phi_\alpha \rangle \,,
\ee
where $\overline{\delta}_{\alpha\beta} = 1 - \delta_{\alpha\beta}$.
Therefore the standard AGS equation for $U_{\beta\alpha}$ reads: 
\be
\label{3N_AGS}
U_{\beta\alpha}= G_0^{-1} \overline{\delta}_{\alpha\beta} + 
\sum_\gamma\overline{\delta}_{\beta\gamma} t_\gamma G_0 U_{\gamma\alpha} \,.
\ee
As in the case of the Faddeev equations an incorporation of a short-range 3NF into the AGS equations
does not lead to serious complications (see e.g. Ref.~\cite{Gl83}). More difficult, however, is the incorporation 
of the long-range Coulomb force in momentum space, although practical high-precision solutions
turn out to be surprisingly simple \cite{DeF05} (instead of a Yukawa cut-off a smoothed step-function
cut-off is taken).

In order to discuss the AGS equations for the 4N system we first introduce some useful notation.
In analogy to the 3N case we encode the six possible pairs [(1,2), (1,3), (1,4), (2,3), (2,4), (3,4)] by
Greek letters. In addition we encode the sub-cluster partitions, 
which are four of the (3+1) type $[(ijk)l$, $(lij)k$, $(kli)j$, $(jkl)i]$ and three of the (2+2) type $[(ij)(kl)$, 
$(ik)(jl)$, $(il)(jk)]$, by capital Latin letters. 

The AGS formalism for the 4N system is reviewed in great detail in Ref.~\cite{Fo87}. 
There it is shown that the two-cluster sub-amplitudes $U_{\beta\alpha}^L$ satisfy the equation
\be
U_{\beta\alpha}^L = G_0^{-1} \overline{\delta}_{\alpha\beta} + \sum_\gamma
\overline{\delta}_{\beta\gamma} t_\gamma G_0 U_{\gamma\alpha}^L \,,
\ee
where pairs $\alpha$, $\beta$ and $\gamma$ must belong to one of the two sub-clusters defined by
partition $L$. Thus, in case of a (3+1) partition they must all belong to the three-body sub-cluster.
A comparison with Eq.~(\ref{3N_AGS}) for the 3N system, shows that $U_{\beta\alpha}^L$ has to be
understood as the AGS 3N $t$-matrix operator embedded in the 4N space. In the case where $L$ encodes
a (2+2) partition $U_{\beta\alpha}^L$ is the AGS $t$-matrix operator for the scattering of two noninteracting
pairs.

As in the 3N case,  in the 4N case an analogy with two-body scattering theory can be made by
defining a transition amplitude $U_{\beta\alpha}^{ML}$ such that the matrix elements of $U_{\beta\alpha}^{ML}$ 
give the appropriate two-cluster to two-cluster transition amplitudes for $L \rightarrow M$ scattering. 
As shown by Alt, Grassberger, and Sandhas~\cite{AlG67} one obtains the following set of equations
\be
\label{4N_AGS}
U_{\beta\alpha}^{ML} = (G_0 t_\beta G_0)^{-1} \overline{\delta}_{LM} \delta_{\alpha\beta} + \sum_{N\gamma}
\overline{\delta}_{MN} U_{\beta\gamma}^N G_0 t_\gamma G_0 U_{\gamma\alpha}^{NL} \,,\label{sandhas}
\ee
where pair $\alpha$ ($\beta$) must belong to one of the two sub-clusters defined by
partition $L$ ($M$). Note that, due to the definition of $U_{\beta\gamma}^N$, pair $\beta$ must also belong to one 
of the two sub-clusters defined by partition $N$, which, due to $\overline{\delta}_{MN}$, is different from 
partition $M$. Thus $M$ and $N$ cannot be both (2+2) partitions. Therefore one may split Eq.~(\ref{4N_AGS})
into two equations
\bea
\label{K'K}
U_{\beta\alpha}^{K'K} &=& (G_0 t_\beta G_0)^{-1} \overline{\delta}_{KK'} \delta_{\alpha\beta} 
+ \sum_{K''\gamma} \overline{\delta}_{K'K''} U_{\beta\gamma}^{K''} G_0 t_\gamma G_0 U_{\gamma\alpha}^{K''K} 
+ \sum_{H\gamma} U_{\beta\gamma}^H G_0 t_\gamma G_0 U_{\gamma\alpha}^{HK} \,, \\
\label{HK}
U_{\beta\alpha}^{HK} &=& (G_0 t_\beta G_0)^{-1}  \delta_{\alpha\beta} + \sum_{K''\gamma}
 U_{\beta\gamma}^{K''} G_0 t_\gamma G_0 U_{\gamma\alpha}^{K''K} \,,
\eea
where $H$ represents (2+2) partitions, while $K$, $K'$, and $K''$ denote (3+1) partitions. Obviously we have
set $L=K$ (initial channel represents a (3+1) partition). Of course the setting $L=H$ is also possible and leads 
for an initial (2+2) channel to a similar set of equations. Selecting a specific 
$U_{\beta\alpha}^{K'K}$ one obtains via Eq.~(\ref{K'K}) couplings to three other transition amplitudes
$U_{\gamma\alpha}^{K''K}$ and to one transition amplitudes $U_{\gamma\alpha}^{HK}$, which then lead to new couplings
via Eqs.~(\ref{K'K}) and (\ref{HK}). Iterating this process one finally arrives at 12 different equations of
type (\ref{K'K}) and 6 different equations of type (\ref{HK}) which together form a set of 18 
coupled equations.  In analogy to what happens for the FY components described in Section~\ref{sec:FY}, 
one has to solve 18 coupled equations for the two-cluster transition operator.

As for the FY case the 18 coupled equations can be reduced further for  identical particles. In fact, using permutation 
operators the equation set is brought down 
to two coupled equations with two transition amplitudes to be determined, one of the $U^{K'K}$ and one of the $U^{HK}$ type.
In case of an initial (2+2) partition state one ends up with two coupled equations of types 
$U^{KH}$ and $U^{H'H}$. 

In order to evaluate scattering observables one has to sandwich the transition operators of Eqs.~(\ref{K'K}) and~(\ref{HK}) 
between the respective components of the initial and
final state two-cluster wave functions. For the (3+1) partition the FY components are constructed from standard Faddeev 
components of the bound cluster, spin and isospin of the single nucleon, and a plane wave for the relative motion.
For the (2+2) partition the FY components are constructed from the deuteron wave functions.

Even though the AGS technique aims at a direct determination of the scattering amplitudes 
the formalism may also be used to calculate bound-state wave functions. In this case one has to solve a
system of 18 homogeneous equations for FY amplitudes,
\be
|\Psi_\alpha^L \rangle = \sum_{M\beta} G_0 t_\alpha G_0 U_{\alpha \beta}^L \overline{\delta}_{LM} |\Psi_\beta^M \rangle \,,
\ee
which again can be reduced to a system of two equations by considering identical nucleons.

The AGS equations do not depend on a specific representation, but they have been worked out
for momentum space calculations only. In this case the relevant equations are projected onto a complete set of
momentum states, which are expressed using Jacobi coordinates. In the 4N case this leads to three
momentum variables plus additional spin and isospin variables. Finally,  
projecting onto a specific state $|J^\pi\rangle$, one obtains an equation system with an infinite number of
channels in three continuous variables. As already discussed for the FY case of Section~\ref{sec:FY} one 
then has to truncate the set of equations appropriately while checking the convergence behavior
of the desired observable.

The reason for the preference of using momentum space for AGS calculations is due to the fact that powerful
approximations can be be incorporated quite easily.
As already pointed out in the beginning of this section
one can employ separable non-local NN potentials. Their use in Eq.~(\ref{sandhas}) 
leads to new equations for so-called subsystem amplitudes, 
which now depend on two rather than three continuous variables. 
Such an approach is usually labeled as [2V] in AGS calculations. In a [2V] calculation one remains with a problem similar to the 
Faddeev case for three particles interacting via local potential models. However, now the number of coupled equations equals  
eighteen rather than three. One can further reduce the problem by making an additional separable ansatz, namely
for the subamplitudes, which then leads to one continuous variable only (usually labeled as [1V]).
The separable ansatz for the subsystem amplitudes should in principle be brought to convergence by increasing
the number of terms, the so-called rank, of the expansion. This however leads to an increase
in the number of coupled channels to be considered. 

In principle it is not necessary to use separable expansions for the (2+2) subsystems.  The set of equations
(\ref{K'K}) and (\ref{HK}) can be reformulated such that the contribution of the (2+2) subsystems is calculated
exactly. This then leads to the so-called convolution method, which, as stated in Ref.~\cite{Fo87}, is particularly advantageous 
when used simultaneously with the [1V] formulation for the (3+1)
partitions, since in this case one has only to calculate
an extra integral over the singularities of a resolvent. Such a procedure is usually labeled as a [1V+C] calculation.

The AGS [1V+C] approach had been considered to be quite promising. In fact the first 4N calculation with a full consideration
of the tensor force in the $^3s_1$-$^3d_1$ NN partial wave has been carried out in this way~\cite{Fo89}.
Various versions of Yamaguchi-like potentials ``Y$n$'' for $^1s_0$ and $^3s_1$-$^3d_1$ NN partial 
waves have been used, while the 3N subsystem has been restricted to the states $j^\pi=1/2^+,\, 3/2^+$. Various 
points have been investigated, among them the convergence concerning the separable ansatz of the (3+1) 
subsystem amplitudes, which we report in Table~\ref{AGS_rank}. One sees that a rather good convergence 
is obtained with rank $N=4$. A comparison to results of [2V] calculations~\cite{GiL76,GiL78}, where the tensor force 
had been taken into account only perturbatively via the so-called $t_{00}$ approximation, shows a tensor force effect of
up to about 0.5 MeV. Later when the same potential models were considered in the above 
mentioned FY calculation~\cite{KaG92a}, only small differences of the order of about 0.2 MeV were observed.
However these calculations did not employ exactly the same channels.
\begin{table}
\begin{center}
\caption{{\label{AGS_rank}} $^4$He binding energy in MeV for NN potential model Y4~\cite{Fo89} obtained in the AGS 
calculations of Ref.~\cite{Fo89} for various ranks $N$ in the separable ansatz of the 
(3+1) subsystem amplitudes (only $j^\pi=1/2^+$ for the 3N subsystem is taken into account).  }
\begin{tabular}{c||c|c|c|c}
\hline \hline 
rank $N$ & $N$=1 & $N$=2 & $N$=3 & $N$=4 \\
\hline
         & 32.370 & 32.368 & 32.675 & 32.666\\
\hline \hline
\end{tabular}
\end{center}
\end{table}

Although modern realistic NN potential models are not of separable form one may use a separable 
representation of the NN $t$-matrix. Usually rank 1 representations are preferred in order
to prevent an increase of the number of channels. As is discussed in Section~\ref{sec:4_had} 
it has been found in the last
decade that such rank 1 representations are not sufficiently precise for all partial waves of the NN $t$-matrix.

\subsection{The Variational Formulation}\label{sec:VAR}

A large variety of approaches to the many-body quantum mechanical problem make use of the Rayleigh-Ritz 
variational theorem~\cite{Ra870,Ri909}.
This theorem is applicable if the solution of a useful equation renders stationary
some proper functional. In particular, one can show that the solution of the
Schr\"odinger equation (for a normalizable state) renders stationary the Rayleigh quotient for H, i.e. the energy
functional
\begin{equation}
 E[\Psi]=\frac{\langle\Psi|H|\Psi\rangle}{\langle\Psi|\Psi\rangle}\,
\end{equation}

A lemma that complements the fundamental variational theorem states that the value of the energy functional 
with any trial function is always greater than the ground-state energy and equal to that, when the trial function 
coincides with the exact ground-state wave function.
Therefore  if one wants to calculate the energy of a system the equation to solve is  
 \begin{equation}
 \delta E[\Psi]=0\,,\label{vareq}
\end{equation}
that for the ground-state energy corresponds to a minimization problem.

Numerous approaches use this variational principle to find the bound-state energy of a many-body system.
In order to maximize the efficiency of
the approach the trial function must have a parametrized functional form that is both convenient and suitable to the problem 
to be solved. With it one calculates the energy functional and searches for the parameters that minimize it.

The different approaches that work with Eq.~(\ref{vareq}) can be divided 
in two groups according to the criteria followed for the choice of the trial function. To the first group belong 
those approaches where the trial function $\Psi_T$ is written as an expansion on a complete (over-complete) set of square
integrable functions $\phi_n$ that respect the symmetries of the Hamiltonian:
\be
\label{Psi_trial}
|\Psi_T(N)\rangle = \sum_{n=1}^N c_n |\phi_n\rangle\,.\label{trial}
\ee
In this way the minimization procedure 
corresponds to finding the solution of a (generalized) eigenvalue problem 
\be
(H-E S)c=0 \,,\label{secular}
\ee
where $H$ and $S$ are $N \times N$ Hermitian matrices of the Hamiltonian ($H_{nm} = \langle \phi_n|  H |\phi_m\rangle$) 
and overlap integrals of the basis functions ($S_{nm} = \langle \phi_n|\phi_m\rangle$),
while $c$ is a $N$-component vector of the linear parameters $c_n$.
With growing $N$ the size of the Hamiltonian matrix 
represented on the chosen basis increases and the true ground-state energy is approached from above. 
In principle the true result would be obtained only for an infinite number of basis functions.
The basis can be complete as the hyperspherical harmonics (HH) or the harmonic oscillator (HO) basis,  or over-complete 
as e.g. in the method described in Section~\ref{sec:GEM}. 
In some cases the trial function is constructed on a basis that does not respect the symmetry of the Hamiltonian, 
and the latter is restored after diagonalization. 
In these cases  the convergence is considerably slower than in the properly symmetrized cases. 
In general, however, one has to consider the trade-off between the much larger basis and 
the complications in symmetrizing the basis functions (see for example the non-symmetrized HH 
approach of Section~\ref{sec:HH}
or the Fermionic molecular dynamics technique of Section~\ref{sec:FMD}).

To the second group belong those approaches where the trial function is chosen according  to more physical considerations. 
For example it can be suggested by a cluster picture of the system like in the resonating group method
(RGM, see Section~\ref{sec:RGM}), or by the form of the potential as in the variational Monte Carlo (VMC) technique
described in Section~\ref{sec:VMC}. In the stochastic variational method (SVM, see Section~\ref{sec:SVM})
the variational procedure does not proceed systematically 
by the diagonalization of a larger and larger Hamiltonian matrix,  
but in a stochastic way (trial and error).

It is clear that the approaches belonging to the first group have the advantage of being more
controlled, in fact the increase of the basis is determined by the growing values of specific quantum numbers 
and in case that convergence for the bound-state energy is reached, one can 
consider it as a precise ab initio result. 

The variational approach also allows one to obtain  a wave function
that can be used to calculate bound-state observables 
other than the energy. However, one has to remember
that the difference between the exact value of the energy and that obtained with the trial function $\Psi_T$
which minimizes the energy functional 
is an infinitesimal of higher order than the difference between the true wave function and $\Psi_T$.

A comment concerning excited states is in order here. While with Faddeev and AGS techniques
the calculation of an excited bound state follows the same line as that of a ground state, i.e. the corresponding
equations have to be fulfilled for the energy of the excited state, the situation is different for the variational
approach.
By diagonalizing the Hamiltonian matrix on a given finite basis of square integrable functions one gets a spectrum of $N$
eigenstates $\Psi_i$ with eigenvalues $E_i$. If the trial function carries the quantum numbers of the ground state
of the system, then, the lowest $E_i$ is the ground-state energy and the ground-state wave function is described by
the corresponding eigenstate $\Psi_i$. The other solutions correspond to excited bound states of the system that
carry the same quantum numbers as the ground state. A converged calculation for the energy of such excited states 
often requires
more basis states than a ground-state calculation. In most cases, however, excited bound states have
quantum numbers different from those of the ground state. For a calculation of such states one 
has to implement the proper quantum numbers into the basis set 
$\Phi_n$ when solving the eigenvalue problem.

For few-nucleon systems only very few, if any, of the excited states obtained by diagonalization, 
 actually
correspond to bound states; most of them lie in the continuum ($E_i>E_{\rm th}$).
Since the basis set $\Phi_n$ consists 
of square integrable
functions the $\Psi_i$ do not have the proper continuum boundary conditions. 
In fact proceeding in such a way one obtains a 
fake discretization of the continuum.

A variational principle similar to that for bound states has been developed by Kohn~\cite{Ko48}
for the phase shifts and elements of the scattering matrix in nuclear collisions. 
Following the original reference the principle can easily be illustrated for the one-dimensional problem.
Considering the collision in $s$-wave of two particles of mass $m$ interacting by a short-range potential with range $R_0$, 
one has to solve the following equation
\begin{equation}
 \bigg[{\frac {d^2}{dr^2}} + k^2 - m V_{12}(r)\bigg] u(r) = 0 \,, 
\end{equation}
where $k$ is the wave number of the relative motion and $V_{12}(r)$ is the interaction potential. The radial 
wave function $u(x)$ fulfills the regularity condition $u(0)=0$, while for $x \gg R_0$ one has
\be
u(r) = A\sin(kr) + B \cos(kr) 
     = A[\sin(kr)+\tan\delta \cos(kr)] 
     = (A^2 + B^2)^{1/2} \sin(kr+\delta)\,,
\ee
where $A$ and $B$ are some constants and the phase
shift $\delta$ is defined by $\tan \delta =B/A$.
If 
$L = u^\prime(a)/u(a)$
indicates the logarithmic derivative of $u(x)$ at $x=a \gg R_0$, then $u$ satisfies the
equation
\begin{equation}
 \int\limits_0^a \,\bigg[-\left({\frac {du(r)}{dr}}\right)^2 + k^2 u^2(r)- m V_{12}(x)u^2(r)\bigg] dr + L  
 u^2(a)=0 \,.
\label{kohn(2.5)}
\end{equation}
For the binding energy the basis for the variational principle lies in the following: for a given value of $L$, 
$k^2$, as calculated from Eq.~(\ref{kohn(2.5)}), is stationary. 
The crucial point is that  one can equally well regard Eq.~(\ref{kohn(2.5)}) 
as providing a stationary expression for $L$ for a given value of $k^2$. Consequently, 
this is the  variational approach more natural in 
collision problems, where the energy of the system is fixed. Having determined $L$ from its 
stationarization one obtains the phase shift $\delta$ from the relation $L= k\cot(ka+\delta)$.
For the general case of a two-fragment scattering the formulation of the variational approach just described 
for $s$-wave scattering in one dimension may appear more involved, but the essence is the same.
(In the literature one also finds different versions of Kohn's original formulation, see e.g. Ref.~\cite{De63}.)

In principle, both  
bound-state and two-fragment scattering calculations (e.g, elastic scattering of $n\,$-$^3$H or $d\,$-$^4$He)
could be tackled 
with any of the various variational approaches discussed in the next subsections. However, scattering 
calculations have been 
carried out only for some of the approaches.

In the following we provide a  description of various present-day variational approaches.
The discussion of two of them will be, however, postponed to the section devoted to the Monte Carlo
technique (see Sections~\ref{sec:MCSMD} and ~\ref{sec:VMC}).

\subsubsection{The Hyperspherical Harmonics (HH) Method}\label{sec:HH}

The HH method is a variational method where the trial function is written as
an expansion on the HH basis. 
The {\it hyper}-spherical harmonics are the $A$-body generalization of the
spherical harmonics $Y_{lm}$.
After Borel~\cite{Bo14} who first employed hyperspherical coordinates in
physical problems, it was Gronwall~\cite{Gr37} who performed an expansion of the Helium 
atom wave function in HH (though not referring to them as such). The first application
in nuclear physics is due to a Ph.D. student of Schwinger, R.E. Clapp who calculated in
his thesis the $^3$H ground state with central and tensor forces~\cite{Cl49}.

The HH  basis is intrinsically translational invariant. In fact the functions are
expressed in terms of the hyperspherical coordinates which are defined by a transformation 
of the $N=(A-1)$ Jacobi vectors ($\veta_1,...,\veta_N$). 
Unaffected by the transformation are the $2N$ polar
angles $\theta_i$ and $\phi_i$ of the Jacobi vectors ($\hat\eta_i=(\theta_i,\phi_i))$. The remaining 
$N$  hyperspherical coordinates, the {\it hyper}-radius $\rho$ and $(N-1)$ 
{\it hyper}-angles $\alpha_n$, contain information from at least two Jacobi 
vectors as can be seen from their definition:
\be
\rho^2_n  = \sum_{i=1}^n \veta_i^{\,2} \,; \,\,\,\,\,\,\,\,\,\,\,\,\,\,
\sin\alpha_n= {\frac {|\veta_n|}{\rho_n}} \,, \,\,\, n=2,...,N \,;
   \,\,\,\,\,\,\,\,\,\,\,\,\,\,  \rho \equiv \rho^2_N  \,.
\ee
Expressed in hyperspherical coordinates the $A$-body kinetic energy operator is a sum of two terms
 \begin{equation}\label{Trho}
  T =  T_\rho + T_K(\rho)\,,\,\,\,\, {\rm with} \qquad T_\rho= - \frac{1}{2m}\Delta_{\rho} \,, \qquad 
T_K(\rho)=\frac{1}{2 m} \frac{\hat{\bf K}_N^2}{\rho^2}\,,
\end{equation} 
where $T_\rho$ and $T_K(\rho)$  are given for a system of identical particles of mass $m$ and using mass weighted 
Jacobi coordinates~\cite{HiD56}.
It is evident that this form, with a $\rho$-dependent Laplacian and a {\it hyper}-centrifugal barrier,
is in perfect analogy to the three-dimensional case.

The {\it hyper}-angular momentum operator $\hat{\bf K}_N$, which depends on all the $(3N-1)$ angles (denoted by 
$\hat\Omega_{N}$), is rather complicated so we refrain from giving an explicit expression here.
The role of $\hat{\bf K}_N$ in an $A$-particle system is an extension of the role of the relative angular 
momentum operator ${\bf l}_1$ in a two-body system. The HH basis set 
${\cal Y}_{[K_N]}(\hat\Omega_{N})$ consists of the  eigenfunctions of  $\hat{\bf K}_N^2$
\begin{equation}\label{K2}
\hat{\bf K}_N^2\,{\cal Y}_{[K_N]}(\hat\Omega_{N})= K_N(K_N+3N-2){\cal Y}_{[K_N]}(\hat\Omega_{N})\,,
\label{eigeneqK2}
\end{equation} 
where the eigenvalues are expressed in terms of the quantum number $K_N$ and 
the subscript $[K_N]$ stands for the total of $(3N-1)$ quantum numbers; for $N=1$ one easily makes the following identifications:
$\hat{\bf K}_{1}^2 \equiv {\bf l}^{\,2}_1$, $K_{1} \equiv l_1$, ${\cal Y}_{[K_1]} \equiv Y_{l_1m_1}$.

Besides the ${\cal Y}_{[K_N]}(\hat\Omega_{N})$ one needs also basis functions for the hyperradial part of the wave function.
Usually one takes expansions in Laguerre polynomials.

For a better understanding of  the HH and their quantum numbers it is instructive to construct them recursively, 
by starting from the $N=1$ case discussed above and adding one particle at a time. 
Following the original paper~\cite{ZeB35} one realizes that the set 
of quantum numbers characterizing these functions is indeed a set of $(3N-1)$ quantum numbers, which consists of: 
{{\bf (i)} the angular momentum quantum numbers  $l_1,l_2...l_N$ relative to the $N$ Jacobi coordinates, 
{\bf (ii)} the corresponding third components $m_i$,
and {\bf (iii)} the $(N-1)$ quantum numbers $n_2,n_3...n_N$ associated with the degree of the Jacobi polynomials 
that enter the solutions of Eq.~(\ref{eigeneqK2}). 

A useful method for constructing the HH is the so-called ''tree`` method proposed by Vilenkin~\cite{ViK65}. 
A didactic presentation of this method can be found in~\cite{Ba97}. 
Following the tree method, the set of $(3N-1)$ quantum numbers characterizing these functions consist of: 
{\bf (i)} the hyperangular quantum number $K_N$, 
{\bf (ii)} the total angular momentum quantum number $L_N$ and the corresponding third component  $M_N$,
{\bf (iii)} the angular momentum quantum numbers $l_1,l_2,...,l_N$ relative to the $N$ Jacobi coordinates, 
{\bf (iv)} the total angular momentum quantum numbers of the $n$-body subsystems $L_{N-1},L_{N-2}...L_2$ 
(note that $L_1$ is missing because it coincides with $l_1$), and 
{\bf (v)} the hyperangular quantum numbers $K_{N-1},K_{N-2}...K_2$ of the $n$-body
subsystems. Note that the following relation holds: 
$K_N=2n_N+K_{N-1} + l_N$.

Constructing the basis functions in such a recursive way makes the evaluation
of any two-body operator easy, since only the matrix element of the two-body
operator between the last two particles needs to be calculated, if the wave functions are properly symmetrized.

The symmetrization of the HH is one of the two main difficulties of the method.
In fact in general the HH basis states possess no special symmetry under particle
permutation. For the 3N and 4N case this is a minor problem since a direct
symmetrization of the wave function is still feasible (see e.g.~\cite{Ef79}). For larger 
systems, it becomes impractical and there is a 
need for  more sophisticated symmetrization methods.
One approach to the HH symmetrization problem consists in constructing  recursively the HH by
realizing irreducible representations not only of the orthogonal group  $O(3N)$ but also of the group $O(N)$ 
($O(3N)\supset O(3)\otimes O(N)$). The recursive construction makes it possible to label the HH  
according to the group--subgroup canonical chain $O(N)\supset O(N-1)\supset...O(2)$~\cite{GeZ50}. 
This is expedient, since
$O(N)$ is the group of the transformations from one set
of Jacobi coordinates to another set, and therefore has the $A$-particle permutation group $S_A$ as subgroup
($O(3)\otimes O(N)\supset O(3)\otimes S_A$).

The classification of the HH with respect to appropriate group--subgroups allows a  model space identified by
$K_N$ to be split into smaller subspaces~\cite{BaN97}. 
An efficient technique has been developed by Barnea~\cite{Ba97} to solve the reduction problem from 
$O(N)$ to $S_A$. It is the application of this technique that has
for the first time allowed a HH ab initio calculation for more than four fermions~\cite{BaL99}.

Another interesting non recursive procedure that automatically guarantees the antisymmetrization 
of the many-body wave function has been suggested in~\cite{Ti02}. There the HH are expanded on
the Slater determinant basis of the HO shell model (SM). The expansion coefficients can be
found by diagonalizing the angular part of the multidimensional Laplacian presented in individual
nucleon coordinates~\cite{Ti08}. The applicability of the
proposed method has been in principle demonstrated for the case of the $^{3-7}$H and $^{4-10}$He
isotopes~\cite{Ti02} and for bosonic systems~\cite{Ti08} by using simple central interactions. 
However further improvements are necessary to obtain convergence.

More recently an approach has been proposed, where the basis is not symmetrized~\cite{GaK11}. 
The physical states are identified by studying the symmetries present in the 
eigenvectors obtained from diagonalization of the Hamiltonian matrix. In principle the drawback of 
such an approach is the large degeneration of the basis. However, this problem is circumvented 
by expressing the Hamiltonian matrix as a sum of products of sparse matrices. This particular representation
of the Hamiltonian is well suited for a numerical iterative diagonalization, where only the action
of the matrix on a vector is needed. Up to now this method has been applied to six nucleons with
a simple  central potential.

The second difficulty in the application of the HH approach is common to all diagonalization methods and concerns convergence.
In fact for potentials with strong short-range repulsion the
convergence rate of the HH basis turns out to be very slow. 
Two methods  have been suggested in order to improve the convergence rate. One of them couples the HH expansion with 
the construction of an effective interaction (EIHH) and will be explained in Section~\ref{sec:EIHH}.
The other one is the so-called correlated hyperspherical harmonics (CHH) method first used by Fenin 
and Efros~\cite{FeE72}. 
{\it Correlation} functions $f(r_{ij})$ are introduced that
{\it correlate} the HH basis according to the characteristics 
of the potential, 
\begin{equation}
{\cal Y}_{[K]}(\hat\Omega_{N})\rightarrow \prod_{i,j} f(r_{ij}) {\cal Y}_{[K]}(\hat\Omega_{N})\,,
\end{equation}
where $r_{ij}$ is
the relative distance of particles $i$ and $j$. The function $f(r)$ can be
obtained from the solution of the two-body Schr\"odinger equation, since at
short distances the role of other particles is rather unimportant.
Though such correlations lead
to a considerable improvement~\cite{FeE72} one still needs in general a rather  
large number of  HH terms in order to reach convergence. The convergence can be 
improved considerably if one introduces state dependent (see Ref.~\cite{EfL07}) or 
longer range correlations~\cite{RoK92}. 
However, the two-body long-range correlations are less under control, because
correlations among more particles become important.
An interesting HH reorganization was made in Ref.~\cite{Ti07}, where HH have been constructed that allow
a better treatment of long-range correlations typical for halo nuclei.
In a toy model of
a loosely bound $^5$He nucleus the clusterization aspect of the system
dynamics could be described.
Unfortunately this approach has not been applied to real nuclear systems.
However, an extension of this idea
to nuclei with two loosely bound valence nucleons,
and in particular to Borromean nuclei, would probably make it possible to achieve a proper
three-body description of such systems at large distances.

The insertion of correlations complicates the calculation of the Hamiltonian matrix elements.
In fact the orthogonality of the HH is destroyed and one has to solve a generalized eigenvalue problem 
involving the metric matrix. This implies fully $3A-3$-dimensional integrals, which can be reduced
to $3A-6$-dimensional integrals using the technique described in~\cite{Ef02}.
Therefore it can be more convenient to use the pair-correlation ansatz with an expansion of the wave function
in Faddeev components
\be
\Psi(\veta_1,...,\veta_N) = \sum_{i<j} f(r_{ij}) \psi_{ij}(\veta_1,...,\veta_N) \,,
\ee 
(where we have dropped again spin and isospin degrees of freedom).
Such a Faddeev approach leads to at most 12-dimensional integrals. 
A calculation along this line has been performed in Ref.~\cite{BaL99}, extending the  integro-differential equation 
approach (IDEA)~\cite{Fa84} 
that limits the expansion to HH of lowest order. 

The first HH calculations for the $^4$He ground state with realistic potential models and with sufficiently 
convergent results have been carried
out by Viviani, Kievsky and Rosati~\cite{ViK95}. In Table~\ref{HH_pisa} we show results 
for the binding energy with the AV14 NN potential~\cite{WiS84}
from their first calculation. Inspecting the table one sees that the result is not yet fully converged. In 
following studies it turned out to be more efficient to work with HH instead with CHH expansions. In Ref.~\cite{Vi98} 
an improved result of 24.18 MeV has been found in a calculation with about 2500 uncorrelated HH channels. This 
contrasts the about 100 channels, which have
been employed for the CHH result with $K_{\rm max}=8$ of Table~\ref{HH_pisa}. In fact for CHH calculations 
much less channels are necessary, while on the other hand one has to 
cope with the loss of the orthogonality of the basis functions. Therefore it cannot be said in general which
of the two methods is advantageous for a specific problem. We should mention that a fully converged HH result
of 24.23 MeV for the  $^4$He binding energy with the AV14 potential is given in Ref.~\cite{KiR08}.
\begin{table}
\begin{center}
\caption{{\label{HH_pisa}} $^4$He binding energy in MeV as function of the maximal value of the grand-angular
quantum number $K$ from a CHH calculation with the AV14 NN potential 
(with Coulomb force)~\cite{ViK95}. }
\begin{tabular}{c||c|c|c|c}
\hline \hline 
$K_{\rm max}$  & 2 & 4 & 6 & 8 \\
\hline
               & 20.60 & 23.12 & 23.71 & 23.85 \\
\hline \hline
\end{tabular}
\end{center}
\end{table}

It is an important fact that the HH expansion is not restricted to local potential models, but can also be applied to
non-local forces~\cite{ViM06}.

\subsubsection{The Gaussian Expansion Method (GEM)}\label{sec:GEM}

The GEM is a variational approach first introduced to study muonic molecules in 1988~\cite{Ka88}. Shortly after,
it was applied to calculate the 3N and 4N nuclear binding energies for a simple potential~\cite{KaK90}.

The trial function is expanded on an over-complete basis.
In order to account for the components generated by the various possible arrangements
of the particles that form the system under investigation, one allows all possible choices of different Jacobi 
coordinates. In this connection one has to distinguish between the cases of non-identical and identical
particles. In the latter case there is no need to employ the Jacobi sets that may be obtained from each other
via permutations of particles. Therefore it is sufficient to use one single set for $A=3$ and for $A=4$ two sets,
for example those which were described in Section~\ref{sec:FY}. 

The GEM wave function for a nuclear state $|J^\pi\rangle$ has the following form~\cite{Ka88}: 
\begin{equation}
|J^\pi\rangle =\sum_{\gamma,\{\lambda N\}} c_{\gamma}^{\{\lambda N\}} \,
{\cal A}\, |\Phi_{\gamma}^{\{\lambda N\}}\rangle\,,
\end{equation}
where the antisymmetrizer is denoted by ${\cal A}$, $\gamma$ enumerates the different Jacobi sets, 
$\{\lambda\}$ represents the angular, spin and isospin quantum numbers of the basis functions $|\Phi_{\gamma}^{\{\lambda N\}}\rangle$,
and the set $\{N\}$ encodes information about the radial basis functions (see below). 
For a better illustration of the ansatz we specialize to the wave function $|J^\pi\rangle$ of the 4N system:

\be
 |J^\pi\rangle =
\sum_{\{\lambda N\}}  c_{K}^{\{\lambda N\}}\, {\cal A}\, |\Phi_{K}^{\{\lambda N\}}\rangle +  
\sum_{\{\lambda N\}}  c_{H}^{\{\lambda N\}}\, {\cal A}\, |\Phi_{H}^{\{\lambda N\}}\rangle \;,  
\ee
where $K$ and $H$ stand for the above mentioned Jacobi coordinate sets. The basis functions are described for 
the $K$ and $H$ sets by 
\bea
\langle{\x}_{12}^K,{\y}_{12,3}^K,{\z}_{123,4}^K|\Phi_K^{\{\lambda N \}}\rangle &=&
 \phi_{n_1}^{\{\lambda \}}\!(x_{12}^K) \,\phi_{n_2}^{\{\lambda \}}\!(y_{12,3}^K) \, 
 \phi_{n_3}^{\{\lambda \}}\!(z_{123,4}^K) 
 \, F_K^{\{\lambda \}}\!(\hat x_{12}^K, \hat y_{12,3}^K, \hat z_{123,4}^K) \\ 
\langle{\x}_{12}^H,{\y}_{34}^H,{\z}_{12,34}^H|\Phi_H^{\{\lambda N\}}\rangle &=&
 \phi_{n_1}^{\{\lambda \}}\!(x_{12}^H) \,\phi_{n_2}^{\{\lambda \}}\!(y_{34}^H)\, 
 \phi_{n_3}^{\{\lambda \}}\!(z_{12,34}^H) 
 \, F_H^{\{\lambda \}}\!(\hat x_{12}^H, \hat y_{34}^H, \hat z_{12,34}^K) \,,
\eea
where we have suppressed the spin and isospin degrees of freedom in the functions $F_{H/K}^{\{\lambda \}}$.
The Jacobi coordinates used in the two equations above are defined in Eqs.~(\ref{K_set}) and (\ref{H_set})
and the quantum number set 
$\{\lambda\}$  indicates a specific channel (due to the different
Jacobi coordinates the quantum numbers for $K$ and $H$ sets are different). The radial functions
are given by
\be
\phi^{\{\lambda\}}_n(r) = r^l \, e^{-(r/r_n)^2} \,,
\ee 
where $l$ belongs to the set of quantum numbers  $\{\lambda\}$.
Furthermore we have the following equality $\{N\}\equiv\{n_1,n_2,n_3\}$.
The Gaussian range parameters are chosen to lie in a
geometrical progression ($ r_n=r_1 a^{n-1}, \, n=1...n_{\rm max})$.
Choosing  the range parameters in this way allows one to 
describe both the short-range correlations and the long-range
asymptotic behavior precisely.
The use of Gaussians allows the analytical calculation of the Hamiltonian matrix elements.

In the GEM approach one proceeds systematically enlarging the basis by increasing the values
of the different quantum numbers until satisfactory convergence is obtained.
In other words the number of channels $\lambda$ has to be increased
to attain sufficient convergence. 
Of course also the various $n_i$ have to be chosen sufficiently large. 

One notes that the GEM calculation exhibits some similarities
to the FY bound-state calculation described in Section~\ref{sec:FY}. 
In fact, $H$- and $K$-type Jacobi coordinates 
and the various quantum number sets ${\{\lambda\}}$ are actually identical, the difference being in the calculation of
the radial parts of the wave function. In the GEM case the radial parts are obtained from a variational calculation,
where the basis consists in a product ansatz of radial functions of single Jacobi coordinates,
as opposed to the FY approach, where the radial parts are determined by a numerical
solution of the FY equations.

\subsubsection{The Stochastic Variational Method (SVM)}\label{sec:SVM}

The SVM is an approach that was first introduced in 1977 by Kukulin and Krasnopol'sk
to calculate the three- and four-body nuclear binding energies for  simple potential models~\cite{KuK77}.
About twenty years later the first calculations with realistic 
potentials were carried out by Varga and Suzuki~\cite{VaS95}.

The strategy of this variational method is somewhat different from the GEM method discussed in 
Section~\ref{sec:GEM}, although the trial function is here also expanded in Gaussian functions.
For the SVM approach {\it correlated Gaussians} are used. That is, the arguments of the
Gaussians  are not simply in terms of single Jacobi vectors, 
but in addition products of different Jacobi vectors. Particular emphasis is given to the 
correlation that must be flexible enough to describe 
the full variety of correlations between the nucleons, e.g., the short-range correlation
due to the strong repulsive force, the clustering typical in some light nuclei, or the long-range 
correlation in light halo nuclei. Correlated Gaussians imply in general many non-linear parameters.
Then again the variational 
basis is nonorthogonal and over-complete, i.e. none of the components is indispensable, and
one can replace a component by a linear combination of others. This fact actually gives an excellent
opportunity to use a stochastic optimization procedure. 
In fact, for a large Gaussian basis the hypersurface of the energy function 
is generally too complicated
to locate an absolute minimum of the energy with respect to the Gaussian parameters. 
However, it is usually enough to find 
a sufficiently low point in energy. In an actual calculation one has to decide whether one wants to spend 
much computational time in a search for the global
minimum of a smaller basis set or, as an alternative, to add more basis functions in order to lower the energy
this way. Finding the right balance between both choices is quite a difficult task.

In the SVM the wave function of the $A$-nucleon system $|J^\pi\rangle$ expanded in correlated Gaussians reads as follows
\be
\label{SVM}
\langle \veta_1,...,\veta_{A-1}|J^\pi\rangle = \sum_{\{\lambda \}k} c_k^{\{\lambda \}}
{\cal A} \{e^{-\frac{1}{2}{\xx}^\dagger B_k^{\{\lambda \}} {\xx}} F_k^{\{\lambda \}}(\hat\eta_1.....\hat\eta_{A-1})\} \,,
\ee
where spin and isospin degrees of freedom are dropped, $\xx$ stands for the set of $(A-1)$ Jacobi vectors 
$\veta_i$,
while ${\{\lambda \}}$ encodes the quantum numbers of the various channels. In practice the index $n$ in Eq.~(\ref{trial}) 
corresponds here to the set $(\{\lambda \},k)$.

The matrix $B_k$ has the dimension $(A-1) \times (A-1)$ and is a positive definite
symmetric matrix of non-linear parameters $B_k\equiv B^k_{ij}$. The explicit form of the multiplication of vectors $\xx$ and 
matrix $B_k$ is defined by
\begin{equation}
\xx^\dagger B_k \xx=\sum_{ij}^{A-1} B^k_{ij} \, \veta_i\cdot \veta_j  \,.
\end{equation}
Note that the trial wave function is not written as a separate product of angular and radial parts. Even though there
is an expansion on angular momenta (contained in the set $\{\lambda\}$), the angular part of the trial function is not
exhausted by this expansion, since there are also angular correlations in the Gaussians. This means that the angular basis
is not a complete basis, but an over-complete one. 

As already pointed out it is not a simple task to 
optimize this large number of parameters towards a minimum.
Fortunately here one works with an over-complete basis and thus one may use a stochastic approach. 
The stochastic
procedure of the SVM is organized in three steps:
{\bf (i)} one generates a trial function by choosing the parameters $B_{ij}$ and the channels  randomly; 
{\bf (ii)} one judges its
utility by the energy gained by including it in the basis, and
either keeps or discards it; {\bf (iii)} one repeats this ''trial and error``
procedure until the basis set leads to convergence.

For a better search of the minimum  after a certain number of new basis functions are accepted,
one may make a {\it cyclic optimization}, 
where one optimizes
the entire basis set by tuning the parameters of only one function at a time~\cite{SuV98}. To optimize the entire basis set 
in such a way is numerically much cheaper than a simultaneous optimization
of all the Gaussian parameters of all basis functions.

In Fig.~\ref{figure_SVM_T}, a result reported from Ref.~\cite{VaS95} illustrates an example
of the convergence rate.
The $^6$Li binding energy with the Volkov potential~\cite{Vo65}
has been calculated  several times starting
from different random points and two of these calculations are shown in the figure. 
The energies obtained from different random paths approach each other after a few initial steps
and converge to the final
energy. The energy difference between two random paths as
well as the tangent of the curves gives  some information on
the accuracy of the method with a given size of the basis.
\begin{figure}[tb]
\begin{center}
\epsfig{file=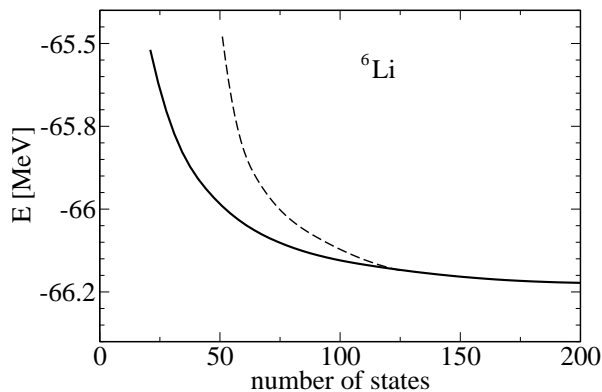,scale=0.3,angle=-90}
\caption{Results of a SVM calculation~\cite{VaS95} for the convergence pattern of the $^6$Li ground-state energy 
with the Volkov potential~\cite{Vo65} on two different random paths.}
\label{figure_SVM_T}
\end{center}
\end{figure}
Besides the large number of non-linear parameters, the
treatment of the increasing number of channels in the
expansion of the wave function also poses a formidable
task. Therefore one can use an alternative approach to 
cope with this problem. Instead of using the partial wave expansion, the angular dependence of the wave function is represented
by a single solid spherical harmonics whose argument
contains additional variational parameters
\begin{equation}
\theta_{LM_L}(\xx) = 
v^{2K+L}Y_{LM_L}(\hat{\bf v})
\end{equation}
with ${\bf v}= \sum_1^{N-1} u_i \xx_i.$
Only the total orbital angular momentum appears in this expression
and the parameters $u_i$ (as well as $K$) may be considered as 
variational parameters.
The factor $v^{2K+L}$ plays an important role in improving the
short-range behavior of the wave function. This form makes
the calculation of the matrix elements for nonzero orbital
angular momentum much simpler.

A substantial improvement in the optimization 
procedure of the stochastic variational approach has been made by the analytic 
gradient optimization method (AGOM)~\cite{BuA06,BuA08}.
In order to speed up the search of the minimum, in the AGOM one calculates 
the energy gradient with respect to the Gaussian parameters in an analytical way.
By indicating with $\alpha_t$ one of the many  non-linear parameters  on which the basis function
$\phi_t$ depends, and by elaborating on the differential of the secular equation~(\ref{secular}), in Ref.~\cite{BuA08} 
it is shown that
the derivative of the total energy with respect  $\alpha_t$ is given by
\be
{\frac {\partial E} { \partial \alpha_t} } = 2 Re \bigg[ c_t^* \sum_{n=1}^N c_n \bigg( {\frac {\partial H_{tn}} { \partial \alpha_t} } 
                                                      - E  {\frac {\partial S_{tn}} { \partial \alpha_t} } \bigg) \bigg]
- c_t c_t^* \bigg( {\frac {\partial H_{tt}} { \partial \alpha_t} } 
                                                      - E  {\frac {\partial S_{tt}} { \partial \alpha_t} } \bigg)  \,.
\ee 
Therefore, by calculating such a derivative for each $\alpha_t$ one can get an analytic expression for the whole energy gradient.
Further details  on the analytic gradient calculation and on its very efficient numerical implementation are 
laid out clearly in Refs.~\cite{BuA06,BuA08}. 

Up to the present the AGOM has only been applied to atomic and molecular few-body systems. 
To illustrate its quality  Table~\ref{AGOM}  shows results from applying AGOM to the positronium molecule 
(molecule of two positronium atoms).  Also shown are results from SVM calculations 
where the above mentioned cyclic 
optimization has been used. In the AGOM calculation the possibility to determine the entire  gradient analytically
allows the optimization of all basis functions simultaneously.
Inspecting Table~\ref{AGOM} one finds only a very small difference between AGOM and SVM results for 100 basis states, whereas
for 200 basis states the AGOM energy is considerably lower than the SVM energy. It is also interesting to see that the AGOM result
with 500 states is somewhat lower than the SVM energy with 1600 terms. Our example shows that the AGOM is a very powerful technique.
\begin{table}
\begin{center}
\caption{{\label{AGOM}}Convergence of the total energy (in a.u.) of the positronium molecule in the 
state with $L=1$ and negative parity }
\begin{tabular}{c||c|c}
\hline \hline
Basis size & AGOM (Bubin/Adamowicz~\cite{BuA08})  & SVM (Varga et al.~\cite{VaU98})  \\
\hline
100   & --0.334 400 893 & --0.334 399 869 \\
200   & --0.334 407 545 & --0.334 405 047 \\
300   & --0.334 408 147 &                \\
400   & $\,\,\,$ --0.334 408 266 3 & --0.334 407 971 \\
500   & $\,\,\,$ --0.334 408 295 5 &               \\
800   &                &  --0.334 408 177 \\
1200  &                &  --0.334 408 234 \\
1600  &                &  $\,\,\,$ --0.334 408 265 8 \\
\hline \hline
\end{tabular}
\end{center}
\end{table}

As already said, the AGOM has  not yet been applied to nuclear few-body systems. While in atomic and molecular systems 
one has to deal with the simple Coulomb force, 
the complicated structure of the nuclear force could  make  the calculation  much more involved. 
However, the method is quite intriguing and deserves being tested in nuclear few-body problems.

\subsubsection{The Resonating Group Method (RGM)}\label{sec:RGM}

The resonating group method consists of an expansion of the nuclear wave function in cluster 
wave functions. In principle these clusters are chosen, based on energy considerations, according to all possible 
fragmentation channels of the compound nucleus~\cite{WiT77,TaL78}. 
For example, $^4$He can be fragmented in five possible ways: ($p+^3$H), ($n+^3$He), ($d+d$), ($d+n+p$), ($p+p+n+n$).
In addition one could also consider other clusters like  $(\tilde d+ \tilde d)$, where $\tilde d$ represents the virtual
quasi-bound state of the $s$-wave $np$ pair with isospin T=1. 

A typical RGM trial wave function can therefore be written as 
\bea
\Psi_T &=& {\cal A} \bigg[ \sum_i \Phi_i^{[2]}\!({\rm A}_i) \, \Phi_i^{[2]}\!({\rm B}_i) \,F_i^{[2]}\!({\veta}^{\,[2]}_{1,i})
 + \sum_i \Phi_i^{[3]}\!({\rm A}'_i) \, \Phi_i^{[3]}\!({\rm B'}_i) \, \Phi_i^{[3]}\!({\rm C'}_i) \,F_i^{[3]}
\!({\veta}^{\,[3]}_{1,i},{\veta}^{\,[3]}_{2,i}) \\ 
 &+& \sum_i \Phi_i^{[4]}\!({\rm A}''_i) \, \Phi_i^{[4]}\!({\rm B}''_i) \, \Phi_i^{[4]}\!({\rm C}''_i) \,
   \Phi_i^{[4]}\!({\rm D}''_i)
 \, F_i^{[4]}\!({\veta}^{\,[4]}_{1,i},{\veta}^{\,[4]}_{2,i},{\veta}^{\,[4]}_{3,i}) + ...
\bigg] \,,
\eea
where ${\rm X}_i$ identifies the fragment and $\Phi_i^{[l]}({\rm X}_i)$ describes its  internal wave function,
$l$ is the number of clusters for a given fragmentation and $i$ enumerates the various $l$-fragment sets
(in the $^4$He example above $l$ would assume only the values 2, 3, and 4).
The functions $F_i^{[l]}\!({\veta}^{\,[l]}_{j,i})$ are the wave functions for the relative motion of the 
fragments and the variables ${\veta}^{\,[l]}_{j,i}$ represent
the corresponding Jacobi coordinates.

 It is evident that the maximal possible value for $l$
is equal to $A$. In few-nucleon physics one normally is limited to $l=2$ thus considering
a fragmentation of the system into sets of two clusters ${\rm A}_i$ and ${\rm B}_i$. The consideration of
$l>2$ is strictly necessary only for scattering states with open $l$-fragment channels. 

The RGM approach is founded on the idea that the Hamiltonian can always be rearranged as a sum of 
well-defined cluster Hamiltonians 
\be
H_{{\rm X}} = T_{{\rm X}} + \sum^{A_{{\rm X}}}_{j<k \in {\rm X}} V_{jk} \,,\label{cluster}
\ee
plus the potential terms among  nucleons belonging to different clusters as well as  the relative kinetic energies 
among the clusters. In Eq.~(\ref{cluster}) 
$T_{{\rm X}}$ is the total kinetic energy of the nucleons in cluster X minus the kinetic energy of 
the CM of the cluster, while $A_{{\rm X}}$ denotes the number of nucleons in the cluster. 
Besides the NN potential $V_{jk}$, a 3NF can also be included in $H_{{\rm X}}$. Apart from the so-called 
cluster model, where one uses effective cluster-cluster interactions and where the clusters are either structureless
particles or described by simple models, in the RGM approach one retains the 
microscopic interaction between the constituents of different clusters, and the cluster wave 
functions,  $\Phi_{\rm X}(1,2,...,A_{{\rm X}}) \equiv  \Phi({\rm X})$, are obtained from the same
microscopic interaction.
 
The RGM strategy consists in a calculation of the various cluster wave functions $\Phi({\rm X})$ as a first step. 
In a second step, the relative motion wave functions are determined 
variationally. Let us consider for example, the simple case of a single fragmentation of the nuclear system into 
clusters A$\equiv$A$_1$ and B$\equiv$B$_1$ with trial wave function
\be
\label{RGM_AB}
\Psi_T = {\cal A} [ \Phi({\rm A})\, \Phi({\rm B}) F({\veta})] \,.
\ee
The factorized form of $\Psi_T$  corresponds to the  division of the Hamiltonian into
\be
H = H_{\rm A} + H_{\rm B} + H^\prime 
\ee
with the two cluster Hamiltonians $H_{\rm A}$ and $H_{\rm B}$ and the Hamiltonian $H^\prime$ for the
relative motion of the two clusters. The latter has the form
\be
H^\prime =  T_{\rm rel} + \sum_{i \in {\rm A}} \sum_{j \in {\rm B}} V_{ij} \,,
\ee 
where $T_{\rm rel}$ is the relative motion kinetic energy. 
A 3NF can also be included in $H_{\rm A}$, $H_{\rm B}$ and $H^\prime$. 

The solution of the  variational problem for the RGM equations can be obtained
 in matrix form (see e.g.~\cite{Ho87}) or by using integral kernels (see e.g.~\cite{TaL78}).

The philosophy of RGM calculations has to be understood as follows. A bound-state
wave function or the interior part of a scattering state  is expanded on a over-complete set of basis 
functions. As shown above such an over-complete basis
consists of two parts: the cluster and the relative motion wave functions. 
As a starting point one could consider all 
possible two-cluster configurations. Additional basis states could include virtual
excitations of the single clusters since these should have the least overlap
with the original basis. 
Let us consider, for instance, a two-cluster state with an arbitrary cluster A and a cluster B consisting of a 
$pn$ pair with isospin T=0.
The lowest energy state of the  $pn$ pair is the deuteron bound state. All virtual excitations of the
$pn$ cluster lie in the continuum and are characterized by relative energies $E_{\rm B}>0$ and quantum numbers 
for relative angular momentum $l$, spin $s$, and total angular momentum $j$. The virtual two-body excitation simulates a
free (1+1) cluster, where, however, none of the two nucleons can escape the nucleus.  For any set of allowed quantum numbers 
$l$, $s$, and $j$ and
any $E_{\rm B}>0$ one has in principle a well-defined continuum solution. However, since here one is concerned 
with short lifetime virtual excitations one may use an expansion in a set of $m$ localized functions. In fact, in RGM 
calculations not only cluster bound states, but also cluster virtual excitations are determined in this way. 
Such expansions lead to a discretization of the continuum. Thus in our example the
$pn$ continuum is represented by $m$ discrete states which can be ordered with respect to growing 
$E_{\rm B}$. 
It should be noted that an increase of $m$ does not only lead to an increase of the number of states in
the nuclear spectrum, but in general also to a shift of the $n$-$th$ state in the spectrum. 
Having fixed the model space of the various clusters one could use as variational parameter
the total energy $E_{\rm tot}$ and allow all two-body cluster states $\Phi({\rm A}_i) \Phi({\rm B}_i)$ with 
$E_{{\rm A}_i} + E_{{\rm B}_i} \le E_{\rm tot}$. 
Then $E_{\rm tot}$ should be further increased until convergence is reached for the observable under investigation.
However, in order to reach convergence, it might be preferable to allow for certain clusters to have higher internal
excitation energies than others, although this makes a controlled expansion a bit difficult.
\begin{table}
\begin{center}
\caption{{\label{RGM_HH}} $^4$He binding energy in MeV for various model spaces from RGM calculation~\cite{HoH97} with  
the BonnR potential~\cite{MaH87} (with Coulomb force). See text. Also given is the corresponding GFMC 
result~\cite{Ca88}. }
\begin{tabular}{c||c|c|c|c|c|c}
\hline \hline 
number of channels  & 3 & 3+2 & 5+60  & 65+82 & 147+80 & GFMC \\
\hline
                    & 14.033 & 15.873 & 24.345 & 25.798 & 25.910 & 25.86(15) \\
\hline \hline
\end{tabular}
\end{center}
\end{table}

>From the discussion above it is evident that in most cases a calculation with the simple ansatz (\ref{RGM_AB}) will 
not lead to very realistic results.
Other fragmentations should be taken into account in addition. To better illustrate the RGM procedure we consider an
example from the literature. In  Ref.~\cite{HoH97} the $^4$He compound system is studied by employing
a realistic $V_{\rm NN}$ (for a more recent calculation with additional consideration of a 3NF 
see Ref.~\cite{HoH08} by the same authors). 
The variational calculation is carried out using Gaussians as basis functions. 
In Table~\ref{RGM_HH} we
display results for the $^4$He binding energy for various model spaces. By including
only the two-cluster states $p\,$-$^3$H, $n\,$-$^3$He, and $d\,$-$\,d$, 
one underestimates the corresponding
GFMC precise ab initio result for the binding energy by 12 MeV. By adding the $d\,$-$\,d$ channels $^5d_0$ and the $\tilde d$-$\tilde d$
channel $^1s_0$-$^1s_0$ one gains almost 2 MeV. In order to reach a high-quality result   
additional {\bf (i)} 30 $p\,$-$^3$H and 30 $n\,$-$^3$He virtual excitations, {\bf (ii)} 82 $np$-$np$ virtual excitations (both pairs with T=0), 
and {\bf (iii)} 80 (NN)-(NN)  virtual excitations (both pairs with T=1) are necessary.

It is interesting to note
that because of the over-complete basis it is not necessary to determine the various cluster wave functions with 
a very high precision, e.g., in Ref.~\cite{HoH97} the deuteron wave function is described by only three Gaussians 
for the $s$-wave and two Gaussians for the $d$-wave. This leads to an underestimation of the deuteron binding energy 
of about 0.3 MeV. Probably more accurate cluster wave functions would lead to a reduction of the number of
virtual excitations, but, of course, one has to search for the best compromise, which makes a convergent expansion
of a RGM calculation a bit tricky. On the contrary,  
for the case of a
scattering calculation with two asymptotic fragments A and B it is desirable to calculate the corresponding  
wave functions of clusters A and B with high precision.

\subsubsection{Fermionic Molecular Dynamics (FMD)}\label{sec:FMD}

The FMD approach has been suggested by Feldmeier  \cite{Fe90} in order to solve the many-body problem of interacting
identical fermions with spin 1/2. The aim has been to describe nuclear ground states and
heavy-ion reactions in the energy regime below  particle production threshold (for a review see \cite{FeS00}).

In FMD the nuclear wave function is a linear combination of Slater determinants of single-particle states $| q_i \rangle$
given by linear combinations of Gaussians with complex parameters ${\bf b}_k$ and $a_k$,
\be
\langle {\bf x} | q_i \rangle = \sum_k c_k \, \exp\left\{ - \left( \frac {{\bf x} - {\bf b}_k}{2a_k} \right)^2 \right\} 
   \, |\chi_k\rangle \, |\xi_k\rangle \,,
\ee
and additional spin and isospin wave functions $|\chi_k\rangle$ and $|\xi_k\rangle$, respectively.
For $|\chi_k\rangle$ all orientations in spin space are possible 
($|\chi_k\rangle = \chi_k^\uparrow | \uparrow\rangle + \chi_k^\downarrow | \downarrow\rangle$). 
While in most of the FMD applications $|\xi_k\rangle$ represents either a proton or a neutron state, 
in more recent studies isospin mixing has been allowed~\cite{BaF08} i.e.
$|\xi_k\rangle = \xi_k^\uparrow | \uparrow\rangle + \xi_k^\downarrow | \downarrow\rangle $.
Such a superposition introduces a charge mixing in single-particle and many-body spaces.
Therefore, in order to conserve charge one has to
project onto a state with a fixed value of the third component of the isospin for the nuclear
system under consideration. 
The long-range correlations in nuclear wave functions originate from a pion exchange between two nucleons.
Besides charge transfer, due to its pseudoscalar nature the pion induces a parity change. 
The FMD states can be chosen not to be good parity states. 
Parity is then restored by a proper parity projection operator. 

The energy minimization is made with respect to all the various single-particle parameters. 
In general, the resulting correlated many-body state breaks translational and rotational invariance. 
Hence, after minimization, the wave function is projected onto zero total momentum and total spin $J$~\cite{NeF08}.
The FMD formalism in its present form relies on an
operator representation of $V_{\rm NN}$ in coordinate space, which limits the number of potential models that
can be used. A possible choice would be the AV18 potential~\cite{WiS95}. It has however a rather strong short-range repulsion,
which makes it very difficult to reach convergence in a FMD expansion. Thus softer versions of AV18 are often used
by employing unitary transformations with the UCOM method~\cite{RoN10}. (Such a potential, however, does not seem to be
ideal for ab initio calculations as discussed in Ref.~\cite{HaP10}). 
Despite all this, the FMD technique might be quite 
promising
and is in principle systematically improvable. 
Up to the present, however, it is difficult to judge the precision of the FMD, since converged benchmark tests
are missing in $A\ge4$ systems. Thus, here we can only report about a case, where a $^4$He FMD calculation has been made
with different UCOM versions of AV18 taking just a single Slater determinant~\cite{BaF08}. In comparison to precise
NCSM and EIHH ab initio results an underbinding between 12.8 and 3.6 MeV is observed~\cite{BaF08}. 

\subsection{The Similarity Transformation Formulation}\label{sec:SIM}

Another way to reformulate the quantum mechanical many-body problem is achieved with the help of similarity
transformations~\cite{Ok54,CoK60,DaS64,SuL80}.
To this end we consider the following mean value
\be
f = \langle\Psi| F|\Psi\rangle \,,
\ee
where $|\Psi\rangle$ is an eigenstate of the Hamiltonian $H$. 
The mean value $f$ is invariant under similarity transformations $e^S$, i.e.
\begin{equation} 
f = \langle\Psi|e^{-S} \,e^{S}\, F\,e^{-S} \, e^{S}|\Psi\rangle
\equiv \langle \bar\Phi|{\cal F} |\Phi\rangle\label{simileq} \, 
\end{equation}
with  
\be
|\Phi\rangle = e^{S}|\Psi\rangle \,, \qquad |\bar\Phi\rangle = e^{-S^\dagger}|\Psi\rangle \,,
 \qquad{\cal F} = e^{S}\,F\,e^{-S}\,.
\ee

One may consider a subspace P of the  Hilbert space  with  eigenprojector $\hat P$ defined by 
\be
\hat P=\sum_{n=1}^N |\phi_n\rangle\langle\phi_n|\,, 
\ee 
where the $|\phi_n\rangle$ are  the eigenfunctions of some convenient Hamiltonian $H_0$.
The residual space Q has a corresponding eigenprojector  $\hat Q=I-\hat P$. Therefore one can write Eq.~(\ref{simileq})
as 
\begin{equation}
f=\langle \bar\Phi|(\hat P + \hat Q){\cal F}(\hat P + \hat Q) |\Phi\rangle =
\langle \bar\Phi|\hat P{\cal F}\hat P+\hat P{\cal F}\hat Q+\hat Q{\cal F}\hat P+\hat Q{\cal F}\hat Q|\Phi\rangle\,.
\end{equation}
Given a $|\Phi\rangle$  entirely contained in the P-space one has
\begin{equation}
f=\langle \bar\Phi|\hat P{\cal F}\hat P|\Phi\rangle\,
\end{equation} 
if  the following decoupling condition is satisfied
\begin{equation}
 \hat Q{\cal F}\hat P = \hat Q e^{S}\, F\,e^{-S}\hat P = 0\,.
\label{decoupeq}
\end{equation}
This means that by solving the decoupling equation (\ref{decoupeq}) it is possible, in principle, to determine $S$ and therefore
calculate $f$ as the mean value of the {\it effective} operator ${\cal F}$ on the P-space.  

Since the similarity transformation $e^S$ represents an infinite sum of operators it can be problematic to deal with it
in actual calculations.
On the other hand there exists a relation  
that is very useful
in applications (see e.g. Section~\ref{sec:CC}), namely the Baker-Campbell-Hausdorff (BCH) expansion 
\begin{equation}
{\cal F} =   F + [F,S] +\frac{1}{2} [[F,S],S]+\frac{1}{3!} [[[F,S],S],S]+\frac{1}{4!} [[[[F,S],S],S],S]+
 \cdot \cdot \cdot \,.\label{Hausdorff}
\end{equation}

Similarity transformations have been used extensively to find the ground and low-lying state energies of a system. 
In this case one has to solve Eq.~(\ref{decoupeq}) replacing $F$ by the Hamiltonian $H$. 
Unfortunately the full solution of Eq.~(\ref{decoupeq})  is a very difficult problem. Various methods have been devised to
approach the solution by limiting the operator $S$, which is in general a sum of $n$-body operators $S^{[n]}$, i.e.
\begin{equation}
S=S^{[1]} + S^{[2]} + S^{[3]}+ \cdot \cdot \cdot +S^{[A]}\,,   
\end{equation}
to the first two or three terms (see Sections~\ref{sec:NCSM} and \ref{sec:CC}).

One strategy to satisfy the decoupling condition~(\ref{decoupeq}) was given in the  work of Okubo~\cite{Ok54} and later
revived by Suzuki and Lee~\cite{SuL80}. It consists in requiring the operator $S$ to have the property 
\begin{equation}
S=\hat QS\hat P\,, \label{SLdec}
\end{equation}
which implies $\hat PS\hat P=\hat QS\hat Q=\hat PS\hat Q=0$.
If $S$ has these properties it allows one to write 
\begin{equation}
e^S=1+S\label{SLexp}\,.
\end{equation}
and the decoupling condition~(\ref{decoupeq}) is also satisfied. 
Finding $S$ via the decoupling condition~(\ref{SLdec}) would permit one to get ground-state and some low-lying state energies 
as the eigenvalues of the effective Hamiltonian ${\cal H}=e^{S}\,H\,e^{-S}$ in a convenient $P$-space. 

In order to work with a Hermitian effective Hamiltonian, 
the similarity transformation $e^S$ must turn into a unitary transformation $U=e^G$ with 
$G=-G^\dagger=arctanh(S-S^\dagger)$~\cite{Ok54,Su82,SuO83}, 
where again $S$  has to satisfy the decoupling condition~(\ref{SLdec}). 
The authors of this formulation called this approach unitary model operator approach (UMOA),
which today, however, has a more specific meaning (see below). Today the approach is more often
referred to as Suzuki-Lee (SL) approach. 

The operator $S$ in Eq.~(\ref{SLdec}) can in principle be determined as follows:
{\bf (i)} let the P space of dimension $d_p$ be spanned by states $|\alpha_p\rangle$,
{\bf (ii)} take a set of the lowest $d_p$ eigenstates $|n\rangle$ 
of the Hamiltonian H, such that $\hat P |n\rangle$
are linearly independent, {\bf (iii)} take a set of $d_q$ states
$|\alpha_q\rangle$ belonging  to the space Q, then the following equality holds
\be
\label{omega_QP}
 \langle \alpha_q | n  \rangle   =  \sum_{p=1}^{d_p}  \langle  \alpha_q |  S |\alpha_p \rangle \langle \alpha_p| n  \rangle\,. 
\ee
By defining matrices $A_Q$ and $A_P$, with elements $\langle \alpha_q | n  \rangle$ 
and $\langle \alpha_p| n  \rangle$ respectively,  one finds the matrix
elements $\langle  \alpha_q |  S |\alpha_p \rangle$ of $S$   via
\begin{equation}
S=A_Q\,A_P^{-1}\,.
\end{equation}
Since a solution of Eq.~({\ref{omega_QP}) at the $A$-body level is not feasible for larger $A$, in practical 
applications it has first been solved at the two/three-body level, which then leads to a two/three-body
effective interaction, which is used to accelerate the convergence of the calculation. 
As will be discussed in the following this has been done in the NCSM as well as in what is today known as the UMOA approach.
In both approaches the unitary transformation has been applied to the two/three-body HO Hamiltonian with
a bare interaction.
The difference between the two approaches lies in their procedures for obtaining the two-body (three-body)  effective
interaction, e.g. in the UMOA case the two-body effective interaction depends on the HO quantum numbers of the center of mass
of the pair. This state dependence makes the calculation more computationally intense. 
Moreover the approach is supplemented by a second step  in which the effective interaction undergoes another
unitary transformation, such that the matrix elements of the new effective interaction between
two-particle-two-hole excitations reduce to zero. In this way the diagonalization of the A-body Hamiltonian
with the new  effective interaction is made only between one-particle-one-hole states. Calculations of
heavier systems, like $^{40}$Ca and $^{56}$Ni, have become possible in this way~\cite{FuO09}, even though at the price of 
a huge numerical effort.

In the EIHH approach the unitary transformation has been applied at the two-/three-body level.
Because of the HH basis the effective interaction turns out to have a considerable dependence on the A-body system.
The complications of the antisymmetrization problem for HH functions, however, has limited its applications to $A<8$.

Here we would like to mention another approach that makes use of similarity
 transformations namely, the similarity 
renormalization group  approach. The latter draws inspiration from the ideas of
renormalization group theory, which
can be paraphrased as follows: when a problem is too hard to be solved in one step one can 
still arrive at the solution by breaking it 
into small steps. So, instead of solving the decoupling condition of Eq.~(\ref{decoupeq}) 
directly in order to obtain an effective 
Hamiltonian ${\cal H}$, one can try to build ${\cal H}$ step by step.

In the approaches described above ${\cal H}$ is obtained from a unitary 
transformation $U=e^G$ with an antihermitian $G$. 
The crucial point of the SRG approach is the introduction of a continuous flow parameter $\alpha$ such that 
\begin{equation}
{\cal H}(\alpha)=U^\dagger(\alpha)H U(\alpha)\label{SRG1} \,,
\end{equation}
however, only when $\alpha$ flows towards a specific value $\alpha_0$ does one have the proper $U$
which leads to the diagonal form of ${\cal H}$. 
The idea is to reach that stage by letting $\alpha$ flow. In fact, $U(\alpha)$ can in principle
be obtained through solving the differential equation, 
\begin{equation}
\label{SRG2}
 \frac{d {\cal H}(\alpha)}{d\alpha}=[\eta(\alpha), {\cal H}(\alpha)]\,,
\end{equation}
which itself is  obtained  by  differentiating Eq.~(\ref{SRG1}) with 
 $\eta(\alpha)$ being an antihermitian operator:
\begin{equation}
\eta(\alpha)\equiv \frac{d U(\alpha)}{d\alpha}U^\dagger(\alpha)\equiv -\eta^\dagger(\alpha)\,.
\end{equation}
 
There are  an infinite number of generators $\eta_i$, since one can construct an infinite number 
of different $U_i(\alpha)$ that tend to the proper $U$ for some value of $\alpha_i$. 
However, an interesting observation is that the relation
\begin{equation}
\eta(\alpha)=[{\cal H}_d(\alpha),{\cal H}(\alpha)]\label{SRG3} \,,
\end{equation}
where ${\cal H}_d(\alpha)$ is the diagonal part of the Hamiltonian in a given basis, 
ensures that the off-diagonal elements of ${\cal H}(\alpha)$ decay exponentially 
with $\alpha$~\cite{We01}. This means that one can solve 
Eq.~(\ref{SRG2}) for ${\cal H}$ 
with $\eta$ given  by Eq.~(\ref{SRG3}). Then, by increasing
$\alpha$ step by step in the numerical solution of Eq.~(\ref{SRG2}) 
one can make ${\cal H}(\alpha)$ more and more (block)-diagonal.

An important remark is in order here. The generator $\eta(\alpha)$ is a many-body operator, 
therefore it will generate an $A$-body effective Hamiltonian and an $A$-body effective potential.  
In most cases it is inconvenient or impossible to implement such an operator in a many-body code. 
What is then often done (see Ref.~\cite{BoF10,RoN10} and references therein) is to express Eq.~(\ref{SRG3}) 
in the  two- or three-body space,   
\begin{equation}\label{SRG4}
\eta^{[2]([3])}(\alpha)=[{\cal H}_d^{[2]([3])}(\alpha),{\cal H}^{[2]([3])}(\alpha)] \label{SRG2[2]} \,,
\end{equation}
and let ${\cal H}^{[2]([3])}(\alpha)$ flow versus a more diagonal form. The resulting
two- or three-body operator can then be implemented in the many-body code. 

In applications to nuclear
forces~\cite{BoF10,RoN10} one normally uses instead of the running ${\cal H}_d(\alpha)$ a different diagonal
Hamiltonian matrix, like e.g. the kinetic 
energy which is diagonal in momentum space.

Contrary to NCSM and EIHH approaches in Sections~\ref{sec:NCSM} and \ref{sec:EIHH}, where one in principle recovers the bare interaction from the effective interaction 
in the limit of an infinite ${\rm P}_{\!\!A}$-space, in the SRG approach the potential is obtained by truncating the flow 
at a two- or three-particle level thus violating the unitarity of operator $U$ 
even for an infinite ${\rm P}_{\!\!A}$ space. Therefore
the violation cannot be compensated in the subsequent many-body calculation by increasing the ${\rm P}_{\!\!A}$ space.

We will end this chapter by mentioning UCOM, a method that allows one to obtain a two-body effective potential, 
phase equivalent to
a realistic $V_{NN}$, but much softer at short range, and therefore suitable to use in A-body calculations. 
In a way this method is similar in spirit to the SRG method when Eq.~(\ref{SRG2}) 
is solved with $\eta^{[2]}$ of Eq.~(\ref{SRG4}).

\subsubsection{The No Core Shell Model (NCSM)}\label{sec:NCSM}

Two methods have facilitated increasing the domain of the few-body field to
nuclear systems with $A>4$, while working with realistic potentials. These are the GFMC and the NCSM approaches.

The name NCSM is due to the fact that all the nucleons of the nucleus are active and explicitly taken into
account as degrees of freedom and thus it is not assumed that there is an inert core. 
The NCSM couples the traditional SM advantages of working in a HO basis with the rigor of an ab initio approach.  
For not so soft potentials convergence can be achieved by using an effective interaction.
In the 1990s NCSM calculations were made with a $G$-matrix approach (see e.g. Ref.~\cite{ZhB95}), whereas in
Ref.~\cite{NaB96} a different and more successful direction was taken, which will be described in the following
and which are based on the SL transformation formulation explained in Section~\ref{sec:SIM}.

There are two versions of the NCSM, which differ in their treatment of the CM degrees of freedom.
In the first version, the Hamiltonian of Eq.~(\ref{Hint})
is modified~\cite{ShF74} by adding a harmonic oscillator CM Hamiltonian $H_{\rm CM}$ to the intrinsic Hamiltonian $H$
 \begin{eqnarray}
H^{[A]}_\Omega &=& H + H_{\rm CM}^{\rm HO} = H +  \frac {{\P}_{\rm \rm CM}^2} {2Am} + {\frac{Am}{2}} \Omega^2 {\bf R}_{\rm CM}^2 \\
 &=& \sum_{i=1}^A \left[ {\frac{{\bf p}_i^2}{2m}}
+\frac{1}{2}m\Omega^2 {\bf r}^2_i
\right] + \sum_{i<j=1}^A \left[ V_{ij}
-\frac{m\Omega^2}{2A}
({\bf r}_i-{\bf r}_j)^2 
\right]
\equiv \sum_{i=1}^A h_i^{\rm HO} + \sum_{i<j=1}^A\tilde V_{ij} \,,
\label{HOmega}
\end{eqnarray}
where ${\bf R}_{\rm CM}$ is the CM coordinate, while 
$\tilde V_{jk}$ is a renormalized potential depending both on the HO frequency $\Omega$ and $A$: 
\be
\label{Vtilde}
\tilde V_{ij}=\left[ V_{ij}
-\frac{m\Omega^2}{2A}
({\bf r}_i-{\bf r}_j)^2\right]\,.
\ee
Naturally, the added CM HO term has no influence on the internal motion. 
This means that once the ground-state energy $E^{[A]}_\Omega$ of $H^{[A]}_\Omega$ is found, then the ground-state energy $E$ of $H$ is obtained
by subtraction of the CM ground-state energy $3\hbar \Omega/2$. For excited states one has to avoid CM excitations.
This can be achieved by the replacement $H_{\rm CM}^{\rm HO} \to \lambda H_{\rm CM}^{\rm HO}$ with sufficiently large $\lambda$ (Lawson term). A good check on 
the convergence of $E$ is through the independence of the result on the HO frequency $\Omega$. 
On the other hand, one may search for the frequency $\Omega$ which exhibits the best convergence pattern.

The calculation of $E^{[A]}_\Omega$  is  performed in a finite
model space P$_{\!A}$. This is the space spanned
by all the $A$-body HO Slater determinants formed by filling the single-particle HO eigenstates
with $N\leq N_{\rm max}$, where $N$ is the total number of single-particle HO quanta. 
The residual space Q$_A$, together with P$_{\!A}$, exhausts the full Hilbert space.

The approach that is chosen to construct the effective Hamiltonian, appropriate to the finite P${\!_A}$-space, 
is that described in Section~\ref{sec:SIM}. However,
if one applies  the unitary transformation to  $H^{[A]}_\Omega$ one  obtains an  effective Hamiltonian 
that is an $A$-body operator. To avoid this complication, an approximation is made in the NCSM. It consists   
in first finding only a two-body effective interaction $\tilde V_{ij}^{\rm [2,eff]}$
which is then used to replace  
the interaction term $\tilde V_{ij}$ of Eq.~(\ref{HOmega}). The effective interaction $\tilde V_{ij}^{\rm [2,eff]}$
is obtained  by applying the decoupling condition of Eq.~(\ref{SLdec}) to a two-nucleon
Hamiltonian $H^{[2]}_\Omega$ that arises from $H^{[A]}_\Omega$ by restricting the sums 
to two nucleons only, keeping however the original mass number $A$ in the interaction term:
\begin{eqnarray}
H^{[2]}_\Omega &=&  \frac{{\bf p}_1^2}{2m}+\frac{{\bf p}_2^2}{2m}
+\left[\frac{1}{2}m\Omega^2 {\bf r}^2_1+\frac{1}{2}m\Omega^2 {\bf r}^2_2
\right] 
+  \left[ V_{12}
-\frac{m\Omega^2}{2A}
({\bf r}_1-{\bf r}_2)^2\right]
\equiv H^{[2]}_{\rm HO} +\tilde V_{12}  \,.
\label{HO2mega}
\end{eqnarray}
The effective Hamiltonian $H^{[2]}_{\rm eff}$ is determined in a subspace of  P, 
the P$_{2}$-space ($\hat P_2+ \hat Q_2 =I_2$), via 
the two-body transformation operator $S^{[2]}=\hat Q_2 S^{[2]}  \hat P_2$. 
Then the two-body effective interaction is obtained 
by subtracting from $H^{[2]}_{\rm eff}$ the two-body HO Hamiltonian, i.e. 
\begin{equation}
\tilde V_{12}^{[2,\rm eff]}  = H^{[2]}_{\rm eff} - H^{[2]}_{\rm HO}\,.
\end{equation}
The corresponding  $V_{ij}^{[2,\rm eff]}$  is then used in Eq.~(\ref{HOmega}).
As is clear from Ref.~\cite{DaS64}, this procedure 
is equivalent to the case where {\bf (i)} the similarity operator $S$ of Section~\ref{sec:SIM} is limited to
a  two-body operator $S^{[2]}$ and {\bf (ii)} the effective Hamiltonian from the 
BCH expansion in Eq.~(\ref{Hausdorff}) 
is truncated at the two-body operator level. 
This is  different from the CCSD approach, described in Section~\ref{sec:CC}. There one has 
$S=S^{[1]}+S^{[2]}$,  without a further truncation of the effective Hamiltonian.
\begin{figure}[tb]
\epsfysize=9.0cm
\begin{center}
\epsfig{file=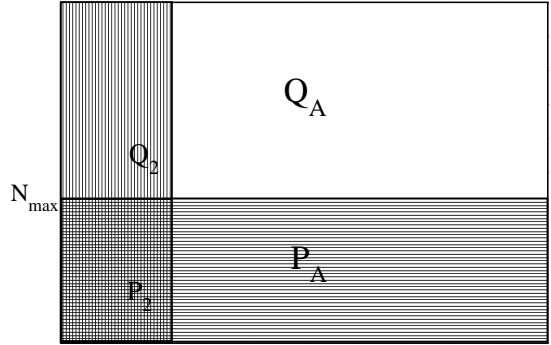,scale=0.3,angle=-90}
\caption{The various P and Q spaces relevant for the construction
of the two-body effective interaction (see text).}
\label{figure_EIHH_box_T}
\end{center}
\end{figure}
Due to the effective interaction the NCSM is no longer a variational method. The $n>2$ terms  $S^{[n]}$ neglected in $S$, as well as 
the $n$-body terms neglected in the corresponding effective Hamiltonian, could either increase
or decrease the binding energy. On the other hand, as the basis space is increased
the result converges to the exact solution. 
This  becomes clear from the illustration 
in Fig.~\ref{figure_EIHH_box_T}. At each P$_{\!A}$,  
the unitary transformation transfers information from a part of the Q$_A$ space, namely the Q$_{2}$ space,
to the P$_2$ space. Thus $E^A_\Omega$ converges much faster when 
performing a calculation in P$_{\!A}$-space 
than in a corresponding calculation with the bare interaction. For a sufficiently large P$_{\!\!A}$, covering practically the 
whole Hilbert space, the effective interaction coincides with the bare one, and one has the exact result. 
>From the figure it is also clear that the same exact result could be achieved by enlarging {\it horizontally} 
the space where the unitary transformation operates, i.e. considering  $S^{[n]}$ terms (as well as $n$-body 
terms in the effective Hamiltonian) with increasing $n$ up to $n=A$.
However, this is much more difficult, since one would need to
know the entire $n$-body spectrum to construct the $n$-body effective interaction. 
It is evident that if three-body forces are present in the original Hamiltonian it is expedient 
to apply the  procedure at least up to $n=3$~\cite{NaO02}.  
 
Here we would like to mention another version of the NCSM, where the $A$-body basis is 
the translationally invariant HO basis. Its application, however, is presently restricted to $A=3,\,4$.
This version starts from the observation that  in Eq.~(\ref{HOmega}) the CM dependence is contained, as a separate HO Hamiltonian,
only in the one-body HO term. 
Therefore, the use of  a translationally invariant basis expressed in terms of $(A-1)$ Jacobi coordinates automatically 
removes the CM contribution to the energy. The natural basis
to use is the HO basis depending on the $(A-1)$ Jacobi coordinates,
$\prod^{A-1}_i \Phi^{\rm HO}_i(\veta_i)$,
supplemented by the spin and isospin parts and totally antisymmetrized.

The procedure of Suzuki~\cite{Su82} discussed in the previous section, which allows one to obtain the Hermitian 
two-body effective Hamiltonian, is applied to
\be
\tilde H^{[2]}_\Omega =  \left( \frac{\vpi_1^2}{2m}+ \frac{1}{2} m \Omega^2 \veta_1^2\right)+ \tilde V_{12} 
= \left[ \frac{\vpi_1^2}{2m}+ \frac{1}{2} m \left (\frac{A-2}{A}\right )\Omega^2 \veta_1^2\right]+ 
V_{12}\,,
\label{H2int_2}
\ee
where $\vpi_1$ is the conjugate momentum of $\veta_1$.
 By using the SL transformation for $\tilde H^{[2]}$ one generates a two-body effective interaction 
\begin{equation}
 \tilde V_{12}^{\rm [2,eff]}=\tilde H^{[2]}_{\rm eff} - \left( \frac{\vpi_1^2}{2m}+ \frac{1}{2} m \Omega^2 \veta_1^2\right)
\label{Veff}
\end{equation}
that replaces $\tilde V_{jk}$  in Eq.~(\ref{HOmega}). 

Alternatively, and distinct from the usual NCSM approach, one could also obtain an effective potential  
$V^{\rm [2,eff]}_{12}$ without an explicit $\Omega$ dependence as follows
\begin{equation}
 V^{\rm [2,eff]}_{12}=\tilde H^{[2]}_{\rm eff} - \left( \frac{\vpi_1^2}{2m}+ 
    \frac{1}{2} m \frac{(A-2)}{A}\Omega^2 \veta_1^2\right) \,.
\end{equation}  
This Jacobi coordinate formulation, implemented in coupled J-scheme, is more advantageous for $A\le 4$ nuclei. 
The single-particle formulation described above and implemented in M-scheme, is more advantageous for $A> 4$. 
In fact the Jacobi coordinates prevent the use of Slater determinants, and antisymmetrization becomes much more difficult.

Fig.~\ref{figure_NCSM_T} illustrates the convergence pattern of the ground-state energy of $^4$He, taken 
from Ref.~\cite{NaK00}. The results are  obtained with
the translationally invariant version of the method and with a realistic potential.
It is interesting to note not only the $\Omega$ dependence of the convergence
but also  how rapid the convergence can be for some $\Omega$ values. Of course all the results have to converge 
to the same value.

Concluding this section we would like to mention that momentum space NN potentials can easily be treated
by using a HO basis in momentum space (see e.g.~\cite{NoN06}).  
A review on recent developments of the method can be found in~\cite{NaQ09}.
\begin{figure}[ht]
\begin{minipage}[b]{0.48\linewidth}
\centering
\includegraphics[scale=0.34,angle=-90]{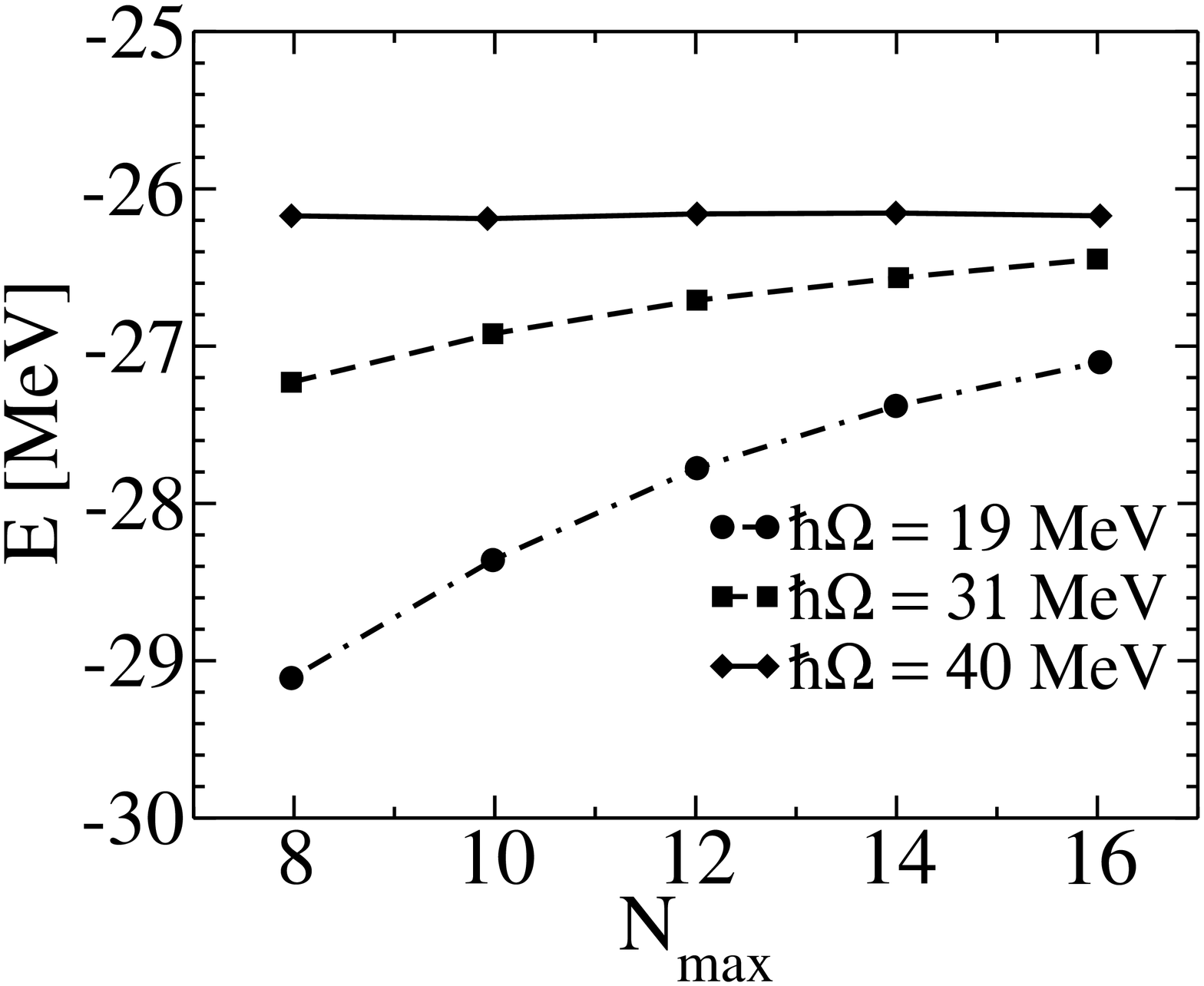}
\caption{$N_{\rm max}$ dependence of the $^4$He ground-state energy for a NCSM calculation with the
CD-Bonn NN potential~\cite{Ma01} and with various values of $\hbar\Omega$ 
(from Ref.~\cite{NaK00}).}
\label{figure_NCSM_T}
\end{minipage}
\hspace{0.5cm}
\begin{minipage}[b]{0.48\linewidth}
\centering
\includegraphics[scale=0.34,angle=-90]{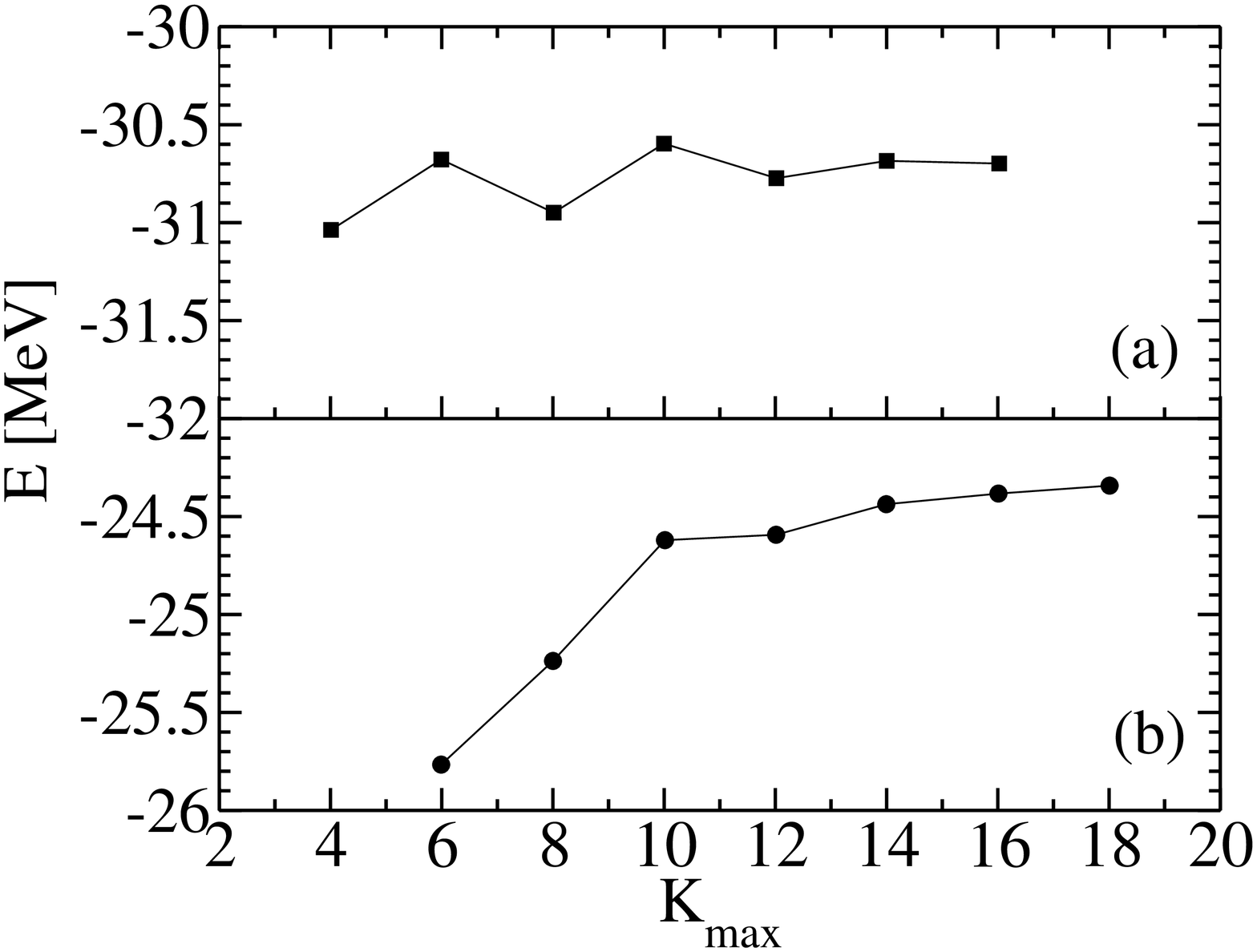}
\caption{$K_{\rm max}$ dependence of the $^4$He ground-state energy for a EIHH calculation with the 
MTI-III~\cite{MaT69} (a) and  AV14~\cite{WiS84} (b) NN potentials (from Ref.~\cite{BaL01}).}
\label{figure_EIHH_T}
\end{minipage}
\end{figure}

\subsubsection{The Effective Interaction for Hyperspherical Harmonics (EIHH)}\label{sec:EIHH}

The EIHH method, inspired by the NCSM approach, was introduced by Barnea and the authors of the present review article 
in Refs.~\cite{BaL00,BaL01}. 
In this method one uses the HH basis  
to construct an effective interaction by applying the SL transformation to the Hamiltonian 
(see Section~\ref{sec:SIM}). 

The idea is very similar to that of the NCSM approach, but with the following differences:
{\bf (i)} the HH basis instead of the HO basis is used, {\bf (ii)} the P-space is defined by a maximal value $K_{\rm max}$ of the 
grand angular quantum number $K_N$ ($N=A-1$), and {\bf (iii)}
the EIHH two-body Hamiltonian that undergoes the similarity transformation, $H_A^{[2]}$, is a {\it quasi} 
two-body Hamiltonian, because it  contains information about the dynamics of the entire $A$-body system. This is not 
the case for the HO two-body Hamiltonian, where the dependence on $A$ in $\tilde V_{ij}$ of Eq.~(\ref{Vtilde}) is merely
fictitious, since it does not add any dynamical information about the $A$-body system.

The HH {\it quasi} two-body Hamiltonian is given by
\begin{equation}
\label{H_quasi}
 H^{[2]}_A(\rho)= T_K(\rho) + V_{A,A-1}(\sqrt{2}\rho \sin({\alpha_N}) \, \hat \eta_N)\,,
\end{equation} 
where we make explicit only the spatial dependence of $V_{A,A-1}$.
The Jacobi coordinates are defined in normalized reversed order with 
$\veta_N = ({{\bf r}_{A-1}-{\bf r}_A)/\sqrt{2}}\,$, and
$T_K(\rho)$ is the collective hyperspherical kinetic energy of the entire $A$-body system 
(see Eqs.~(\ref{Trho})). 

The unitary transformation is applied separately for each value of the hyperradius $\rho$,
hence the effective interaction becomes a function of $\rho$.
In addition to its $\rho$-dependence, the HH effective interaction
also depends on some quantum numbers of the
residual system. For the more complicated case of a noncentral force we refer  
to Ref.~\cite{BaL01} for an explicit expression, while here we illustrate the simpler case of a central potential model, where 
the matrix element of $H^{[2]}_A$ is given by
\be
\langle [K_N] | H^{[2]}_A(\rho) | [K'_N]\rangle = \delta_{[K_N][K'_N]}{\frac {1}{2m}} {\frac {K_N(K_N+3N-2)}{\rho^2}} 
    + \delta_{[K_{N-1}][K'_{N-1}]} V^{K_{N-1} L_{N-1}}_{K_N L_N l_N, K_{N'} L_{N'} l_{N'}} (\rho) \,,
\ee 
(the various quantum numbers are defined in Section~\ref{sec:HH}). 
One sees that the interaction depends on $K_{N-1}$ and on $L_{N-1}\,$, which are quantum number of
the residual system, and on the quantum numbers
$K_N$, $L_N$, and $l_N$, which characterize the entire $A$-body system. 
 The various, just mentioned, additional many-body information contained in the HH effective interaction 
(obtained by subtracting the hypercentrifugal term)
leads to a fast convergence of the HH expansion as can be seen in Fig.~\ref{figure_EIHH_T}.

Further improvements of the HH effective interaction have been made in Ref.~\cite{BaL03} by introducing {\bf (i)}
 a three-body effective interaction, and {\bf (ii)} by including the hyperradial kinetic energy
$T_\rho$ of  Eq.~(\ref{Trho}) in the two-body Hamiltonian of Eq.~(\ref{H_quasi}). An extension
of the HH effective interaction to non-local NN potentials  has been
carried out in Ref.~\cite{BaL10}, while an additional consideration of a 3NF is described in Ref.~\cite{BaL04}.

\subsubsection{The Coupled Cluster (CC) Method}\label{sec:CC}

The coupled cluster (CC) method originates from the field of nuclear physics~\cite{Co58} 
in the late 1950s.  The approach was further developed during the 1970s.
 A summary of the status of the field in the late 1970s can be found in Ref.~\cite{KuL78}.
Because of the difficulties posed by the strong short-range repulsion and by the 
tensor term in the nuclear interaction there were few nuclear physics applications then, see  Ref.~\cite{BiF90b}.
The most advanced of these were
those of Mihaila and Heisenberg~\cite{HeM99} who, at the end of the 1990s, were able to calculate
the electron scattering elastic form factor of $^{16}$O with realistic two- and three-body potentials.
More successful, however, has been the use of the CC method in quantum chemistry~\cite{Ci66}, 
a field it was exported to 50 years ago and where it
has had an enormous success. Today it has become a standard approach in quantum chemistry
(for a review see~\cite{BaM07}).
Recently, starting with the work of Ref.~\cite{DeH04}, the  CC theory has been reimported to nuclear physics.

A first important observation is that in its original formulation the CC theory is not translationally invariant.
Later, translationally invariant versions were formulated~\cite{BiF90b}. For clarity of presentation 
we will, in the following, first sketch the original idea and later comment on the problem of the CM 
spurious effects.

One can understand the principal point of the CC method starting from what in quantum chemistry is called 
configuration interaction (CI) method. This is a  variational expansion method.  The variational wave function is expanded on 
the complete basis of $\nu$-particle-$\nu$-hole Slater determinants formed by Hartree-Fock or simply HO single-particle  
wave functions.
In the language of second quantization the CI variational wave function is written as 
\be
\label{PsiM}
|\Psi\rangle_N = (1+ C^{[1]}+C^{[2]}+C^{[3]}+\cdot\cdot\cdot + C^{[A]})|\Phi_0\rangle \,,\label{PsiN}
\ee
where $|\Phi_0\rangle$ is for example an $A$-body Slater determinant of the $A$ lowest single-particle states. 
The $\nu$-particle-$\nu$-hole creation
operator $C^{[\nu]}$ with $\nu$ variables $ijk...$ is defined by
\be
C^{[\nu]} |\Phi_0\rangle =
\sum_{i>j>k..,a>b>c..}  C^{abc...}_{ijk...} |\Phi^{abc...}_{ijk...}\rangle
\ee
with 
\begin{equation}
|\Phi^{abc...}_{ijk...}\rangle=a^\dagger_a a^\dagger_b a^\dagger_c...a_k a_j a_i\, |\Phi_0\rangle\,,
\end{equation}
where  $i \le A$ and $A < a \le N$.
The restriction $a \le N$ means that the highest state that can be occupied is the $N$-${th}$ single-particle
state. In principle only when  $N \rightarrow \infty$ is the full Hilbert space is covered.  In practical applications
one has only few terms $C^{[\nu]}$ in Eq.~(\ref{PsiM}) and in most cases $\nu$ is limited to two or three. Convergence 
both in $\nu$ and N is very slow. 

In order to speed up the convergence for the ground state $|\Psi_0\rangle$ one uses the
similarity transformation, of Section~\ref{sec:SIM}, applied to the Hamiltonian operator,
\begin{equation}
E_0 = \langle\Psi_0|H|\Psi_0\rangle = \langle\Psi_0|e^S \,e^{-S}\,H\,e^S \, e^{-S}|\Psi_0\rangle
\equiv \langle \bar\Phi_0|{\cal H} |\Phi_0\rangle\,,
\end{equation}
where ${\cal H} = e^{-S}\,H\,e^S $ is the {\it effective} Hamiltonian whose mean value in the model space P,
represented by $|\Phi_0\rangle$, coincides with $E_0$.
As already discussed in Section~\ref{sec:SIM}, the most general form of the operator $S$ is a 
sum of a one-body, a two-body  up to an 
$A$-body operator, i.e.  $S=S^{[1]} + S^{[2]}+S^{[3]}+ \cdot \cdot \cdot +S^{[A]}$. In the language of second quantization
the action of the various operators $S^{[1]}$ and $S^{[2]}$ on $|\Phi_0\rangle$ is given by
\begin{eqnarray}\label{snu1}
S^{[1]}|\Phi_0\rangle&=&\sum_{i,a} s_i^a|\Phi_i^a\rangle=\sum_{i,a}s_i^a a^\dagger_a a_i |\Phi_0\rangle\\
S^{[2]}|\Phi_0\rangle&=&\sum_{i>j,a>b} s_{ij}^{ab}|\Phi_{ij}^{ab}\rangle=
\sum_{i>j,a>b}s_{ij}^{ab}a^\dagger_a a^\dagger_b a_j a_i|\Phi_0\rangle  \,, \label{snu2}
\end{eqnarray}
and for $S^{[\nu]}$ with $\nu>2$ accordingly. 
While in the CI the unknowns are the coefficients $C^{abc...}_{ijk...}$, here they are the coefficients $s^{abc...}_{ijk...}$. 
There is clearly a
relation among them. In fact, expanding the similarity transformation as
$(1+S +{S^2}/{2!}+{S^3}/{3!}+...)$
one can show that 
\begin{eqnarray}
C^{[1]}&=& S^{[1]} \,, \,\,\,\,\,\,\,\,
C^{[2]}= S^{[2]}+\frac{1}{2}(S^{[1]})^2 \,, \,\,\,\,\,\,\,\,
C^{[3]}= S^{[3]}+S^{[1]}S^{[2]}+\frac{1}{3!}(S^{[1]})^3\\
C^{[4]}&=& S^{[4]}+\frac{1}{2}(S^{[2]})^2+S^{[1]}S^{[3]}+ \frac{1}{2}(S^{[1]})^2(S^{[2]})+\frac{1}{4!}(S^{[1]})^4 \,,
\end{eqnarray} 
where these equations can easily be extended to higher $C^{[\nu]}$. One notes that   
already a limitation of the expansion of $S$ up to $S^{[2]}$ (CCSD) or to $S^{[3]}$  (CCSDT)  
induces correlations  into $|\Psi_0\rangle$, which range from two-body (two-particle-two-hole) up to $A$-body correlations 
($A$-particle-$A$-hole). This property is called {\it size extensivity}. 
However, from the equations above one also sees that a consideration of all operators $S^{[\nu]}$ up to order $\mu$ takes into
account all correlations induced by operators $C^{[\nu]}$ with $\nu \le \mu$, but includes only a part of the correlations induced
by operators $C^{[\nu]}$ with $\nu > \mu$, 
namely those limited to the so-called {\it disconnected} components (the name refers to the graphic representation of
the various correlations).

In the CC method the problem is then to find the values of  $s^{abc...}_{ijk...}$. This is achieved by applying  
the decoupling condition $Q{\cal H}P=0$, i.e by projection onto a sufficient number of excitations
\be
\label{CC_last}
\langle\Phi^a_i| {\cal H} | \Phi_0\rangle =0 \,,\,\,\,\,\,
\langle\Phi^{ab}_{ij}| {\cal H} | \Phi_0\rangle =0\,,\,\,\,\,\,
\langle\Phi^{abc}_{ijk}| {\cal H} | \Phi_0\rangle =0
\ee
and further equations accordingly. Presently, in nuclear physics the CC calculations are limited to CCSDT, where the above three equations have to be
considered.

Here we would like to point out that it is expedient to use the BCH expansion of Eq.~(\ref{Hausdorff})
 in representing ${\cal H}$, 
since the expansion terminates after fourfold commutators when $H$ has no more than two-body operators 
(after sixfold commutators when $H$ has no more that three-body operators etc.).

In order to illustrate the convergence rate  
and the role of doublets and triplets in a typical CC calculation, we show in Fig.~\ref{figure_CC_T} 
the results for the $^4$He ground-state energy  
by Hagen~et~al.~\cite{HaD07}. A few comments are in order here. Convergence is difficult to  obtain  for hard-core potentials, 
therefore in this example a very soft low-momentum interaction $\vlowk$~\cite{BoK03} is used 
(the latter is obtained by implementing in momentum space the 
renormalization group evolution described in Section~\ref{sec:SIM}). 
Since a CCSDT calculation is at
present  prohibitively expensive compared to CCSD, a fifth order perturbative treatment of the triplets CCSD(T)
has been implemented. It has been shown that for a specific choice of a soft potential, the addition of
this contribution results in  agreement with the FY precise ab initio result.
Recently, a more sophisticated way of including the triplets, known as 
the $\Lambda$-CCSD(T),
has been developed~\cite{TaB08}. Performing such a $\Lambda$-CCSD(T) calculation~\cite{HaP10} 
with the Idaho-N3LO (I-N3LO)~\cite{EnM03} the 
$^4$He binding energy came with a difference of about 0.15 MeV very close to the corresponding FY result.

Now we turn to the problem of spurious CM effects in CC calculations. 
This is an old problem that was discussed already in the 
work of Zabolitzky~\cite{Za81},
who used the common approach, already mentioned in Section~\ref{sec:NCSM}, of adding a CM
HO Hamiltonian to the intrinsic Hamiltonian. A different approach was used by Bishop et al. 
who formulated the problem and its solution in coordinate space~\cite{BiF90b}. 
Recently, it has been found numerically by Hagen, Papenbrock, and Dean~\cite{HaP09} that in a sufficiently
large model space, even when the model space
is not a complete $N\hbar\Omega$ space, the center-of mass
wave function is approximately a Gaussian whose
width varies little with the frequency $\Omega$ of the underlying
HO basis. The reported results on the $^{16}$O nucleus open the door for
a verifiable description of translationally invariant states
for a large variety of model spaces and many-body methods.
\begin{figure}[ht]
\begin{minipage}[b]{0.48\linewidth}
\centering
\includegraphics[scale=0.34,angle=-90]{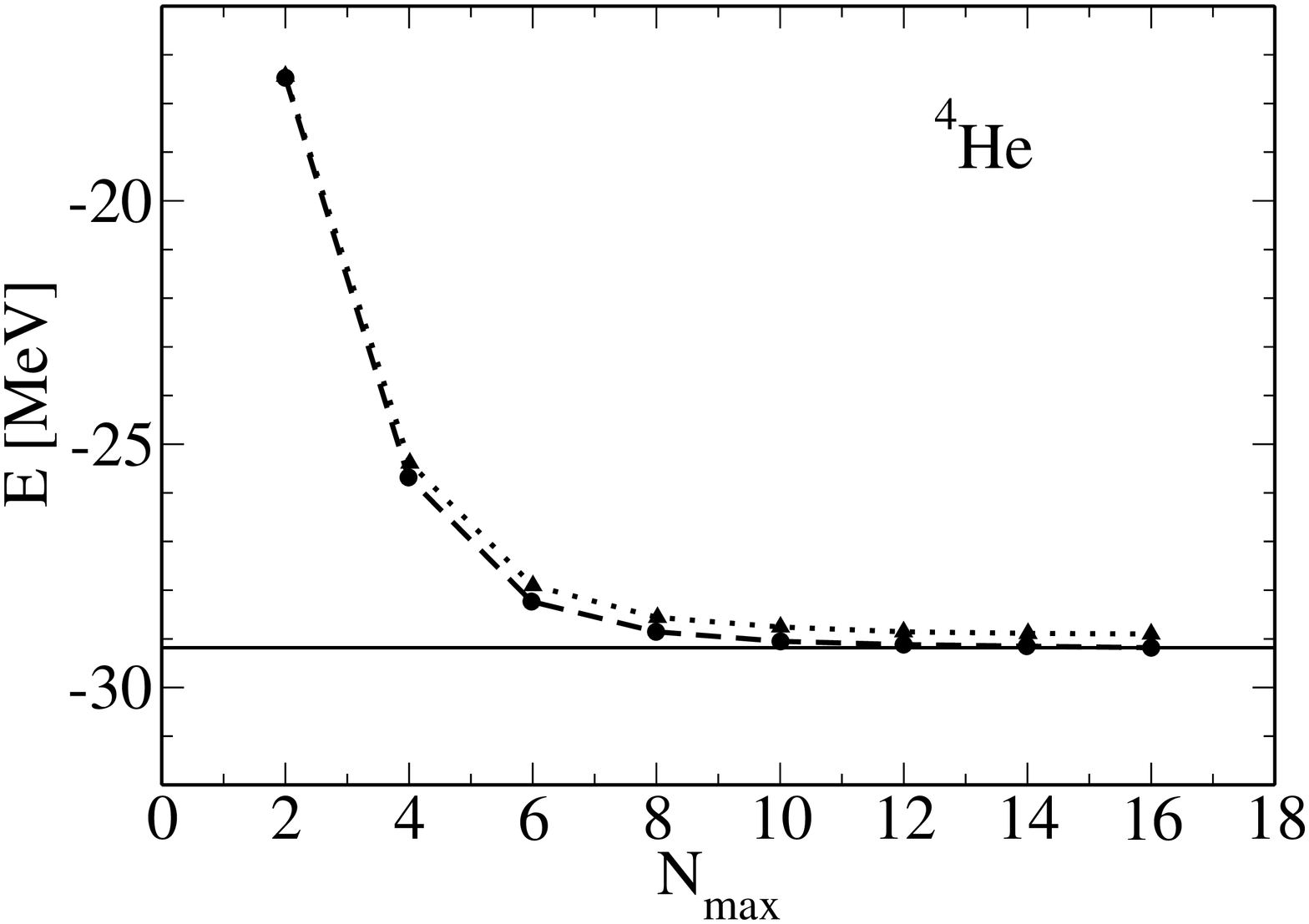}
\caption{$N_{\rm max}$ dependence of the $^4$He ground-state energy for a CC calculation with a $\vlowk$ NN potential;
shown are CCSD (dotted) and CCSD(T) (dashed) results, also given FY result~\cite{NoB04} (full) (from Ref.~\cite{HaD07}).}
\label{figure_CC_T}
\end{minipage}
\hspace{0.5cm}
\begin{minipage}[b]{0.48\linewidth}
\centering
\includegraphics[scale=0.34,angle=-90]{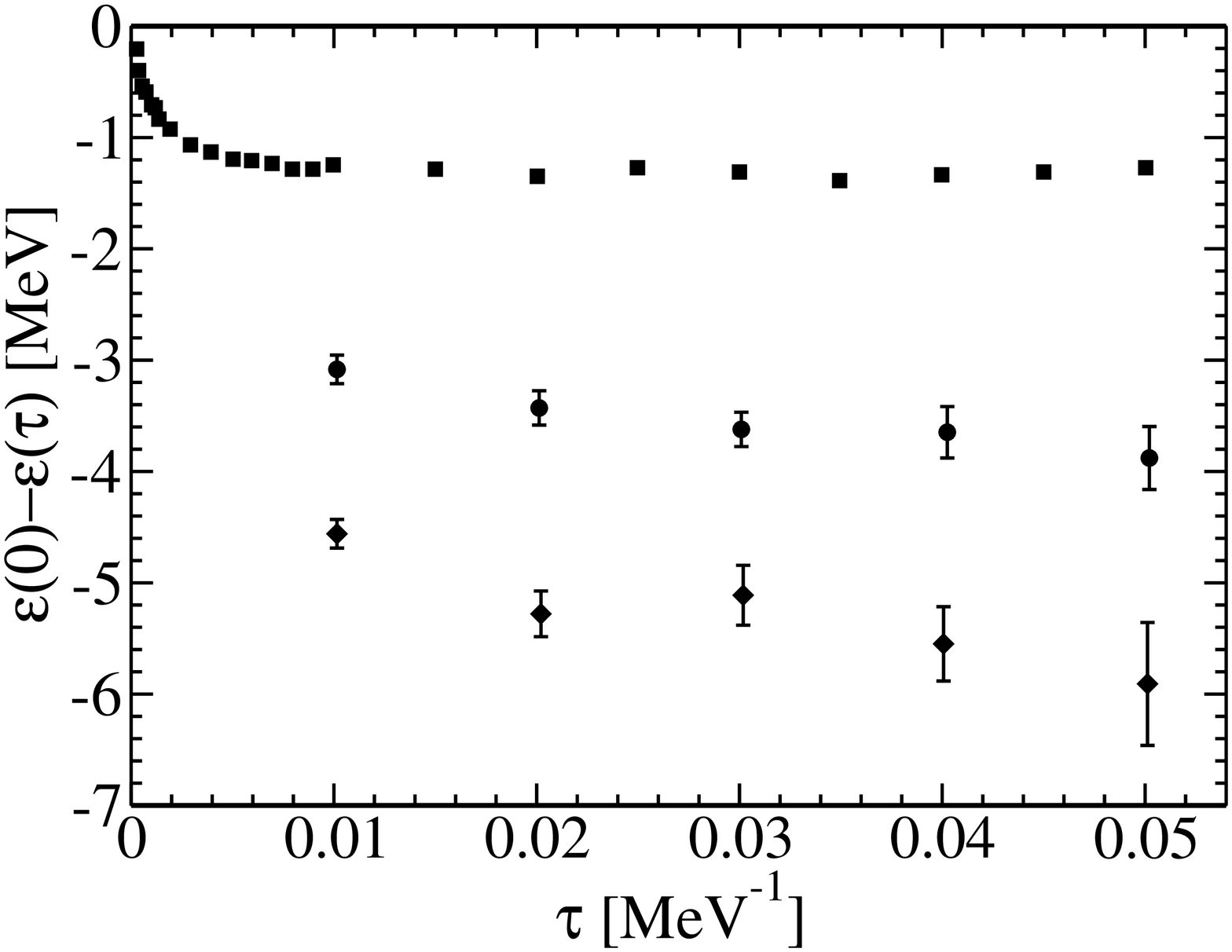}
\caption{Results for $\epsilon(\tau)-\epsilon(\tau = 0)$ from a GFMC ground-state calculation for
$^4$He (squares), $^6$He (circles) and $^7$Li (diamonds) with the AV18 NN potential~\cite{WiS95} and the Urbana IX 
3NF~\cite{PuP97} (from Ref.~\cite{PuP97}).}
\label{figure_GFMC_T}
\end{minipage}
\end{figure}

With the CC method it is possible to study excited states by making use of
the equation-of-motion (EOM)
formulation~\cite{Ro68}. Essentially this consists in writing an excited
state $|\Psi_k\rangle$ in terms
of an excitation operator $\Omega_k$ acting on the ground state, i.e.
$|\Psi_k\rangle=\Omega_k|\Psi_0\rangle$.
By starting from the following two Schr\"odinger equations,
\begin{eqnarray}
 H |\Psi_0\rangle &=& E_0|\Psi_0\rangle \,,\\
 H |\Psi_k\rangle &=& E_k\Omega_k | \Psi_0\rangle \,,
\end{eqnarray}
and subtracting the former from the latter one obtains the EOM expression
\begin{equation}
 [H,\Omega_k]| \Psi_0\rangle=(E_k-E_0) \Omega_k| \Psi_0\rangle \,.
\end{equation}

In applying the EOM idea to CC theory, one starts from the CC ground state
and writes the operator $\Omega_k$ as
a linear combination of the $\nu$-particle-$\nu$-hole $S^{[\nu]}$ operators (see Eqs.(~\ref{snu1}),~(\ref{snu2})).
Therefore since $\Omega_k$ commutes with $S$ 
one is then allowed to write
\begin{equation}
 [{\cal H},\Omega_k]| \Phi_0\rangle=(E_k-E_0) \Omega_k | \Phi_0\rangle\,,
\end{equation}
or, equivalently,
\begin{equation}
{\cal H}\Omega_k|\Phi_0\rangle = E_k\Omega_k|\Phi_0\rangle\,.
\end{equation}
Therefore one can construct the ${\cal H}$ matrix on the $\nu$-particle-$\nu$-hole basis
and then diagonalize it to obtain the excitation spectrum. Special care
has to be paid in the last two steps,
since ${\cal H}$ is non-Hermitian.

\subsection{A Formulation Suitable for  Monte Carlo (MC) Approaches}\label{sec:MC}

In this section we describe four methods for which the formulation of the quantum mechanical many-body problem
lends itself to a stochastic approach.  These methods, which are generally
referred to as Monte Carlo techniques, are
the Green's function Monte Carlo (GFMC), the chiral effective field theory on a lattice (LCEFT),
the Monte Carlo shell model diagonalization (MCSMD), 
and the variational Monte Carlo (VMC). 
The GFMC method is  based on the path integral formulation of quantum mechanics, as is the 
diffusion Monte Carlo (DMC).  The differences between the GFMC and DMC
methods are actually very small so that in the literature they are often interchanged.
The LCEFT is essentially a DMC approach as well, except that the dynamical
degrees of freedom are fields rather than particles.
The GFMC and LCEFT methods are based on the Euclidean time (imaginary time) evolution of the system.  
On the contrary the MCSMD, although inspired by the imaginary time formulation,
is still effectively a variational method. In fact MCSMD starts formally from an  
imaginary time evolved trial function, which after some manipulation, leads to an expression that suggests 
a way of constructing a variational SM basis.
The imaginary time becomes just one of the non-linear parameters.
The fourth approach, VMC, is a fully variational method, where the MC technique is used to evaluate 
the many-dimensional energy functional integrals.  The variational wave function that is obtained with this method  
is very important since it serves as starting trial function for the GFMC imaginary time evolution.

\subsubsection{The Green's Function Monte Carlo (GFMC) Method}\label{sec:GFMC}

The first application of a GFMC idea is due to Kalos~\cite{Ka62} who in 1962 calculated the ground state
of three- and four-body nuclei using some simple central interactions like square well and Gaussian potentials.
The application of the GFMC algorithm to nuclear Hamiltonians
containing spin and isospin operators was proposed in 1987 by Carlson~\cite{Ca87}.
To deal with a large number of nucleons and Hamiltonians
containing spin and isospin operators the auxiliary
field diffusion Monte Carlo (AFDMC) method, which uses the Hubbard-Stratonovich (HS) transformation~\cite{Hu59,St57}
to linearize quadratic terms in the evolution operator, was introduced by Schmidt and Fantoni~\cite{ScF99}.

With the GFMC method  it is possible to aim directly at the calculation of ground-state and low-lying energies.
In fact it is shown below that the ground-state energy $E_0$ can be obtained by 
 \begin{equation}
E_0=\lim_{\tau\to \infty}\epsilon(\tau)\label{GFMCeq}  \,,
\end{equation}
where $\tau$ is the so-called ``imaginary time`` defined as $\tau\equiv i t$ and
\begin{equation}
\epsilon(\tau)  = 
\frac{\langle \Psi | H | \Psi(\tau)\rangle}
     {\langle \Psi   |   \Psi(\tau)\rangle} \,.
\label{eq:mixed}
\end{equation}
Here $|\Psi (\tau)\rangle$ represents the evolution in imaginary time of a
function $|\Psi\rangle$ (in principle any function, called a
{\it trial} function)
\begin{equation}
|\Psi (\tau)\rangle = N e^ { - i H t } |\Psi\rangle  = N e^ { -  H \tau } |\Psi\rangle  \,,
\end{equation}
where the normalization $N$ is chosen as $N= e^{ \tau E_0}$. Therefore, instead of solving the Schr\"odinger 
equation to calculate the 
ground-state energy of a system, one  determines $E_0$ via 
\begin{equation}
 E_0=\lim_{\tau\to \infty}\frac{
   \langle \Psi|H e^{ - (H - E_0)  \tau}   | \Psi\rangle}
     {\langle \Psi| e^{ - (H - E_0)  \tau}   | \Psi\rangle}\,.\label{Epath}
\end{equation}

The proof for this assertion is rather simple. For this purpose one
formally writes the trial function as a linear combination of the eigenfunctions $|\Psi_\gamma\rangle$ of $H$:
\begin{equation}
| \Psi\rangle= \sum\!\!\!\!\!\!\!\int d\gamma \, c_\gamma |\Psi_\gamma\rangle \,.
\end{equation}
Inserting this expression in Eq.~(\ref{eq:mixed}) one finds
 \be
\nonumber
\epsilon(\tau) \,=\,  
\frac{\displaystyle\sum\!\!\!\!\!\!\displaystyle\int \!\!\!\!\!\!\!\! \displaystyle\int \, d\gamma^\prime d\gamma \, 
   \langle \Psi_{\gamma^\prime} |c_{\gamma^\prime}^* 
     e^{ - (E_{\gamma^\prime} - E_0)  \tau}  E_\gamma c_\gamma| \Psi_\gamma\rangle}
     {\displaystyle\sum\!\!\!\!\!\!\displaystyle\int \!\!\!\!\!\!\!\! \displaystyle\int \, d\gamma^\prime d\gamma \,  
\langle \Psi_{\gamma^\prime} |c_{\gamma^\prime}^*
     e^ {- (E_{\gamma^\prime}- E_0)  \tau} c_\gamma |   \Psi_\gamma\rangle}
 \,=\,
\frac{ \displaystyle\sum\!\!\!\!\!\!\!\displaystyle\int d\gamma \,|c_\gamma|^2 e^ { - (E_\gamma - E_0)  \tau } E_\gamma} 
     { \displaystyle\sum\!\!\!\!\!\!\!\displaystyle\int d\gamma \,|c_\gamma|^2 e^ { - (E_\gamma - E_0)  \tau}  } \,.
\label{epstau}
\ee
When the imaginary time tends to infinity and under the condition that the trial function is not orthogonal to the ground state,  
the only contribution to the sum/integral that survives is the ground-state energy:
 \begin{equation}
\epsilon(\tau\to \infty) =  E_0 \,.
\end{equation}

The reason why this formulation takes the name of {\it Green's Function} MC is clear if one represents
in configuration space the function $|\Psi(\tau)\rangle$,
necessary to calculate  $\epsilon(\tau)$ numerically. 
In fact in configuration space one has
\bea
\nonumber
|\Psi(\tau) \rangle &=& e^ { -  (H-E_0) \tau } |\Psi\rangle\\
&=&\int\int\ d{\bf R} d{\bf R}'|{\bf R}\rangle\langle {\bf R}| e^ {- (H- E_0)  \tau}
|{\bf R}'\rangle\langle {\bf R}'|\Psi\rangle
\,\equiv\, \int\int d{\bf R} d{\bf R}'|{\bf R}\rangle
 G({\bf R},{\bf R}',\tau)\langle {\bf R}'|\Psi\rangle\,,
\label{greenf}
\eea 
where ${\bf R} $ represents all the particle coordinates and $G({\bf {\bf R},{\bf R}'},\tau)$ is just 
the Green's function of the imaginary time evolution operator. 

In order to evaluate $|\Psi(\tau) \rangle$ one writes the evolution operator as   
\be
e^ {- (H - E_0)\tau}=\left[e^ {-( H - E_0)\Delta\tau}\right]^n \label{deltatau}
\ee
and therefore the Green's function becomes
\be
G({\bf R},{\bf R}',\tau)=\int \cdot\cdot\cdot \int d{\bf R}_n d{\bf R}_{n-1} \cdot\cdot\cdot 
d{\bf R}_1 G({\bf R},{\bf R}_n,\Delta\tau)G({\bf R}_n,{\bf R}_{n-1},\Delta\tau) \cdot\cdot\cdot
G({\bf R}_2,{\bf R}_1,\Delta\tau)G({\bf R}_1,{\bf R}',\Delta\tau) \,.
\label{GR}
\ee
At each time step the evolution is implemented by the propagation of a finite collection of points ($\delta$-functions)
approximating the function in ${\bf R}$.

Dividing the evolution into small imaginary time steps is crucial in order to write the explicit expression 
for $G({\bf R},{\bf R}_n,\Delta\tau)$ as~\cite{An74}.
\be
G({\bf R},{\bf R}',\Delta\tau)=\left(\frac{1}{2\pi\Delta\tau}\right)^{\frac{3N}{2}}{\rm exp}
\left\{-\frac{m ({\bf R}-{\bf R'})^2}{2\Delta\tau}-\left[{\frac {1}{2}} \left( V({\bf R})+V({\bf R'}) \right)-E_0\right]\Delta\tau\right\}\,.\label{GG}
\ee 
Such an expression is correct only to the order $\Delta\tau$. 
In order to better understand the above expression one can 
consider the  Schr\"odinger equation in imaginary time as equivalent to a diffusion equation with an absorption term
governed by the Hamiltonian. In the zero mass limit, i.e. when the kinetic energy becomes dominant, 
the solution of this diffusion equation is 
\be
\Psi({\bf R},\tau) = \int d{\bf R'} G({\bf R}-{\bf R}',\tau)\Psi({\bf R'},0)
\ee
with the  Green's function given by
\be
G({\bf R}-{\bf R}',\tau)=\left(\frac{m}{2\pi\hbar^2\tau}\right)^{\frac{3N}{2}}{\rm exp}
\left\{-\frac{m ({\bf R}-{\bf R}')^2}{2\hbar^2 \tau}\right\}\,.
\ee 
On the other hand, the consideration of the infinite mass limit gives the term $[(V({\bf R})+V({\bf R'}))/2-E_0]$ 
proportional to  $\Delta\tau$ in the argument of the Gaussian.
It is clear that the sum in the exponent of a kinetic and a potential term  is exact only for small increments 
of the imaginary time, since the two operators do not generally commute.

There are a few delicate points in the numerical application of the GFMC procedure. One of them has to do with
the choice of the normalization constant $N= e^{ \tau E_0}$, 
which is just a function of the ground-state energy itself.
Such a choice is crucial to keep the denominator in Eq.~(\ref{eq:mixed}) equal to a finite value 
(actually equal to one) in the infinite $\tau$ limit.
So one of the delicate points  consists in keeping the normalization under control without knowing the value of $E_0$. 

Another important point is the so-called ''sign-problem`` that arises in the case of fermions.
In fact the  time evolution leads in principle to the 
state of minimal energy, which is the bosonic ground state.  There are various ways to cure this 
problem approximately. The most widespread is the  {\it fixed node approximation}. 
It is based on the fact that an antisymmetric continuous  wave function  must become zero on 
some hypersurface of dimension (3N-1). 
Since one does not know this hypersurface, i.e. one does not know the correct nodal structure of
the exact fermion wave function, one imposes that the evolved function has the same nodal structure as the antisymmetric
trial function. The error made by this approximation is not known. One can only prove that the value of the
energy obtained starting from a variational trial function (like e.g. the VMC discussed in Section~\ref{sec:VMC}) 
lies between the variational result and the exact one. We refer
to Ref.~\cite{FoM01} for a comprehensive discussion of the sign problem. 

The GFMC method requires a very good trial function as a starting point, in order to be able 
to reach  convergence, while increasing the imaginary time. Therefore the VMC result
is often used as a trial function.
To have an idea of the quality of the results for $E_0$ in the  $\tau$ infinity limit and of the rate of convergence
in Fig.~\ref{figure_GFMC_T} we show  one of the results from Ref.~\cite{PuP97}. 
As expected the errors grow with $\tau$ and $A$. 
It is interesting to note that the $^4$He points do not show visible errors. This is due to the fact that 
the sign problem is avoided in this case. In fact for $^4$He the antisymmetry of the wave function resides 
entirely in the spin-isospin Slater determinant. Therefore the spatial part is symmetric and the problem is equivalent 
to a boson problem.

The GFMC technique can be applied to excited states as well. If they have quantum numbers different from
the ground state, the procedure is exactly as described above, provided that the            
trial function has the desired excited state quantum numbers.
It is also possible to treat a few excited states with the 
same quantum numbers as the ground state or any other reference state~\cite{PiW04}. 
In this case one proceeds as follows. 
One first constructs a $\tau$-dependent Hamiltonian matrix $H (\tau)$ between normalized states 
\begin{equation} 
 H_{ij}(\tau) = \frac {\langle \Psi_i(\tau/2)| \ H \ | \Psi_j(\tau/2) \rangle}  
 {|\Psi_i(\tau/2)| |\Psi_j(\tau/2)|} \,. 
\end{equation}  
This is possible because in GFMC one can compute {\it mixed} expectation values starting   
from states $|\Psi_i\rangle$ and $|\Psi_j\rangle$ obtained by diagonalizing a VMC Hamiltonian and
propagated separately. 
One then solves the generalized eigenvalue problem for $H(\tau)$.  
For large values of $\tau$ one recovers the exact discrete spectrum and a discretized continuum spectrum. 

A different technique is largely used in atomic, molecular and cluster physics. Here one starts
from the imaginary time evolution of a state which is prepared applying a projection operator $\Pi$
on an initial trial function that selects the proper quantum numbers. 
More explicitly, what is calculated by GFMC for many values of $\tau$ is
the quantity
\begin{equation}
{\cal L}(\tau)=\langle \psi_i|\Pi^\dagger e^{ - (H - E_i)  \tau}  \Pi | \psi_i\rangle \,.
\end{equation}
It can be easily shown that ${\cal L}(\tau)$ is equivalent to the Laplace transform
of the response function $R(E)$ of a system, which is initially in a state $\psi_i$, to a  perturbation
induced by $\Pi$ (see Eqs.~(\ref{R_E}) and (\ref{Phi})).
This means that knowing $R(E)$ one also knows the excitation spectrum of $H$ induced by $\Pi$. This is recovered
from its Laplace transform by means of different algorithms (see e.g.~\cite{BlL97}). Due to the ill-posed
nature of the inverse Laplace transform, however, these algorithms may have a scarce efficiency (see also
Section~\ref{sec:LIT}). 

The GFMC technique can also be applied to low-energy scattering, where only a single two-fragment channel is open. 
As described in Ref.~\cite{NoP07} one 
proceeds as follows. First a VMC calculation is
carried out for a given scattering energy $E_0$ in such a way that at a large
distance, $R_0$, the wave function has the typical asymptotic form
consisting of internal wave functions for the two fragments and a relative
wave function for a given angular momentum, where the radial part is
a proper superposition of Bessel and Neumann functions. This leads to
a well-defined logarithmic derivative $\gamma_0$ at $R_0$. In a subsequent GFMC calculation the VMC wave function is used as trial function. The GFMC
simulation is restricted to a volume corresponding to cluster separations less than
$R_0$ and the logarithmic derivative $\gamma_0$ is used as boundary condition. The GFMC procedure then leads to the proper energy $E$ corresponding
to the boundary condition. Having determined
$R_0$, $\gamma_0$, and $E$ the phase shift is uniquely defined.  

The applicability of the method is limited to local potentials in configuration 
space. Potentials in momentum space, and in general non-local potentials, can generate great problems in 
finding an expression for the propagator $G({\bf R},{\bf R}_n,\Delta\tau)$ that can be sampled. 
Therefore phenomenological local configuration space potentials have been constructed ad hoc for application
of the  GFMC approach  to nuclear systems; the modern realistic AV18 potential is one of them.

\subsubsection{Lattice Chiral Effective Field Theory (LCEFT)} \label{sec:LCEFT}

Effective field theory provides a systematic approach to studying low-energy phenomena in few- (and many-body) systems. 
Like in all effective field theories, there is a wide disparity between the long-distance 
scale of low-energy phenomena and the short-distance scale of the underlying interaction. One can characterize 
the low-energy phenomenology according to 
exact and approximate symmetries and low-order interactions according to some hierarchy 
of power counting. In the case of nuclear physics and
for nucleon momenta comparable to the pion mass, both the contribution from nucleons and pions 
must be included in the effective theory.

Recently an innovative method that combines the MC formulation with the chiral effective field theory of the 
nuclear force has been introduced (for a review see Ref.~\cite{Le09}). The approach is similar to lattice QCD, but 
the fundamental fields are replaced by effective nucleon and pion fields. 
The nuclear interaction is described by an effective chiral potential and the inverse of the lattice spacing serves 
as an ultraviolet regulator of the theory. The ground-state energy of the nucleus is also here filtered out at 
large imaginary time.

In practical applications one builds up the action $\mathcal{S}$ in terms of nucleon 
 and pion fields. If the action is given by the functional $\mathcal{S}$ of field configurations, 
then the time ordered vacuum expectation value of a functional $H$ is given by
\begin{equation}
  \left\langle H\right\rangle=\frac{\int \mathcal{D}\phi H[\phi] e^{ i  \mathcal{S}[\phi]}}{\int\mathcal{D}\phi 
  e^{i\mathcal{S}[\phi] }}, \label{pathint}
\end{equation}
where the symbol $\int \mathcal{D}\phi$ 
represents the infinite-dimensional integral over all possible field configurations on the entire space-time. 

In view of an evaluation of $\left\langle H\right\rangle$ on the  lattice one can express the time integral 
of the Lagrangian  $\mathcal{L}$ in the action $\mathcal{S}$ as a sum of N small time intervals $\Delta t$, and 
therefore $i \mathcal{S}$ as a sum of N small Euclidean (imaginary) time intervals $\Delta \tau$:
\begin{equation}
 i \mathcal{S}= \sum_k i  \mathcal{L} \Delta t_k = \sum_k  \mathcal{L} \Delta\tau_k    \,.
\end{equation}
In a four-dimensional lattice evaluation of $\left\langle H\right\rangle$  the time interval $\Delta \tau$
corresponds to the imaginary time lattice spacing $a_\tau$.

>From the technical point of view one then proceeds in essentially the same way as in the GFMC
(compare Eq.~(\ref{pathint}) with Eq.~(\ref{Epath})). Due to the appearance of quadratic forms in the fields 
in the so-called {\it transfer matrix}, i.e. of the normal-ordered exponential of the Hamiltonian over one 
temporal lattice spacing, one makes use of auxiliary fields, like in the AFDMC.
The main difference between the LCEFT approach and the GFMC (or AFDMC) consists in the different choice of 
the dynamical degrees of freedom: particles in the case of GFMC or AFDMC, fields in LCEFT. 

In principle potentials derived in the EFT framework
could also be used in the GFMC method  provided that 
their representation  in configuration space is given.

\subsubsection{Monte Carlo Shell Model Diagonalization (MCSMD) Method} \label{sec:MCSMD}

There are two different methods that can be confused because they are both referred to as {\it Monte Carlo 
shell model}.
The first, described in Ref.~\cite{KoD97} is an extension of  Eq.~(\ref{GFMCeq}) to
large SM problems with core.
The second, MCSMD, introduced in Ref.~\cite{HoM95} and 
illustrated in~\cite{OtH01}, is in some way similar to the SVM discussed in Section~\ref{sec:SVM}, in that it 
is  based on a stochastic generation of the basis.

The MCSMD method  has been inspired by the AFDMC, but it does not make use of
Eq.~(\ref{GFMCeq}). Instead the method consists in  creating  a basis starting formally from an  
imaginary time evolved trial function (taken as a HO slater determinant $|\Psi_0\rangle$)
\begin{equation}
|\psi \rangle = e^{-\tau H} |\Psi_T\rangle\,,
\end{equation}
and using the HS transformation~\cite{Hu59,St57}. This linearizes the two-body 
interaction by auxiliary fields and allows $|\psi \rangle$ to be rewritten as 
\begin{equation}
|\psi \rangle = \int d\sigma N(\sigma) e^{-\tau \sigma \hat h} |\Psi_0\rangle\,,
\end{equation}
where the operator $\hat h$ is now a one-body operator.
At this point, however, one does not proceed to the $\tau \to \infty$ limit,
but instead one observes that the vectors 
\begin{equation}
|\phi(\sigma) \rangle= N(\sigma) e^{-\tau \sigma \hat h} |\Psi_0\rangle\,,
\end{equation}
could be good candidate states for a basis. 
In order to explain the MCSMD procedure, let us assume to have already chosen a certain number of basis states by random
$\sigma$ values. The ground-state energy is calculated. Another vector is added to the basis by choosing a new random 
value of $\sigma$, and after having orthogonalized it with respect to the other basis vectors the Hamiltonian is diagonalized again. 
Only if the ground-state energy shows a significant decrease the state is kept.  
The procedure is repeated up to convergence.
During the MCSM generation of the basis vectors, symmetries, e.g. rotational and parity
symmetry, are restored before the diagonalization. 

\subsubsection{The Variational Monte Carlo (VMC) Method} \label{sec:VMC}

Another variational method that uses the stochastic approach is the variational Monte Carlo (see e.g. Ref.~\cite{PiW01}). 
In this case the trial function,
written in configuration space, consists of a sum of differently correlated  antisymmetric single-particle states. 
Correlations of a various nature are obtained by operating with  a certain number 
of two- and three-body {\it correlation operators}. They reflect the  operator form of the potential and have an influence on the short-range part of the wave function.  
Appropriate boundary conditions are imposed at long range.
The stochastic nature of the approach lies in the Metropolis Monte Carlo evaluation of the energy functional 
multi-dimensional integral.

More specifically the variational trial function has the form
\begin{equation}
     |\Psi_T\rangle = \left[1 + \sum_{i<j<k} W_{ijk} \right] 
                      \left[ {\mathcal S}\prod_{i<j}(1+U_{ij}) \right]
                      |\Psi_J\rangle \ .
\label{eq:psit}
\end{equation}
The $U_{ij}$ and $W_{ijk}$ are non-commuting two- and three-nucleon
correlation operators
\begin{equation}
     U_{ij} = \sum_{p} u_p(r_{ij}) O^p_{ij} \ ,
\label{eq:uij}
\end{equation}
where the $O^{p}_{ij}$ are the same  operators
that appear in the NN potential used.
The $W_{ijk}$ have the spin-isospin structure of the dominant parts
of the 3NF, as suggested by perturbation theory,
${\mathcal S}$ is a symmetrization operator  and $\Psi_J$ is 
the fully antisymmetric wave function that has already been Jastrow correlated. For example, for
$s$-shell nuclei one writes
\begin{equation}
     |\Psi_J\rangle = \left[ \prod_{i<j<k}f^s_{ijk} \right]
                      \left[ \prod_{i<j}f_s(r_{ij}) \right]
                     |\Phi_A(JMTT_{3})\rangle \ ,
\label{eq:jastrow}
\end{equation}
where $|\Phi_A(JMTT_{3})\rangle$ is a simple spin-isospin Slater determinant and the various $f$ are correlation functions.

For a generic nucleus $|\Psi_J\rangle$ is more complicated because it is a linear combination of $N$ channel basis functions 
characterized by total orbital angular momentum $L$, total spin $S$ and principal quantum number set $[n]$.

The overall wave function is translationally invariant. This is clear when
the single-particle basis is a simple spin-isospin Slater determinant, or a linear combination of them, since
the correlation functions are functions of distances between particles. However, even when the single-particle basis 
is a Slater determinant of single-particle wave functions from a parametrized one-body potential, one can implement  
translational invariance numerically. In fact at each step of the Metropolis sampling the values of the $A$ particle positions ${\bf r}_i$ are substituted 
by the positions with respect to the CM $ {\bf r}_i\,'= {\bf r}_i - 1/A \sum_i {\bf r}_i$ and the kinetic energy is renormalized 
in an analogous way. 

In Table~\ref{Wi90_7} we show VMC results for various $^4$He ground-state quantities in comparison to GFMC results.
One notes that the ground-state energy underestimates the precise ab initio result by only 1 MeV, whereas the differences
amount to about $\pm7$ MeV for expectation values of kinetic energy and NN interaction. The results for
the 3NF differ by 1 MeV. Summarizing one can conclude that the VMC is already rather
close to the precise ab initio results and thus supplies a very good trial wave function for the GFMC calculation.
\begin{table}
\begin{center}
\caption{{\label{Wi90_7}} $^4$He ground-state expectation values in units of MeV for ground-state energy $E$, kinetic 
energy $T$, $V_{ij}$ (AV14 NN potential~\cite{WiS84}), and $V_{ijk}$ (Urbana VIII 3NF~\cite{ScP86}) from 
VMC~\cite{Wi91} and GFMC~\cite{Ca88} calculations (GFMC result as given in~\cite{Wi91}). }
\begin{tabular}{c||c|c|c|c}
\hline \hline 
 & $E$ & $T$ & $V_{ij}$ & $V_{ijk}$ \\
\hline
VMC  & --27.2(2) & 106.6(8) & --129.7(7) & --4.84(9)   \\
GFMC  & --28.3(2) & 113.3(20) & --136.5(20) & --5.8(3)  \\
\hline \hline
\end{tabular}
\end{center}
\end{table}

\subsection{Bound-State Formulations of the Continuum Problem}\label{sec:bs-cont}

Various formalisms discussed in the previous sections have  been applied in the past
to ab initio calculations of nuclear observables for nuclei with $A>3$.
However, only  energies below the threshold for three-body break-up were considered. 
The solution of the quantum mechanical problem in the far continuum i.e. where more than two fragments exist 
is much more complicated, since one then has 
to deal with the full many-body scattering problem. Several methods have been proposed to reduce the
continuum state problem to a bound state one~\cite{NuC69,Ef85,EfL94,UzK03}, which have in common  
the use of a complex expression 
for the resolvent operator. In Refs.~\cite{UzK03,DeF12} a numerical extrapolation to the real axis is performed.
In the following we will discuss the two  methods~\cite{EfL94,SuH10,HoZ12}, where one remains on the complex plane. 
They enable calculations of  perturbation induced reactions to the continuum and at the same time 
avoid an explicit determination of continuum wave functions.

\subsubsection{The Lorentz Integral Transform (LIT) Method}\label{sec:LIT}

The LIT method was first formulated for perturbation induced reactions involving transitions 
to the  continuum~\cite{EfL94}.
The success of the method has been such that it has   lead to  ab initio results for observables of  $A=4,6,7$ nucleon systems 
beyond the three-body break-up threshold and even in the full continuum.  

To explain the LIT method consider a typical observable for a perturbatively-induced reaction, namely
the inclusive response function 
\be
\label{R_E}
R(E) = \sum\!\!\!\!\!\!\!\int d\gamma \, |\langle \Psi(E_\gamma) 
| {\cal O} | \Psi(E_0) \rangle|^2 \,\delta(E_\gamma-E_0) \,,
\ee
where ${\cal O}$ is an operator inducing a transition from the ground state $\Psi(E_0)$ with ground-state energy $E_0$
to a state $\Psi(E_\gamma)$ with energy $E_\gamma$. Performing an integral transform with a well-defined and smooth
kernel $K(\sigma,E)$ one has
\be
\Phi(\sigma) = \int dE \, R(E)\, K(\sigma,E) \,\label{LIT}
\ee
which by inserting the relation (\ref{R_E}) becomes
\bea
\nonumber
\Phi(\sigma) &=& \sum\!\!\!\!\!\!\!\int\, d\gamma \, \langle \Psi(E_0) | {\cal O}^\dag | \Psi(E_\gamma) \rangle \, K(\sigma,E) 
\, \langle \Psi(E_\gamma) | {\cal O} | \Psi(E_0) \rangle \\
  &=&  \sum\!\!\!\!\!\!\!\int d\gamma \, \langle \Psi(E_0) | {\cal O}^\dag K(\sigma,H)| \Psi(E_\gamma) \rangle \,
  \langle \Psi(E_\gamma) | {\cal O} | \Psi(E_0) \rangle\,.
\eea
Use of the  closure property of the eigenstates of $H$,
\be
\sum\!\!\!\!\!\!\!\int\, d\gamma \,|\Psi(E_\gamma) \rangle \langle \Psi(E_\gamma)| = 1 \,,
\ee
then leads to
\be
\label{Phi}
\Phi(\sigma) = \langle \Psi(E_0) |K(\sigma,H)| \Psi(E_0)\rangle \,. 
\ee
The principal idea of the integral transform method consists in the calculation of the transform 
$\Phi(\sigma)$ without knowing $R(E)$
and next in the inversion of the transform in order to obtain it. 
For this purpose one needs an appropriate kernel $K(\sigma,E)$. In the past the Stieltjes transform with
$K(\sigma,E)=1/(E+\sigma)$ with $\sigma > 0$, proposed by Efros \cite{Ef85}, and the Laplace transform 
with $K(\sigma,E)=\exp(-\sigma E)$,
used in Ref.~\cite{CaS92},  were considered in the past for nuclear physics problems first.
Both transforms suffer a drawback in that information about the response
$R$ at a given energy $E$ is dispersed over a rather large $\sigma$ range. 
This makes
the inversion of the transform rather unstable as discussed in detail in Ref.~\cite{EfL93} for the Stieltjes
transform.

Considerable improvement of the inversion results from kernels that distribute the strength over only a
relatively small $\sigma$ range. Such a request is satisfied by kernels that are representations of the 
$\delta$-function. Therefore Efros and the authors of the present review article have proposed   
the Lorentz kernel~\cite{EfL94}
\be
K(\sigma,E)= {\frac {1}{(E-E_0-\sigma_R)^2 + \sigma_I^2} } 
           = {\frac{1} {E-E_0-\sigma_R - i\sigma_I} }\, {\frac{1} {E-E_0-\sigma_R + i\sigma_I} } \,,
\ee
where $\sigma_I$ is a constant chosen to be sufficiently small and $\sigma_R>0$. Insertion of this kernel into 
Eq.~(\ref{Phi}) leads to the following result for the LIT:
\be
L(\sigma_R,\sigma_I) = \langle\tilde\Psi | \tilde\Psi\rangle \,.
\ee 
The localized LIT function $\tilde\Psi$ is obtained from the LIT equation:
\be
\label{LITeq}
(H-E_0-\sigma_R-i\sigma_I) |\tilde\Psi\rangle = {\cal O} |\Psi(E_0)\rangle \,.
\ee
The LIT equation represents a bound-state like equation and thus can be solved by applying one of the usual 
methods for a solution of a bound-state problem. Once $L(\sigma_R,\sigma_I)$ is calculated $R(E)$ can be 
obtained by inverting Eq.~(\ref{LIT}). The reduction from an originally continuum-state to a bound-state like problem
is an enormous simplification and allows the calculation of electroweak observables which otherwise,
until the present, cannot be calculated. 

A benchmark test for the triton total photoabsorption cross section in unretarded 
dipole approximation  has been performed in Ref.~\cite{GoS02}. 
For this purpose the transform $L(\sigma_R,\sigma_I)$ has been calculated for $\sigma_R= 0,$ 1, 2, ..., 300 MeV 
and at a fixed $\sigma_I$ of 10 MeV.
This allowed the integral transform to be inverted with great accuracy
so that the unretarded dipole response function $R_{D, \rm unret}(E_\gamma)$ could be determined with high precision.
Finally by multiplying $R_{D, \rm unret}(E_\gamma)$ with a trivial kinematical factor one obtains the  
total photoabsorption cross section in unretarded dipole approximation. In Fig.~\ref{figure_LIT_T} the small inversion
error range is represented by the space between the two almost overlapping curves, while the dots represent
the Faddeev result calculated at various energies. A nice agreement between the 
LIT and Faddeev results can be observed.

The LIT approach can also be applied to exclusive processes, for which details we refer to Ref.~\cite{EfL07}.
Applications have been made for the reactions 
$^4$He$(\gamma,$N$)^3$X~\cite{QuB04}, $^4$He$(e,e'$p$)^3$H~\cite{QuE05}, and $^4$He$(e,e'd)^2$H \cite{AnQ06}.
The applicability of the LIT to photon scattering has also been established in Ref.~\cite{BaL11}.

A remark concerning the inversion of the LIT is in order here. The
inversion represents a so-called ill-posed
problem. The mathematical term {\it ill-posed} is somewhat misleading,
since ill-posed problems can be solved reliably.
For a detailed discussion of this issue we refer to Ref.~\cite{BaE09},
where it is pointed out that the LIT method has to be understood as
an approach with a {\it controlled resolution}.
\begin{figure}[tb]
\begin{center}
\epsfig{file=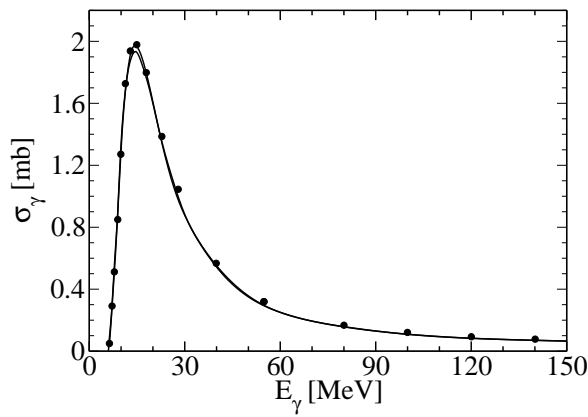,scale=0.3,angle=-90}
\caption{Results for the total $^3$H photoabsorption cross section with the AV18 potential
from LIT/CHH (full) and Faddeev (dots) calculations in unretarded dipole approximation
(the two curves represent the uncertainty of the inversion of the LIT); from Ref.~\cite{GoS02}.  }
\label{figure_LIT_T}
\end{center}
\end{figure}

The solution of Eq.~(\ref{LITeq}) can be obtained by matrix inversion
after  expanding the function $|\tilde\Psi\rangle$
on localized functions. However, in practical calculations one can
avoid the matrix inversion. In fact one can
rewrite $L(\sigma_R,\sigma_I)$ as
\be
\label{lin_resp}
L(\sigma_R,\sigma_I)= -\frac{1}{\sigma_I}Im\left\{\langle \Psi(E_0) | {\cal O}^\dagger
\frac{1}{\sigma_R + E_0 - H + i\sigma_I}{\cal O}|
\Psi(E_0) \rangle\right\} \,,\label{lanczos}
\ee
and apply the Lanczos algorithm with pivot ${\cal O}| \Psi(E_0) \rangle$~\cite{MaB03}.

It is interesting to note that if the Lorentz kernel tends to the $\delta$-function ($\sigma_I\to 0,$ $\sigma_R +E_0 = E$),
Eq.~(\ref{lanczos}) corresponds to the relation
between $R(E)$ and the so-called {\it linear response} $\chi(E)$,
\be
\label{resp}
R(E)=-\frac{1}{\pi}Im\{\chi(E)\}=-\frac{1}{\pi} Im\left\{\langle \Psi(E_0) | {\cal O}^\dagger
\frac{1}{E-H+i\epsilon}{\cal O}| \Psi(E_0) \rangle\right\} \,.\label{R(E)}
\ee
We would like to stress that in the LIT method the continuum problem reduces to a bound-state like problem,
due to the fact that the resolvent 
$1/(E-H+i\epsilon)$ acquires a finite imaginary part in the denominator. 

\subsubsection{The Complex Scaling Method (CSM)} 

Another approach whose goal is the rigorous reduction of continuum problems to bound-state-like problems 
is the complex scaling method. 
This method appears to have been first introduced in 1944  by D.R. Hartree et al. for solving problems 
relevant to air defense research and development~\cite{CoS83}. It was rediscovered in the late 
sixties/early seventies~\cite{NuC69,BaC71,Si73} to study scattering amplitudes and resonances
and is still an active field of current research in atomic and molecular physics (for a review  see Ref.~\cite{Mo98}).

The method is based on a similarity transformation performed by the complex scaling operator 
\be
\hat S(\theta)= e^{i \theta r \partial / \partial r}
\ee
such that
\be
\hat S(\theta) f(r)=f(r e^{i \theta})
\ee
for any analytical function $f(r)$.
This corresponds to a  transformation of $\r$ and $\p$ operators such that
$\r\to e^ {i\theta} \r$ and $\p \to e^ {-i\theta} \p$. 

The complex scaling is particularly suited for studying resonances since 
it associates the resonance phenomenon with the
discrete part of the spectrum of the complex-scaled Hamiltonian. 
(The complex eigenvalues  do not depend on $\theta$, although
the resonance complex-scaled eigenfunctions do). 

In nuclear physics ab initio calculations the CSM has
only been applied in Ref.~\cite{LaC11} for calculating   elastic and three-particle break-up observables in the 3N system,
and in Ref.~\cite{LaC05a} for investigating   four-neutron resonances. Here, however, we would like to mention 
an interesting application 
of the complex scaling idea to the direct calculation of $R(E)$.

In Ref.~\cite{HoZ12} the complex scaling transformation has been used in the expression of $R(E)$,
\be
\label{respcc}
R(E)= -\frac{1}{\pi}Im\left\{\langle \Psi(E_0) | {\cal O}^\dagger\hat S^\dagger(\theta)
\frac{1}{E-H(\theta)+i\epsilon}\hat S(\theta) {\cal O}| \Psi(E_0) \rangle\right\} \,,
\ee
where $H(\theta)=\hat S(\theta) H \hat S^\dagger(\theta)$.
In this way the resolvent acquires a finite imaginary part.
Therefore, within a suitable range of positive 
$\theta$, one is allowed to solve the eigenvalue problem for $H(\theta)$  by an expansion on square integrable functions.
If sharp resonances exist, $\theta$ has to cover the resonance poles on the
complex energy plane~\cite{Mo98}. 

A very nice feature of the CSM calculation of R(E) is that one has a direct
access to  it  and  no inversion is required. On the other hand, 
while in the LIT case the stability of the inversion has to be studied, in the CSM
the stability of $R(E)$ with respect to $\theta$ has to be examined,

The CSM approach  
 appears to be very promising for calculating perturbative inclusive reactions 
 and deserves to be investigated further.
 
\section{Discussion of Results}\label{sec:RECRES}

In this section we  present an overview   of results obtained from ab initio calculations for nuclear systems 
with $A \ge 4$. 
A complete overview of the field   is beyond our scope. Instead, here our aim is, on the one hand to illustrate  
the development of the field from a historical perspective and on the other hand to assess today's state 
of the art, paying particular attention to
 benchmark calculations.

It goes without saying that modern-day methods and techniques in few-body calculations could not have been 
developed without the continuous progress made over, at least, the last fifty years.  Therefore we  think 
that,  in addition to reporting  more recent results, we have 
also to pay tribute to those researchers who have made the first attempts to solve the nuclear 
many-body problem with  ab initio calculations using the most realistic potentials of the time, 
 thus laying out the 
basis for further developments. 
 
Here we are dealing with nuclear systems with $A \ge 4$, but one should not forget 
that progress in treating the 3N system had an important influence on the development of the field of 
more complex few-body nuclei (for reviews see Refs.~\cite{GlW96,CaS98,GoS05}). 

One should note that the spectra of nuclear systems 
with $A \ge 4$ are generally much richer than those of the 2N and 3N systems. Let us take as example the 
4N system 
with isospin $T=0$: One has the $^4$He bound state and five different continuum channels with five different
thresholds, namely two (3+1) channels ($^3$H-$\,p$, $^3$He-$\,n$), one (2+2) channel ($d\,$-$\,d$), one (2+1+1) 
channel ($d$-$np$), and a (1+1+1+1) channel ($2n2p$). This can be compared to the 3N system in the 
isospin $T=1/2$ channel with projection $T_z=1/2$, where one has one bound state  ($^3$He), 
but only two different continuum channels (($dp$) and ($npp$)). The comparison shows that the difference between the 3N and 
4N systems does
not just consist in one additional nucleon, but also in a more complex structure of the excitation spectrum.

Our discussion will show that whereas enormous progress has been made over the years
in few-nucleon calculations, still quite some work remains to be done.
The most visible advancement has been obtained for bound-state 
calculations,  where even   $A>10$ systems  have been tackled. On the other hand continuum-state problems are considerably more 
complicated. For example, with the exception of the LIT method, common ab initio 
methods have not yet succeeded in controlling the complete 4N continuum up to pion threshold.
Rather, the energy has been limited to the 
range below the three-body break-up threshold. Thus, it is not surprising that also for the $A>4$ nuclear 
systems such ab initio methods are presently only able to treat the energetically lowest two-fragment break-up
channels.

The following discussion  is organized as follows. In Sections \ref{sec:4_hist} we
describe the main aspects in the historical 
development of $A=4$ calculations for the time period up to the late 1980s when a new phase in the 4N calculations 
had its beginnings. This modern period  is then discussed by  considering first
bound-state calculations in 
Section~\ref{sec:4_bound} and then continuum-state calculations 
in Section~\ref{sec:4_scatter}, where we will
distinguish between hadronic and perturbation-induced reactions.

\subsection{\it 4N Problem: The Early Phase}
\label{sec:4_hist}

Probably the beginnings of serious $\alpha$ particle bound state calculations
using semi-realistic NN potentials can be placed
near the end of the 1960s. Variational calculations~\cite{TaH67,AfT68,FaP70} were performed 
with local central NN potentials, with and without spin dependence (some  of them also fit to describe the NN 
$S$-wave phase shifts). The trial functions were constructed using Jastrow type correlation
functions. 

In the following years non-local separable potential models became more popular. In fact,
for the two-body problem, it was long known \cite{YaY54} that non-local separable NN potential models, like
the Yamaguchi NN potential models \cite{YaY54}, reduce the problem to an algebraic equation. Somewhat later it was 
also realized (see e.g. Ref.~\cite{Mi62}) that the three-body problem 
leads to an effective two-body problem by use of separable NN potentials. The generalization of such a procedure,
namely the reduction of the problem from $A$ to $A\,$--$\,1$, was pointed out for the AGS formalism in Ref.~\cite{AlG67}
(see also the discussion in Section~{\ref{sec:AGS}).  
The equations obtained for the ($A\,$--$\,1$)-system then led to quasi-potentials, which again can be described by an
expansion in separable terms. As was shown in Ref.~\cite{AlG67} a repeated application of such a procedure
increasingly reduces 
the complexity of the equations. 

A first 4N calculation along these lines was performed by Alt, Grassberger, and Sandhas in Ref.~\cite{AlG70}
who used Yamaguchi potentials for the $^3s_1$ and $^1s_0$ NN partial waves and took separable approximations
for the (3+1) and (2+2) subsystem amplitudes. The resulting AGS equations were however solved only rather 
approximately. Similar reductions with separable models were also made for the FY equations by Kharchenko and 
Kuzmichev \cite{KhK72} and by Narodetsky, Galpern, and Lyakhovitsky \cite{NaG73}.
Spin and isospin degrees of freedom were included in the work of Ref.~\cite{NaG73} but not in that of Ref.~\cite{KhK72}.
The corresponding 
integral equations were solved with the Yamuguchi potential as a NN interaction, and by further assuming that the
(3+1) and (2+2) subamplitudes are dominated by s-waves. In addition, the convergence of the calculated 
observables with respect to the separable expansion of the subamplitudes was checked. Shortly after, Tjon \cite{Tj75}
performed essentially the same calculation as in Ref.~\cite{NaG73}, but used instead of a separable NN potential, a 
local s-wave potential of Yukawa type  (MTI-III potential \cite{MaT69}) represented by a separable expansion. 
A rather realistic result of 29.6 MeV for 
the $^4$He binding energy was obtained. In addition, by using various approximations to the local interaction,
Tjon could show that there was a linear relationship between the binding energies of $^3$H and $^4$He. This was 
later confirmed for many other interactions and the linear relationship received the name {\it Tjon-line}. 
On the other hand it was not yet clear whether the separable ansatz for the amplitudes of the subsystems was reliable.
This was investigated by Gibson and Lehmann \cite{GiL76} by directly solving 
the coupled two-variable integral equations resulting from the Schr\"odinger equation for the 
Yamaguchi NN potential, and, in fact the results of Ref.~\cite{NaG73} could be reproduced.

Here we would like to point out that more realistic NN potentials models than those mentioned above were 
also in use.
Demin, Pokrovsky, and Efros \cite{DeP73} applied the HH method (potential harmonics approximation for $K>6$) and 
obtained a rather good convergence of the expansion with the EH \cite{EiH71} and GPT \cite{GoP70} potentials, 
which led to binding energies of 23.03 and 26.76 MeV, respectively.
This range of binding energies for realistic NN models is confirmed in GFMC and Faddeev calculations
(see below and also Table~\ref{KaG}). Applying the CC 
technique, Zabolitzky presented 
a calculation \cite{Za74} with the Reid potential \cite{Re68} with a full and partial consideration of 
the potential in the two-body (CCSD) and three-body (CCSDT) parts, respectively. A $^4$He binding 
energy of about 21 MeV, and, in an improved calculation, of 23.7 MeV \cite{KuL78}, were obtained. 
It should be mentioned, however, that Zabolitzky's approach was not completely translationally 
invariant for parts of the calculation. 

With time FY calculations became somewhat more realistic. In the late 1970s tensor force effects were included 
in an approximate way by Tjon~\cite{Tj78}. He 
calculated the $^4$He binding energy with a truncated Reid potential retaining in the NN $t$-matrix only the 
$s$-wave amplitudes, but such that the $^3s_1$-$^3d_1$ tensor force was included perturbatively (so-called $t_{00}$
approximation). This led to a 
binding energy of 19.5 MeV. 
Shortly after, the relative good quality of such a truncated $t$-matrix approximation was shown for the 3N case 
by Gibson and Lehman \cite{GiL78}, and there, in addition,   the same approximation was used to calculate 
$^4$He binding energies with separable NN potential models. 

One focus in the early literature of FY/AGS 4N calculations concerned the best choice for the expansion method for the
subsystem amplitudes. This was discussed in quite a few publications by various authors 
\cite{SoF77,PeS77,SoM79,SoF80,CaH82,FoH83}. Still another point of discussion concerned the necessity of introducing 
separable expansions not only for the (3+1), but also for the (2+2) subsystem. As mentioned by Narodetsky, Galpern, and 
Lyakhovitsky \cite{NaG73} the expansion was made because of convenience and that 
an exact method, the convolution method, exists for the (2+2) case. An application of the convolution method is found 
in the field theoretic approach~\cite{FoS76}. By applying the convolution method to the AGS
formalism it was shown by Haberzettl and Sandhas \cite{HaS81} that the field theoretic formalism
of \cite{FoS76} corresponds to the AGS formalism in the case that only single-term separable approximations for NN and (3+1) amplitudes 
are employed. 

In the early 1980s other groups appeared on the scene. HH calculations with central NN potential models were 
performed by Ballot \cite{Ba81}. At the same time the first more realistic GFMC calculations in nuclear physics appeared.
These were carried out by Zabolitzky, Schmidt, and Kalos \cite{ZaK81,ZaS82}, who considered 3N and 4N systems  
and employed the MTV potential \cite{MaT69} in addition to a simple central 3NF model \cite{CaP81}. 
Nearly concurrently  variational
calculations were initiated for $^4$He from the Urbana-Argonne group. The trial functions were constructed variationally
from various choices of pair correlation operators and/or of three-body correlation operators 
(see also Section~\ref{sec:VMC}). NN potentials, like MT models and the V8 version of the Reid potential, were considered 
without \cite{LoP81}, and with the above mentioned simple 3NF \cite{CaP81}. The models for the NN potentials and for 
the 3NF were improved in subsequent variational calculations by Carlson, Pandharipande, and Wiringa \cite{CaP83} and
by Schiavilla, Pandharipande, and Wiringa \cite{SchP86}.
\begin{table}
\begin{center}
\caption{{\label{MTV}} $^4$He binding energy in MeV for the MTV potential~\cite{MaT69} with various methods
as published in Ref.~\cite{ZaS82} in 1982.  }
\begin{tabular}{c||c|c|c|c}
\hline \hline
Method & GFMC~\cite{ZaK81}  & VMC~\cite{CaP81} & CC~\cite{Za81}) & ATMS~\cite{NaA80,AkS74} \\
\hline
   & 31.3(2) & 31.19(5) & 31.24 & 31.3\\
\hline \hline
\end{tabular}
\end{center}
\end{table}
It is worthwhile mentioning that due to the above mentioned developments a benchmark for the MTV potential was 
established in the early 1980s. The corresponding results are listed in Table~\ref{MTV}.
The various results agree quite well. Though the potential model is rather simple one may regard the agreement,
obtained with four different methods, as a milestone in the history of the $^4$He binding energy calculations.
The advantage of
such established precise ab initio results is self-evident: they serve as a standard for future calculations. 
Later the various methods that appear in Table~\ref{MTV} were further improved and applied to more 
realistic potential models. Only the
variational ATMS method, to our knowledge, has never been applied to more realistic interactions.

Improvements in AGS calculations were made by allowing more exact representations of the $s$-wave 
(3+1) amplitude \cite{Fo84}. Articles, in a common book, by
Ferreira~\cite{Fe87}, Oryu~\cite{Or87}, and Plessas~\cite{Pl87} clarify the important
role played by separable expansion techniques for both subamplitudes and NN potentials at
that stage of the treatment of few-body problems.

All the above mentioned FY/AGS calculations tackle the 4N problem in momentum space. Configuration space FY calculations
were initiated by Merkuriev, Yakovlev, and Gignoux \cite{MeY84}. They derived the differential equations for the
FY components and, in addition, described their asymptotic form. 
Four-body binding energies were calculated with local $s$-wave NN potential models. A rather good 
agreement with results from other calculations \cite{Tj75,Ba81} was obtained.

Important progress for the GFMC calculations was made by extending the technique to spin-dependent forces~\cite{Ca87}. 
Calculations for potentials of V6 form (central potential with spin and isospin dependences plus 
tensor force) and simpler models were performed.
A comparison with Faddeev calculations for the 3N system showed that precise results could be obtained with the GFMC
technique, whereas variational results were not of such good quality. In addition the following results for
the $^4$He binding energy were found for the $V6$ potential: 24.79(20) MeV (GFMC) and 22.75(10) MeV (VMC). 

The first FY/AGS calculation that went beyond the truncated NN $t$-matrix approximation by fully considering the
$^3s_1$-$^3d_1$ tensor force (NN potential of Yamaguchi form) was achieved by Fonseca \cite{Fo89} (see also 
Table~\ref{AGS_rank}). The results
showed that a large part of the tensor force effect to the $^4$He binding energy was already contained in a truncated
$t$-matrix approximation.

In the early phase of the 4N problem various ab initio methods were also applied to 4N scattering. 
We do not discuss the corresponding results in detail, but, at least, give a list of these early 4N
scattering studies: \cite{ReT71} (RGM), \cite{PeP72} (HH), \cite{KhL76} (FY), \cite{Tj76} (FY), 
\cite{Fo79} (AGS-like), \cite{HoZ81} (RGM).

Here we just mention a few points. Tjon~\cite{Tj76} made an interesting finding, namely that the effect of $p$-waves 
in the 3N subamplitudes is not negligible for low-energy 4N scattering. This was later confirmed in an AGS 
calculation~\cite{Fo86}, where it was shown that the effect is particularly strong in $p$-$^3$He elastic cross
sections, but also noticeable in $dd \rightarrow p\,$-$^3$H and $dd \rightarrow dd$ reactions. These studies of
low-energy 4N scattering reactions by Fonseca with the AGS approach were continued in Ref.~\cite{Fo89b}, and, as
we will see in Section~\ref{sec:4_had}, have been further pursued till today.

\subsection{\it 4N Problem: The Modern Phase}
\label{sec:4_mod}

The more modern phase of 4N calculations started at the end of the 1980s, when not only the first GFMC
calculations with realistic NN potential models emerged, but, in addition  in the early 1990s quite a variety 
of different groups applying different methods appeared on the scene. Actually, the list of methods is quite long:
{\bf (i)} GEM~\cite{KaK90,KaK89}, 
{\bf (ii)} DMC~\cite{BiF90a}, 
{\bf (iii)} CC~\cite{BiF90b}, 
{\bf (iv)} IDEA~\cite{OeS90}, 
{\bf (v)} CHH~\cite{RoK92}, 
{\bf (vi)} FY in configuration space~\cite{SchS92},
and {\bf (vii)} FY in momentum space~\cite{KaG92a} (these methods are discussed or at least mentioned in
Section.~\ref{sec:FB}). 
All these various techniques were employed in $^4$He bound-state calculations, whereas a new onset of
4N scattering-state calculations came somewhat later as will be discussed below.

\subsubsection{\it The Four-Nucleon Bound State}
\label{sec:4_bound}

A first GFMC calculation with realistic NN potential models was carried out by Carlson~\cite{Ca88}
using the BonnR~\cite{MaH87} potential and the Reid~\cite{Re68} potential reduced to V8 form. 
He obtained an underbinding of about 5 MeV, which is
similar to the results of previous realistic calculations~\cite{KuL78,DeP73}. In Ref.~\cite{Ca88} 
it was also shown that plausible models for the 3NF can remove the underbinding. With time, improved models 
for NN potentials
and 3NFs evolved. Since the GFMC technique cannot treat momentum dependent interactions, (see Section~\ref{sec:GFMC})
the method
remained restricted to $r$-space interaction models. Thus a GFMC $^4$He binding energy result for a modern 
realistic force model exists only for the AV18~\cite{WiS95} NN potential. With an additional inclusion of the 3NF
model UIX~\cite{PuP97} a binding energy of 28.30(2) MeV was obtained~\cite{PuP97}, which is in agreement with the 
experimental result. 

\begin{table}
\begin{center}
\caption{{\label{MTV_2}} $^4$He binding energy in MeV for the MTV potential with various methods.  }
\begin{tabular}{c|c|c|c|c|c}
\hline \hline
Method & GEM~\cite{KaK90,KaK89}  & IDEA~\cite{OeS90} & DMC~\cite{BiF90a}) & FY~\cite{KaG92a} & CHH~\cite{RoK92}\\                       
\hline
       & 31.357 & 30.98 & 31.5 & 31.36 & 31.355\\
\hline \hline
\end{tabular}

\end{center}
\end{table}
In the momentum space FY calculation of Ref.~\cite{KaG92a} quite a variety of NN potential models were considered: 
local and non-local
$s$-wave models with and without spin dependence, local central 
models, separable models with $^3s_1$-$^3d_1$ tensor force, $^1s_0$ and $^3s_1$-$^3d_1$ interactions from Reid potential. 
An extensive comparison with other results in the literature was made. In Table~\ref{MTV_2} we show the results 
with the MTV potential, where a benchmark had already been established before (see Table~\ref{MTV}). 
Comparing with the benchmark one sees that most of the techniques lead to similar results. 
One could even say that the precision of the benchmark result was further increased (compare GEM,
FY, and CHH results from Table~\ref{MTV_2} and ATMS result from Table~\ref{MTV}). At this point it should also
be mentioned that the convergence of the FY calculation is shown for an increasing number of channels and it is
evident that excellent convergence is established (the maximal number of channels was 92, note that the MTV
potential reduces the 4N problem to a four-boson problem with a considerably smaller number of channels than given
in Table~\ref{channels} for the four-fermion case). Of all the results
listed in  Table~\ref{MTV_2} the IDEA binding energy shows the strongest deviation, but
IDEA is not a full ab initio method, since it can be seen as equivalent to a HH expansion limited to
the lowest HH state (see Section~\ref{sec:HH}).
It should also be mentioned that the FY result for the MTI-III interaction agreed very well with the high-precision
result from the coordinate space FY calculation of Ref.~\cite{SchS92}. 

Kamada and Gl\"ockle extended their FY calculations to other potential models. 
In Ref.~\cite{KaG92b} the $^1s_0$ and $^3s_1$-$^3d_1$ interactions from various
realistic NN potentials were considered. In comparison to
Ref.~\cite{KaG92a} the precision of the calculation was improved by increasing the number of channels considered 
(Table~\ref{KaG} shows examples for some of the NN potentials). In addition  
it was shown that potentials that are fit to describe $np$ or $pp$ scattering 
data lead to somewhat different Tjon lines. A further improvement was made in Ref.~\cite{GlK93a} by 
taking into account NN forces in all NN partial waves with $j \le 3$ (results given in the last column of Table~\ref{KaG}).
The resulting binding energies ranged from 23.45 to 28.11 MeV. It is interesting to note that the lowest value,
which arises from the Reid potential, is close to the CC results from the 1970s  mentioned in Section~\ref{sec:4_hist}.
The formalism to incorporate 3N forces into the 4N Yakubovsky scheme was laid out in Ref.~\cite{GlK93b}. 
Fully converged momentum space FY calculations with modern realistic forces and various 3NF models
were performed in Refs.~\cite{NoK02,NoK00}. The variation of the $^4$He binding energy with modern realistic NN forces was found to be considerably smaller than with the non-modern realistic models used in Ref.~\cite{GlK93a}. 
It was observed in Ref.~\cite{PuP97} that by adjusting the 3NF in order to describe the triton binding energy
one is led to a $^4$He binding energy rather close to the experimental value. This apparently diminishes
the need for a 4NF. 
The smallness of the 4NF is also confirmed in Ref.~\cite{RoG06}, where the leading contribution of a chiral 4NF
to the $\alpha$-particle binding energy was estimated to be of the order of a few 100 keV.

\begin{table}
\begin{center}
\caption{{\label{Bench_1}} The $^4$He expectation values $\langle T \rangle$ and $\langle V \rangle$ of kinetic and potential
energies, the binding energy BE, and the root-mean-square (rms) radius with AV8$^\prime$ NN potential~\cite{PuP97} 
calculated with various methods (from Ref.~\cite{KaN01}).  }
\begin{tabular}{c|c|c|c|c}
\hline \hline
Method & $\langle T \rangle$ [MeV]  & $\langle V \rangle$ [MeV] & BE [MeV] & $\sqrt{\langle r^2 \rangle}$ [fm]\\
\hline
FY  & 102.39(5) & --128.33(10) & 25.94(5) & 1.485(3) \\
GEM & 102.30    & --128.20     & 25.90    & 1.482 \\
SVM & 102.35    & --128.27     & 25.92    & 1.486 \\
HH  & 102.44    & --128.34     & 25.90(1) & 1.483 \\
GFMC& 102.3(1.0)& --128.5(1.0) & 25.93(2) & 1.490(5) \\
NCSM& 103.35    & --129.45     & 25.80(20)& 1.485 \\
EIHH& 100.8(9)  & --126.7(9)   & 25.944(10) & 1.486 \\    
\hline \hline
\end{tabular}

\end{center}
\end{table}
During the 1990s not only was great progress made with the GFMC and FY calculations, but also with some other of the 
above mentioned techniques (see e.g. CHH \cite{ViK95}, GEM calculations for 4-body hypernuclei~\cite{HiK99}). 
In addition other players entered the game with further methods:
the SVM~\cite{VaS95,SuV98}, the NCSM~\cite{NaK00,NaB99}, and
the EIHH methods~\cite{BaL00,BaL01}. A manifestation of all this progress
was the benchmark calculation of Ref.~\cite{KaN01}, where seven different theory groups applied seven different techniques to solve
the $^4$He bound-state problem with the realistic AV8$^\prime$ potential~\cite{PuP97}. Besides the binding energy various 
calculational results were compared: expectation values for kinetic and potential energies, separate expectation
values for various potential terms, the NN correlation function, and the probabilities of the total angular 
momentum components. In Tables~\ref{Bench_1} and \ref{Bench_2} we list a few of these results.
One sees that a very good agreement between the various methods was obtained.
A closer inspection of Table~\ref{Bench_1} shows that deviations for the binding energy and 
rms radius are within 0.5$\%$.
Except for NCSM  and EIHH similarly good agreement is found for $\langle T \rangle$ and $\langle V \rangle$. However, the
NCSM results are still within 1$\%$ and EIHH within 1.5$\%$ of the others, but one should note that the EIHH results were calculated
with the bare operators. The probabilities of the various wave function components displayed in Table~\ref{Bench_2}
also show very good agreement apart from a small excursion in NCSM. 
Like the benchmark for the simple MTV potential from the early 1980s (see Table~\ref{MTV}), the benchmark for the 
realistic AV8$^\prime$ of Ref.~\cite{KaN01} can be considered as a milestone in the development of ab initio calculations for the 
4N system. 
\begin{table}
\begin{center}
\caption{{\label{Bench_2}} Probabilities of $^4$He total angular momentum components in $\%$ for AV8$^\prime$ potential 
calculated with various methods (from Ref.~\cite{KaN01}).  }
\begin{tabular}{c|c|c|c}
\hline \hline
Method & $s$-wave & $p$-wave & $d$-wave \\
\hline
FY  & 85.71 & 0.38 & 13.91 \\
GEM & 85.73 & 0.37  & 13.90  \\
SVM & 85.72 & 0.368 & 13.91   \\
HH  & 85.72 & 0.369  & 13.91 \\
NCSM& 86.73 & 0.29   & 12.98 \\
EIHH& 85.73(2) & 0.370(1)  & 13.89(1) \\    
\hline \hline
\end{tabular}
\end{center}
\end{table}

\begin{table}
\begin{center}
\caption{{\label{AV18_UIX}} $^4$He binding energy in MeV for AV18+UIX Hamiltonian 
calculated with various methods.  }
\begin{tabular}{c|c|c|c|c}
\hline \hline
Method & GFMC~\cite{WiP00} & FY~\cite{NoK02} & HH~\cite{ViK05} & EIHH~\cite{GaB06} \\
\hline
  & 28.34(4) & 28.50 & 28.46 & 28.42 \\
\hline \hline
\end{tabular}
\end{center}
\end{table}
More recently, a similar agreement was achieved considering different NN potentials and 3NFs.
As an example we show in Table~\ref{AV18_UIX} results for the interaction model AV18+UIX.
  
A different approach for describing the nuclear force, namely without a 3NF, is
made via a coupled channel calculation, where N and $\Delta$ degrees of freedom are treated on
the same level by using an appropriate potential like the CD-Bonn+$\Delta$ potential \cite{DeM03}. Such a model, 
however, led to a slight underbinding in three- and four-nucleon systems implying that many-nucleon forces not
accounted for by the $\Delta$ isobar make a non negligible contribution to the
binding energies \cite{DeF08}.

\subsubsection{\it The Four-Nucleon Scattering State}
\label{sec:4_scatter}

Continuum wave functions describing the 4N system in scattering states can be studied both in hadronic reactions
and in reactions with external electroweak probes, i.e. photons, electrons and neutrinos. It is worthwhile
pointing out that the information supplied by one type of reaction 
does not render redundant information gained from the other. Specific effects could
have different relative sizes in the various types of reactions. In this context it is important that
in electron scattering, as in principle also in neutrino scattering, energy transfer $\omega$ and momentum transfer
$q$ can be varied independently. Thus, for a given $\omega$, i.e for a fixed continuum state,
one can study the $q$ dependence of various effects. 
In addition one should note that, due to the existence of two-body current operators, 
two completely phase equivalent potential models can nonetheless lead to different responses to external probes.

\subsubsection{\it Hadronic Reactions}
\label{sec:4_had}

Starting in the late 1990s considerable progress was made in 4N continuum calculations. The zero-energy
scattering of $n\,$-$^3$H and $p\,$-$^3$He was studied in a HH calculation by Viviani, Rosati, and Kievsky~\cite{ViR98}
through application of the Kohn 
variational principle (see Section~\ref{sec:VAR}). Together with the Coulomb force
various versions of the Argonne potentials and various 3NF models were
considered. The convergence of the results was checked by increasing the
number of channels in the HH calculation. Almost simultaneously Ciesielsky, Carbonell, and 
Gignoux began a study of the 4N scattering problem in a coordinate space FY calculation~\cite{CiC98b}. This
enabled a successful 
comparison of HH and FY results for the $n\,$-$^3$H singlet and triplet scattering lengths for the AV14 potential. 
Concerning other results in Ref.~\cite{ViR98}, 
it was found that the 3NF decrease the $n\,$-$^3$H total zero-energy scattering cross section by about 10\%, 
which then led to a good agreement with experimental data. A somewhat inferior agreement,
with a difference of about 3\%, was 
obtained for the coherent scattering length. 

The already above mentioned FY study~\cite{CiC98} was an extension, by inclusion of spin and isospin
degrees of freedom, of the four-boson configuration space calculation of Ref.~\cite{MeY84}. 
The Grenoble group used the MTI-III NN potential without consideration
of the Coulomb force in their first 
study. For energies below the three-body break-up threshold they calculated phase shifts for the elastic (3+1), 
as well as  phase 
shifts for the inelastic (3+1)$\rightarrow$(2+2) reaction. A particularly interesting finding of their calculation
was the good 
description of the resonant structure ($p$-wave resonance) of the $nt$ total elastic cross section obtained
with the simple MTI-III 
potential (see Fig.~\ref{figure1_R}a). This fact initiated a very fruitful and interesting discussion further on. In an improved calculation 
\cite{CiC99} various $^1s_0$ and $^3s_1$-$^3d_1$ interactions from realistic NN potential models were used. In addition a 
phenomenological 3NF was fit to describe the binding energy of $^3$H. Since the number of channels considered was not 
sufficiently high the calculation was reasonably well but not fully converged (a comparison of a corresponding calculation of
the $^4$He binding energy with precise ab initio results showed that less than 1 MeV was missing). We list the main findings of 
Ref.~\cite{CiC99}: {\bf (i)} only a rather weak dependence of the results on the choice of the NN potential model, 
{\bf (ii)} effect of the 3NF is very important at very low energies (as already found in Ref.~\cite{ViR98}, it led to a very good 
agreement with the zero-energy experimental cross section for elastic $nt$ scattering), 
{\bf (iii)} in contrast to the MTI-III case 
a good description of the cross section in the $p$-wave resonance region was not obtained 
with realistic NN potentials (see Fig.~\ref{figure1_R}b), 
and {\bf (iv)} 3NF effects are very small in the resonance region. 
\begin{figure}[ht]
\begin{center}
\epsfig{file=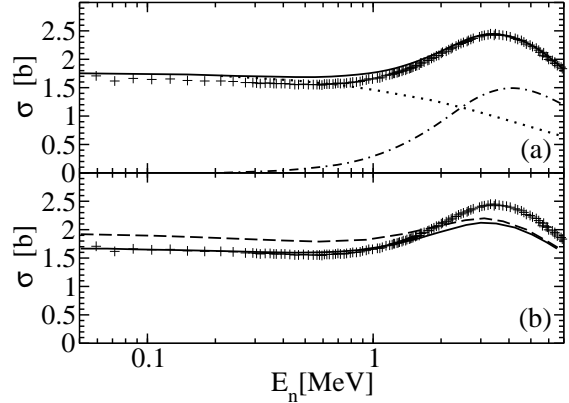,scale=0.3,angle=-90}
\caption{n-$^3$H total cross section: FY results~\cite{CiC99,CiC98}. (a): for MTI-III,
with 4N-$s$- (dotted), $p$- (dash-dotted), and $(s+p)$-wave (full); 
(b): with 4N-$(s+p)$-wave for AV14 (dashed) and AV14+3NF (full).
Experimental data (crosses)~\cite{PhB80}.}
\label{figure1_R}
\end{center}
\end{figure}

The FY calculation for $n\,$-$^3$H elastic scattering of Ref.~\cite{CiC99} was further improved by Lazauskas and
Carbonell~\cite{LaC04} by including the NN force in NN partial waves with $j\le3$, and additionally also in $^3f_4$-$^3h_4$. 
As interactions both the
AV18+UIX and the non-local NN potentials INOY \cite{DoB03} were used (the INOY non-localities were constructed such that
an additional 3NF becomes obsolete). It was shown that the non-local model leads to rather good descriptions of the $^4$He binding 
energy and of the zero-energy $n\,$-$^3$H elastic scattering cross section,  
but fails, as does the AV18+UIX interaction model, to describe the cross section in the resonance region. 

At the end of the 1990s, the AGS approach for 4N scattering (see Section~\ref{sec:4_hist}) was also improved~\cite{Fo99}. 
The calculation was formally very similar to the previous work in Ref.~\cite{Fo89} but contrary to the latter the 
interaction model became more realistic by inclusion of the interaction in NN $p$-waves.
However, like the $s$-wave interactions the $p$-waves were also represented by separable expansions.
In addition the Coulomb force was not included. 
Low energy elastic $n\,$-$^3$H, $(3+1) \rightarrow (2+2)$, and $dd$ scattering were discussed.
Here we only mention two results: {\bf (i)} the calculated analyzing power $A_y$ in the $^3$He$(\vec p,p)^3$He reaction was
considerably smaller than the experimental $A_y$, and {\bf (ii)} the effect of the NN $p$-waves on the $n\,$-$^3$H elastic scattering cross 
section was found to be different from Ref.~\cite{CiC99}. 

In Ref.~\cite{HoH97} Hofmann and Hale studied the $^4$He compound system, i.e. the 4N system in the isospin $T=0$ channel,
in the continuum at low energies. They performed both, a R-matrix analysis and a RGM calculation. In Ref.~\cite{PfH01} their 
work is continued and applied to $n\,$-$^3$H and $p\,$-$^3$He scattering, where among other nuclear forces they also
include the AV18+UIX model
in the RGM calculation (interaction included for NN $s$-, $p$-, and partially also $d$-waves). Also in this case we 
want to mention two of the results: {\bf (i)} the elastic $n\,$-$^3$H cross section
agrees with data in the very low-energy region, but as in Ref.~\cite{CiC99} underestimates the resonance peak; 
and {\bf (ii)} similar to Refs.~\cite{Fo99,ViK01} the experimental $A_y$ in $p\,$-$^3$He scattering is strongly underestimated 
(note that Refs.~\cite{PfH01} and~\cite{ViK01} were published almost simultaneously;
further discussion of Ref.~\cite{ViK01} is given below). 
\begin{figure}[ht]
\begin{minipage}[b]{0.425\linewidth}
\centering
\includegraphics[scale=0.3,angle=-90]{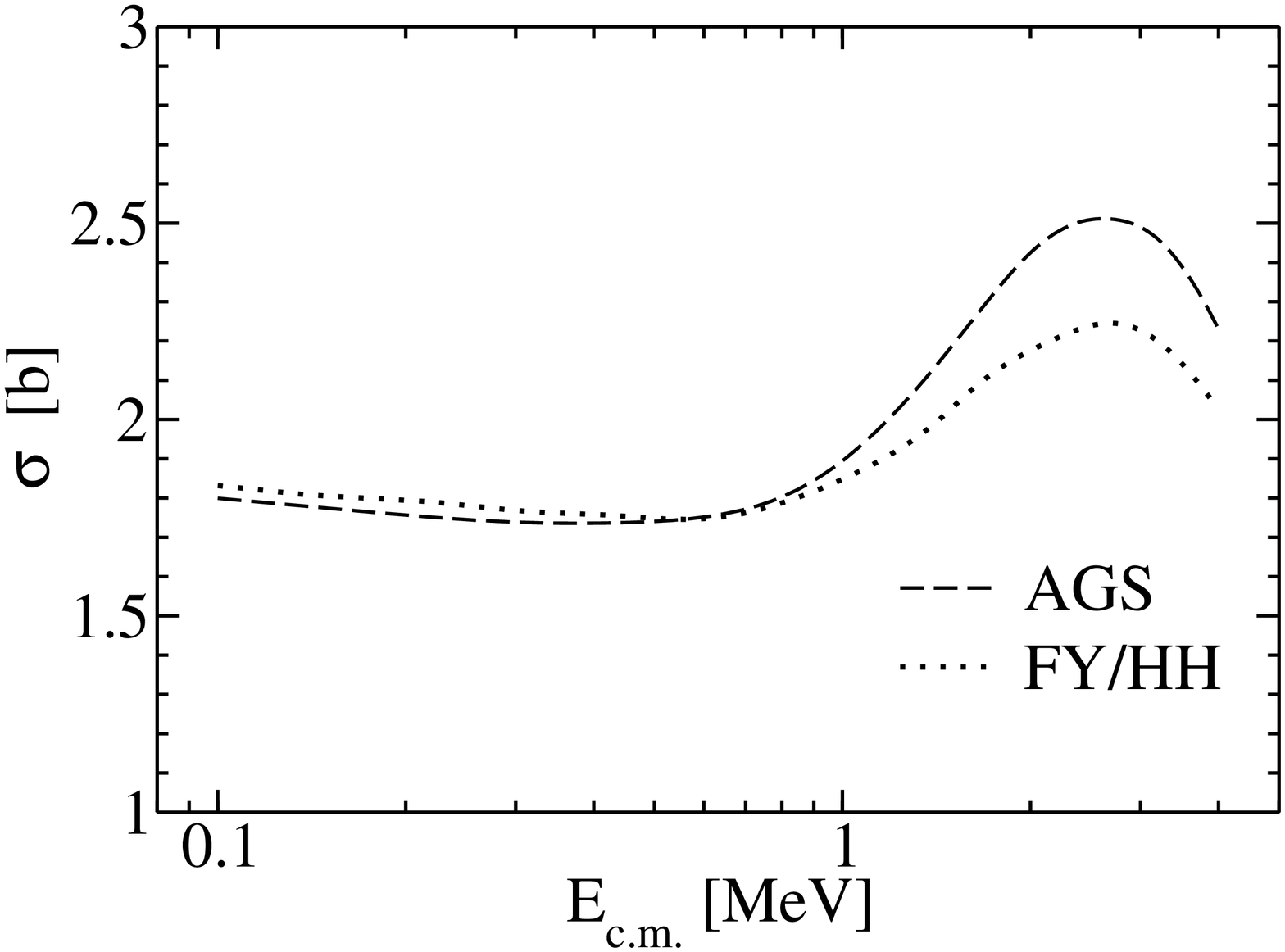}
\caption{As Fig.~\ref{figure1_R} but for AV18 NN potential calculated using three different
methods, almost identical results for FY and HH calculations (dotted) and AGS results
(dashed); from Ref.~\cite{LaC05b}.}
\label{figure2_R}
\end{minipage}
\hspace{0.25cm}
\begin{minipage}[b]{0.55\linewidth}
\centering
\includegraphics[scale=0.33,angle=-90]{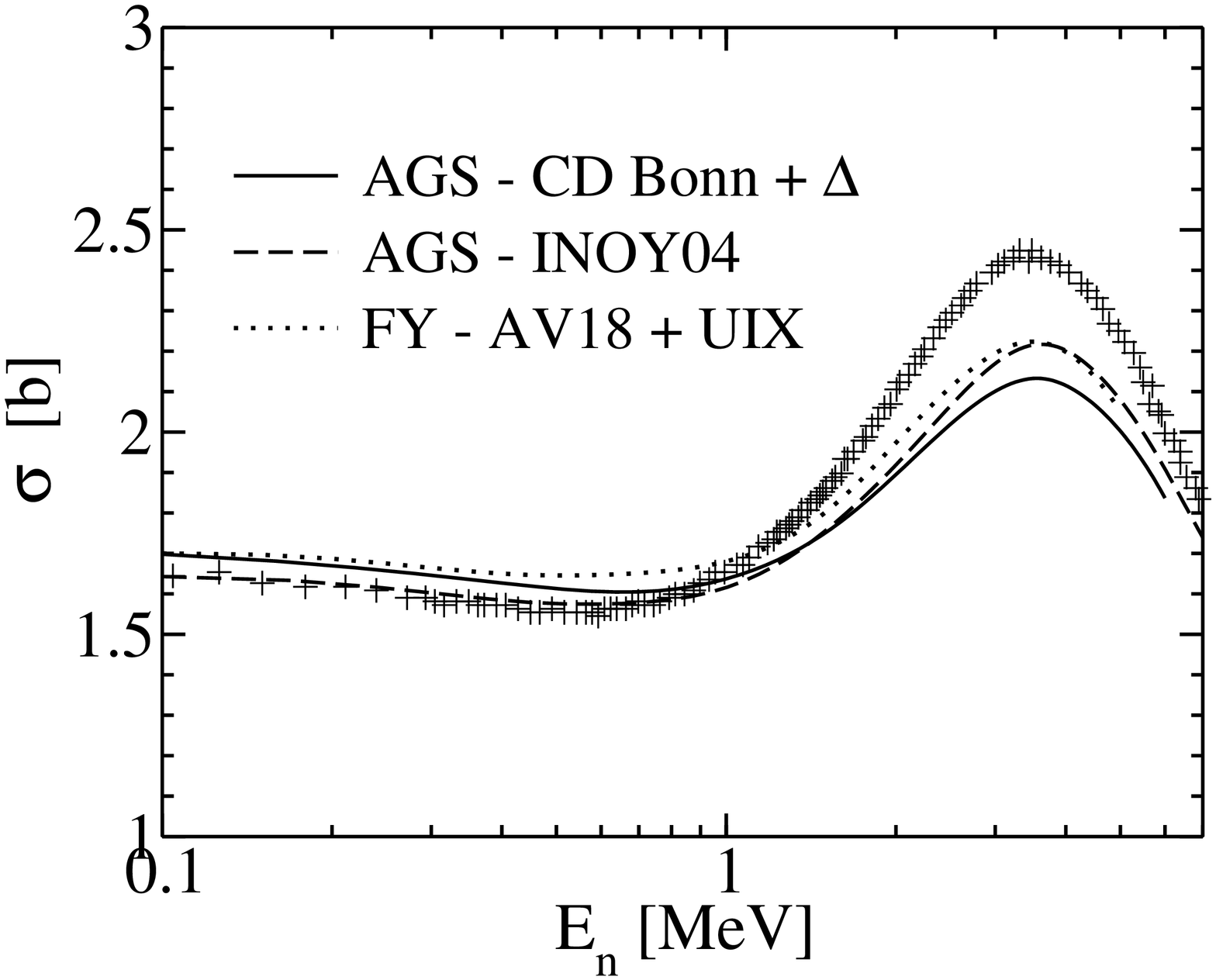}
\caption{As Fig.~\ref{figure1_R} but calculated with various methods
and potentials:
AGS result~\cite{DeF08} with CD-Bonn+$\Delta$~\cite{DeM03} (full),
AGS result~\cite{DeF07a} with INOY04~\cite{Do04} (dashed),
FY result~\cite{LaC05b} with AV18+UIX (dotted). Data as in Fig.~\ref{figure1_R}. }
\label{figure3_R}
\end{minipage}
\end{figure}

As a matter of fact the $n\,$-$^3$H elastic scattering results obtained with various methods (AGS, FY, HH, RGM)  
showed a large variance among each other for some observables (e.g. resonance cross section in $n\,$-$^3$H scattering). 
Improvements in the AGS, FY, and HH calculations were made in order to produce a benchmark~\cite{LaC05b}, wherein 
various scattering observables were calculated by the three participating groups. In Fig.~\ref{figure2_R} we show the cross section
for elastic $n\,$-$^3$H scattering as an example. It is evident that FY and HH results agree perfectly, whereas the 
AGS cross section is 
somewhat different. Similar findings could be made for a few others of the cases studied. It was then concluded
that probably the AGS separable expansion of the NN potentials with just one separable term is not sufficient for $p$-waves.

The above example of a benchmark test calculation is of course only possible for ab initio calculations, since only in this
case does a unique ab initio result exist. The conclusion to be drawn from the benchmark case above was quite clear: 
improvements were necessary for the AGS calculation. In fact not much later the AGS calculation was
considerably improved. In Ref.~\cite{DeF07a} Deltuva and Fonseca abandoned the various separable expansions usually made in
the AGS formalism. Instead, they treated the NN $t$-matrix exactly and solved the resulting three-variable integral equations
without any further reduction. In addition the number of NN, 3N, and 4N partial waves was increased up to convergence. 
The new calculation for low-energy elastic $n\,$-$^3$H scattering confirmed in principle the results of Refs~\cite{CiC99,CiC98} 
concerning the resonance peak, although the cross section results were slightly 
higher, probably due to better convergence in the AGS calculation. Agreement was also obtained with the phase shift calculations 
of the Grenoble and Pisa groups given in the benchmark of Ref.~\cite{LaC05b}. 

A few years earlier the Pisa group extended the HH calculations for the zero-energy $p\,$-$^3$He scattering~\cite{ViR98} 
to low-energy scattering employing the AV18+UIX nuclear force. Their results were published in Ref.~\cite{ViK01} 
together with  new accurate data of the proton analyzing power $A_y$. It was found that 
the theoretical results for $A_y$ were up to 40\% smaller than the data,
thus leading to a situation which seemed to be similar to the well-known $A_y$ puzzle in 3N scattering. The 4N
$A_y$ problem was confirmed in a later study~\cite{FiB06}, where new accurate experimental results were presented
together with new HH calculations of the Pisa group. Shortly afterwards Deltuva and Fonseca came out with a fully 
converged AGS calculation for $p\,$-$^3$He scattering~\cite{DeF07b}. Various NN potentials, including
the non-local INOY potential, were employed. The inclusion of the Coulomb force was achieved by using a 
screened Coulomb force according to the methodology developed in \cite{DeF05}. Large Coulomb effects were found for 
all the studied observables. The calculation described experimental data quite well, except for the proton $A_y$, where 
the discrepancy between theory and experiment, found in \cite{ViK01,FiB06}, was confirmed. 

The AGS calculations were extended to 4N scattering in the isospin $T=0$ channel~\cite{DeF07c}, which makes the
consideration of 
couplings between the $n\,$-$^3$He, $p\,$-$^3$H, and $d\,$-$\,d$ channels necessary. The same NN potential
models as in Ref.~\cite{DeF07b} were used. Observables pertaining to the six different elastic and transfer reactions were obtained
and compared to experiment. The agreement between theory and experiment was fairly good, but for
some observables stronger differences were present. In order to include many-body forces the study was continued
using the CD-Bonn+$\Delta$ two-baryon model~\cite{DeF08}. However, the inclusion of the $\Delta$ was unable to 
resolve the existing discrepancies with experimental data and the effect of the 4NF turned out to be very small. 
In Fig.~\ref{figure3_R} we show as an example the $n\,$-$^3$H elastic scattering cross section with various interaction models.
One sees that all of them fail to describe the $p$-wave resonance peak.

In a further AGS calculation various observables of nucleon transfer reactions in low-energy $d\,$-$\,d$ 
scattering were studied~\cite{DeF10}.
For example, it was found that $dd$ fusion with spin-aligned deuterons is significantly suppressed at a few MeV
but not in the keV regime (see Fig.~\ref{figure4_R}).

In continuation of their work Hofmann and Hale \cite{HoH08} carried out another, more complete, RGM calculation for the 
$^4$He compound system using as nuclear force the model AV18+UIX.
The resulting phase shifts were compared to a comprehensive $R$-matrix analysis. The agreement between calculation
and analysis was rather good in most cases and 3NF effects were found to be quite large for some specific cases. The calculation
was restricted to a value of $L=3$ for the maximal orbital angular momentum between the various two-body clusters,
which, however, for some observables is not sufficient as pointed out later in Ref.~\cite{DeF10}. 

Another RGM calculation was carried out by Quaglioni and Navr\'atil~\cite{QuN08}. They considered neutron scattering
on $^3$H (and also on $^4$He and $^{10}$Be, see Section~\ref{sec:RGM}) as well as proton scattering on $^3$He (and also on $^4$He). 
As opposed to the RGM calculations mentioned above their nuclear wave function consisted in a fragmentation of the
form  $^3$X-N, with the 3N cluster in its ground state.
The calculation is based on the NCSM, with corresponding HO expansions for the various wave function pieces to be determined. 
Several NN potentials were used, but a 3NF was not considered.
The obtained phase shifts were compared to AGS results of the Lisbon group~\cite{DeF07a,DeF07b} for the
same interaction model. A fairly good agreement was found except for $n\,$-$^3$H $p$-waves, where significant differences 
were observed at $E_{CM} > 1.5$ MeV. Certainly space for further improvement remained,
which could be obtained by extending the RGM calculation beyond the single-fragment approximation. 
On the other hand we should mention
that the 4N calculation merely served as a testing ground, because the principle aim expressed in Ref.~\cite{QuN08} was the 
determination of phase shifts for elastic $n\,$-$^4$He, $n\,$-$^{10}$B, and $p\,$-$^4$He scattering.

In further calculations of the Pisa group various nuclear force models were considered, among them a chiral force 
(I-N3LO NN potential~\cite{EnM03}, N2LO-3NF~\cite{EpN02,Na07}). Their work consisted of 
high-precision calculations of the 4N bound state, the zero-energy scattering states~\cite{KiR08}
and of $p\,$-$^3$He scattering~\cite{ViG10}. 
In this study it was found that use of the chiral nuclear force gave improvements when comparing
to experimental results.  In fact a considerably
better description of the analyzing power $A_y$ in $p\,$-$^3$He scattering was obtained. 
However, some differences between experimental and theoretical results remained as can
 been seen in Fig.~\ref{figure5_R}, where
we show $A_y$ results with conventional and chiral nuclear forces. One notes that there is a relatively strong increase
of $A_y$ due to the chiral 3NF. 

The $^4$He compound system exhibits a $0^+$ resonance between the thresholds for $p\,$-$^3$H  and $n\,$-$^3$He break-up.
This resonance was studied by Lazauskas in a FY calculation for elastic $p\,$-$^3$H scattering~\cite{La09}. Various
nuclear force models ($V_{NN}$ + 3NF + Coulomb force) were used and effects from charge independence and charge 
symmetry breaking in the strong part of the nuclear interaction were taken into account. The convergence of the
calculation was checked and a convergence level better than 0.5\% was obtained.
It was found that none of the tested Hamiltonians were 
able to reproduce simultaneously the shape of the resonance and the $^4$He binding energy.

Recently another benchmark~\cite{ViD11}, a follow-up of that in Ref.~\cite{LaC05b}, was performed for $n\,$-$^3$H and 
$p\,$-$^3$He scattering using AGS, HH, and FY techniques and
taking various realistic NN potentials. The agreement among the three different approaches for the considered 
observables (phase shifts, differential cross sections, analyzing powers) was found to be better than 1\%. 
Thus another milestone in the development of ab initio calculations in the 4N system was set.

Elastic $n\,$-$^3$H and $p\,$-$^3$He scattering above the four-body break-up threshold was considered in a RGM-NCSM 
calculation limited, however, to a fragmentation into a nucleon and the 3N system in its ground state~\cite{FrL11}. 
Nonetheless, a fairly good agreement with experimental results was obtained (differences were particularly present
at forward angles). Only very recently a breakthrough was achieved by Deltuva and Fonseca~\cite{DeF12} 
in an AGS calculation of the elastic $n\,$-$^3$H scattering, considering the full four-body continuum.

Finally we would like to mention that recently phase shifts of elastic N$\,$-$^3$He scattering 
were calculated with realistic
NN potential models for energies below the $^3$He break-up threshold by applying bound-state like methods~\cite{KiV12}.
\begin{figure}[ht]
\begin{minipage}[b]{0.42\linewidth}
\centering
\includegraphics[scale=0.29,angle=-90]{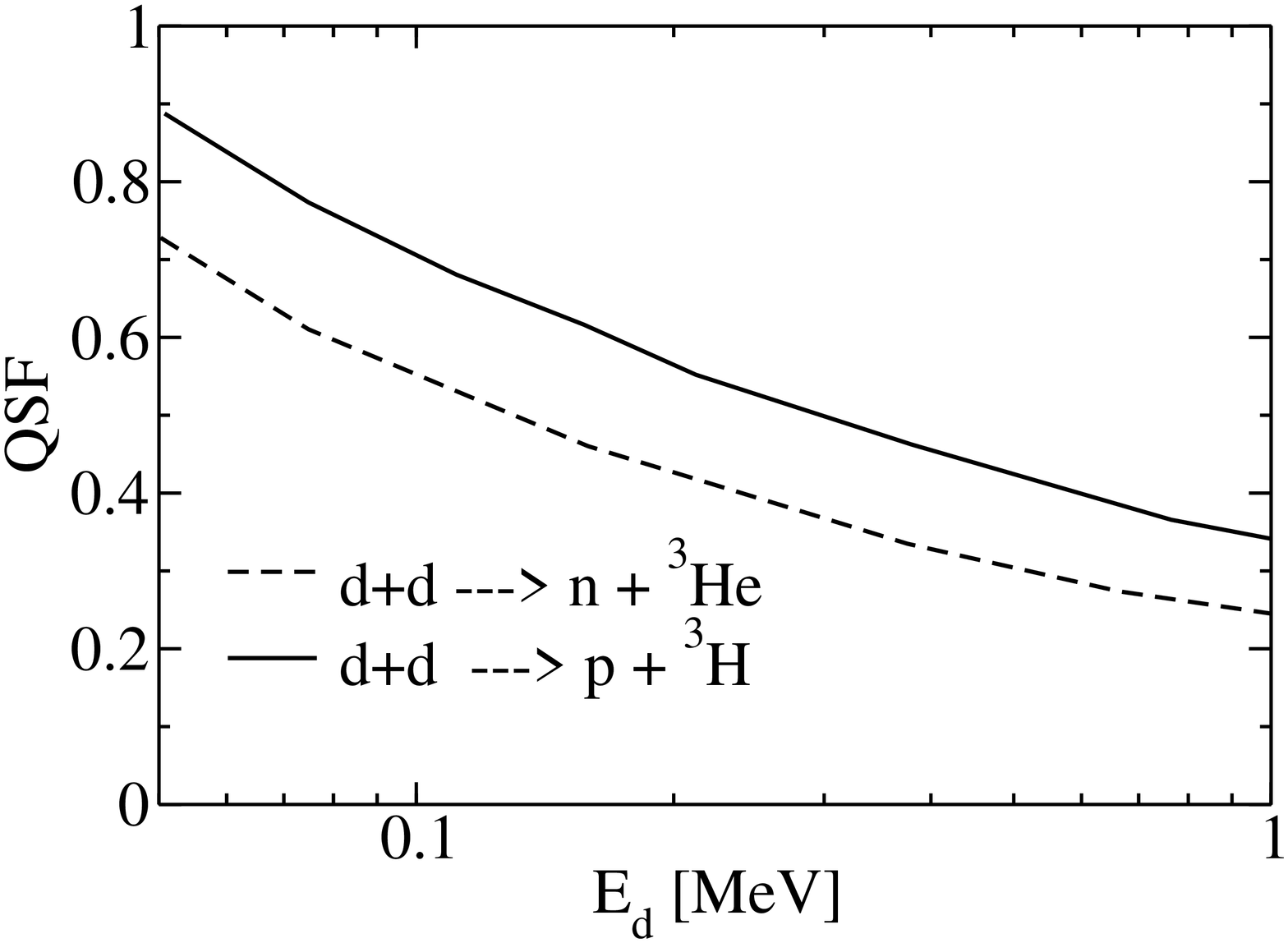}
\caption{The quintet suppression factor (QSF)
for  $d+d \to p+^3$H and  $d+d \to n+^3$He reactions
as function of the deuteron energy in the laboratory system from an AGS calculation with the INOY04 potential~\cite{DeF10}.}
\label{figure4_R}
\end{minipage}
\hspace{0.5cm}
\begin{minipage}[b]{0.54\linewidth}
\centering
\includegraphics[scale=0.3,angle=-90]{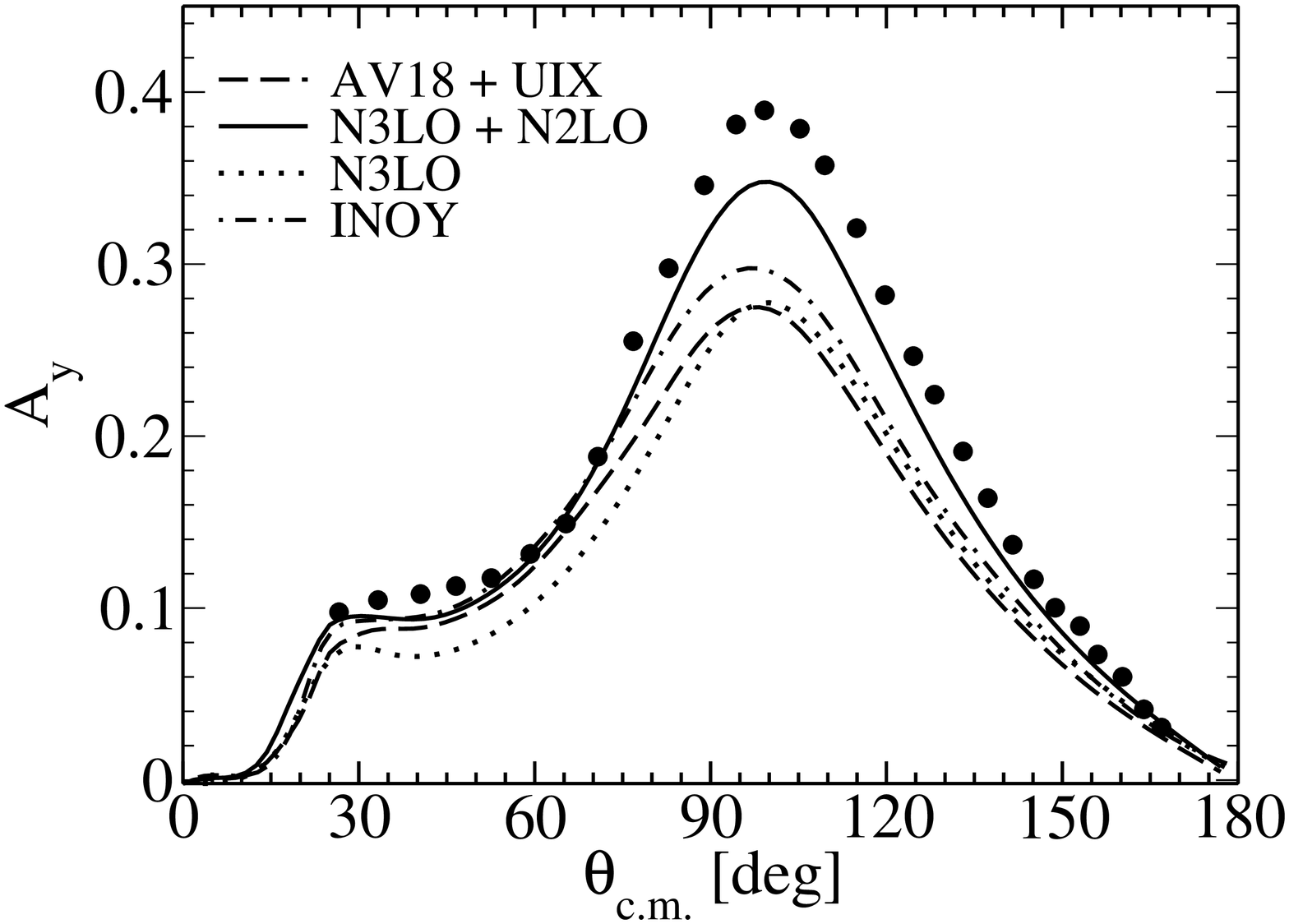}
\caption{$A_y$ of $\vec p$-$^3$He scattering at $E_p$=4 MeV calculated with various interactions.
HH results~\cite{Ki11}: AV18+UIX (dashed), I-N3LO+N2LO (solid);
AGS results~\cite{DeF07b}: I-N3LO (dotted) and INOY (dash-dotted).
Experimental data from Ref.~\cite{AlK93}.} 
\label{figure5_R}
\end{minipage}
\end{figure}

\subsubsection{Reactions with External Probes} 
\label{sec:EXT_PR}

Among the various perturbation-induced reactions with $^4$He, elastic electron scattering was naturally
considered first in ab initio calculations. Since the total angular momentum of the $^4$He ground state is zero,
there exists only an elastic $^4$He charge form factor.

For inelastic perturbation-induced reactions one has to include the final state interaction (FSI)
of the disintegrated 4N system. This can be accomplished either by a direct calculation
of the 4N continuum state or by the methods described in Section~\ref{sec:bs-cont}.

A calculation of the elastic $^4$He charge form factor with a nuclear one-body charge operator was 
carried out by Tjon in a FY calculation in 1978~\cite{Tj78}. By taking for the NN interaction a 
truncated Reid potential in $t_{00}$ approximation (see Section~\ref{sec:4_hist}) he obtained a rather 
good agreement with experimental data up to almost the first form factor minimum at about $q^2=10$ fm$^{-2}$.
A more recent calculation was made in Ref.~\cite{ViS07}, where various modern nuclear Hamiltonians were
considered and the $^4$He wave function was computed via a HH expansion.
Besides the one-body charge operator, two-body operator corrections (TBOC) were also taken into account.
The dependence of the form factor on the force model was found to be rather weak. As shown in Fig.~\ref{figure6_R}
there is a sizable TBOC contribution which then leads to good agreement with experimental data in the entire 
$q$ range considered. This demonstrates that the modern nuclear force models describe the internal structure of
$^4$He with great precision. 
\begin{figure}[ht]
\begin{minipage}[b]{0.48\linewidth}
\centering
\includegraphics[scale=0.3,angle=-90]{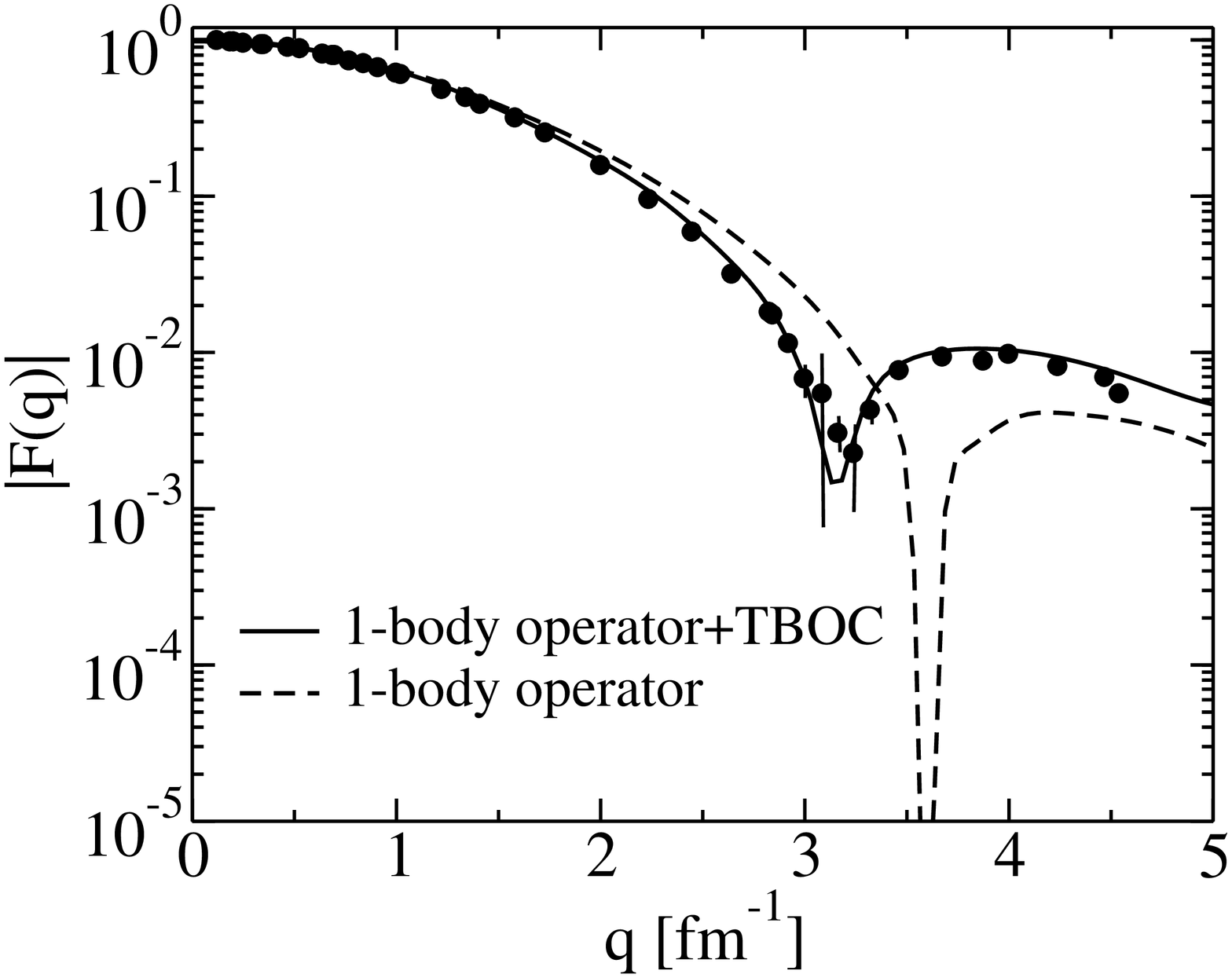}
\caption{$^4$He charge form factor: HH results~\cite{ViS07} with AV18+UIX potential;
experimental data from~\cite{Fretal68}.}
\label{figure6_R}
\end{minipage}
\hspace{0.5cm}
\begin{minipage}[b]{0.48\linewidth}
\centering
\includegraphics[scale=0.3,angle=-90]{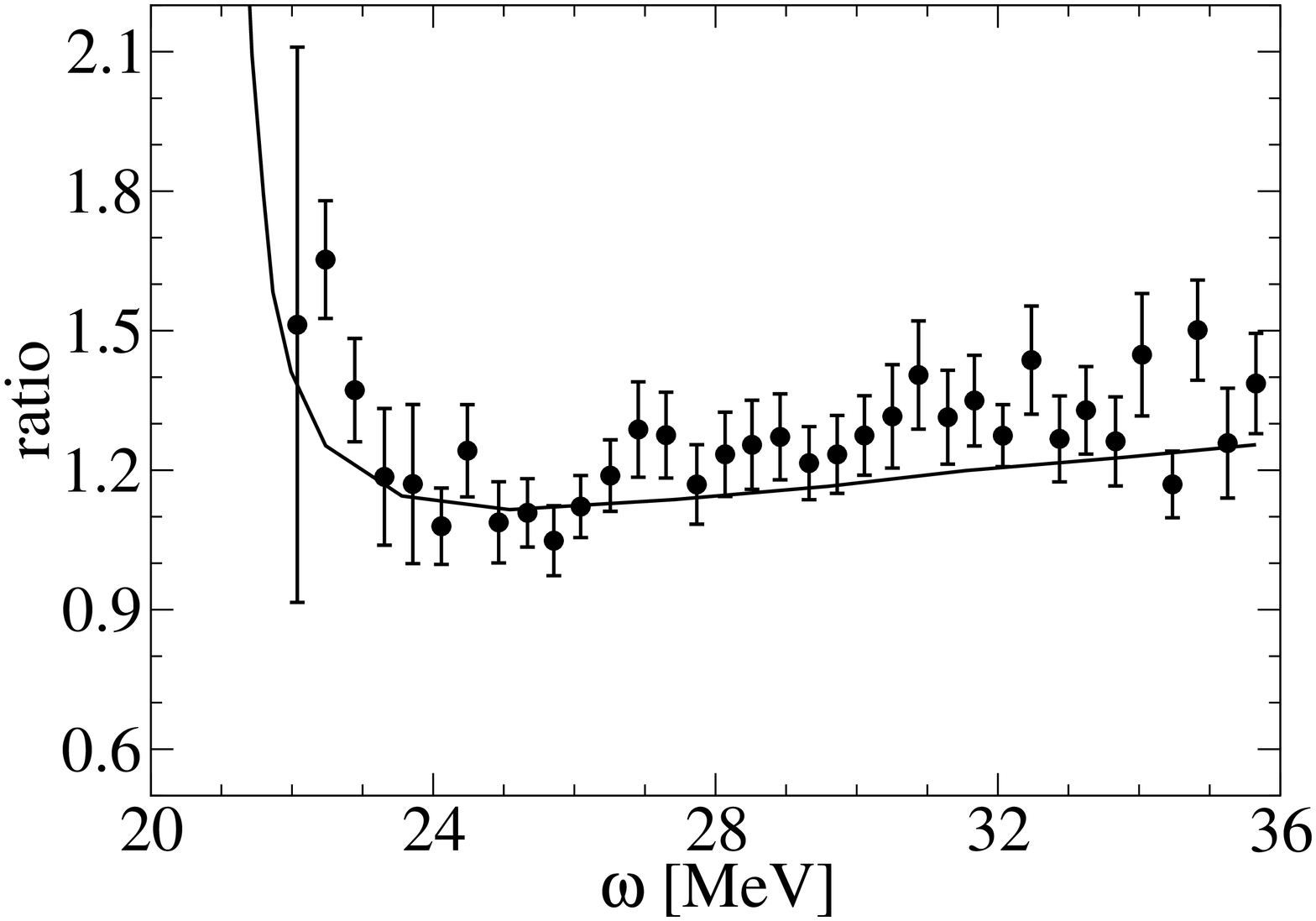}
\caption{Cross section ratio $^4$He$(e,e'p)^3$H/ $^4$He$(e,e'n)^3$He at $q$=77 MeV/c.
RGM results (see text) and experimental data from~\cite{SpK89}.}
\label{figure7_R}

\end{minipage}
\end{figure}

Of all the inelastic perturbative-induced reactions with $^4$He, perhaps most attention was given to 
photodisintegration. Calculations were made for various low-energy break-up reactions, i.e. 
$^4$He$(\gamma,p)^3$H, $^4$He$(\gamma,n)^3$He, and $^4$He$(\gamma,d)^2$H. For the latter case,
there had already been early interest in the cross section at very low energies
in order to determine the astrophysical $S$-factor.
Because the reaction is of isoscalar nature, E1 transitions 
are strongly suppressed and the reaction is mainly driven by E2 transitions. A review of the field up to 1988 is 
given by Weller and Lehman~\cite{WeL88}. Later more ambitious calculations were carried out by applying the
RGM technique (see Ref.~\cite{SaU04} for a summary). With regard to the $^4$He low-energy photodisintegration into the
(3+1) channel, abundant experimental data are available although some of the cross section 
results are controversial. It is beyond the scope of the present review to describe the rather complicated situation, instead we refer 
to Ref.~\cite{RaT12} (see also the discussion below). 

An important astrophysical reaction in the 4N system is the hep process ($p\,+ ^3$He$\,\rightarrow\,^4$He$\, + \,e^+\,+\nu_e$).
In this case the $S$-factor was calculated by Marcucci, Schiavilla, and the Pisa group using HH expansions for bound and 
scattering states~\cite{MaS00,PaM03}. A calculation of the thermal neutron 
capture on $^3$He was pursued along similar lines~\cite{GiK10}. In all these cases the nuclear force model AV18+UIX was used, 
but it should be pointed out that the electroweak operators were determined
in two different ways, namely using a conventional nuclear physics approach~\cite{MaS00,GiK10}} and 
a chiral expansion~\cite{PaM03,GiK10}. The following $S$-factors were obtained $S_{\rm hep}(0)=9.64\cdot10^{-20}$ 
keVb~\cite{MaS00} and $S_{\rm hep}(0)=(8.6\pm1.3)\cdot10^{-20}$ keVb~\cite{PaM03}. The error estimate in the 
chiral calculation stems from the cut-off dependence inherent in the chiral approach (in Ref.~\cite{AdG11} a doubling of
the error due to the cut-off dependence is recommended). For the thermal neutron capture
on $^3$He the chiral expansion was carried out up to the 4th order (N$^3$LO) and showed a rather surprising convergence pattern
(strong effect of the N$^3$LO term). A very good agreement with experimental data was attained. On the other hand one should mention
that in order to determine the chiral M1 current operator four of the so-called LECs (low-energy constants) were fitted 
to electromagnetic {\it nuclear} observables (magnetic moments of $^{2,3}$H, $^3$He and $np$ thermal capture cross section).
As shown by Girlanda~et~al.~\cite{GiK10} the conventional nuclear physics approach with a standard, though rather complicated,
current operator, obviously without any open fit parameter, also leads to a result rather close to the experimental data.

Low-energy processes with $^4$He were studied in electron scattering as well. In three experiments the $0^+$ resonance,
which is located closely above threshold (see also Section~\ref{sec:4_had})
was investigated in the momentum transfer range 100 $ \le q \le 380$ MeV/c~\cite{FrR68,Wa70,KoO83}. 
By treatment of the resonance as a bound state a GEM calculation~\cite{HiG04}  with the AV8$^\prime$ NN potential and 
an ad hoc 3NF resulted in fairly good agreement with the experimental
data. A further experimental study at low energy was made in Ref.~\cite{SpK89} for the $^4$He$(e,e'p)^3$H and $^4$He$(e,e'n)^3$He 
reactions for excitation energies $E_x$ in the range $22 \le E_x \le 36$ MeV and with $q=77$ MeV/c. 
The experimental results were compared to those of 
an RGM calculation, which was carried out with a charge symmetric nuclear Hamiltonian with a semi-realistic $V_{\rm NN}$ 
consisting of central, spin-orbit, and tensor components. The Coulomb force was included and
electromagnetic multipoles were considered up to the order $l=2$. Even though the calculation did not consider
three- and four-body break-up reactions (thresholds at $E_x=26.1$ and 28.3 MeV), a very good agreement of theoretical and 
experimental cross sections, both for proton and neutron knock-out, could be obtained. Because the uncertainty of the measured 
absolute cross section amounted to $\pm$15\%, it was safer to compare the cross section ratio $R_e=\sigma(e,e'p)/\sigma(e,e'n)$.
Of course very good agreement between theory and experiment was found for $R_e$ (see Fig.~\ref{figure7_R}).
This result for $R_e$ certainly
did not leave much need for charge symmetry breaking terms in the nuclear Hamiltonian. This was 
in contrast to findings for the corresponding ratio $R_\gamma$ obtained in $^4$He photoabsorption,
but, as already mentioned above, the $^4$He photodisintegration cross sections were not yet sufficiently settled.

Next we turn our attention to reactions beyond the low-energy regime. There,
only very few theoretical ab initio studies exist and in none of them does one find 
ab initio solutions of the 4N scattering problem beyond the three-body break-up threshold.
It is true that continuum wave functions were calculated with AGS~\cite{ElS96} and  RGM~\cite{UnH92}
techniques up to rather high energies, but the many-body break-up was neglected in these calculations. 
Due to the recent success of Ref.~\cite{DeF12} in treating the full four-body scattering, one can envisage in the 
not too far future a similar
progress for the perturbation induced break-up of $^4$He.
For $A>4$   the  proper implementation of the full continuum is presently out of reach with ab initio methods, 
where the continuum wave function is calculated explicitly. 
Using, however, the approach discussed in Section~\ref{sec:LIT} one reduces the continuum-state problem 
to a much simpler bound-state like problem, which then can be solved with standard bound-state techniques.

A first use of integral transforms in few-nucleon ab initio calculations was made by Carlson and Schiavilla.
They calculated the Laplace transform (Euclidean response) of the $^4$He inclusive longitudinal response function
$R_L(q,\omega)$ via the GFMC technique~\cite{CaS92} for various $q$-values. As nuclear forces the V8 form of the AV14 potential 
and the UVIII-3NF~\cite{Wi91} were employed. Comparisons with experimental results were made for the Euclidean 
response, which became possible by making Laplace transforms of the experimental $R_L(q,\omega)$ at a constant $q$
and assuming in addition a specific high-energy behavior of $R_L(q,\omega)$.  
For a more direct comparison of theory and experiment it is necessary to invert the integral transform.
Since the GFMC calculation contains statistical errors, and as any numerical calculation also other small errors, 
the inversion of the Laplace transform is rather unstable. A similar instability for the inversion was found for the 
Stieltjes transform~\cite{Ef85}. 
As described in Section~\ref{sec:LIT}, among the various proposed transforms the LIT turned out to be the most
adequate one, since stable inversion results could be obtained. 

In the following years the LIT method was applied to quite a 
number of perturbation-induced reactions with three- and four-body nuclei~\cite{EfL07}. Bound and LIT states were 
determined via HH expansions of CHH and/or EIHH type. The two very first LIT applications were made for $^4$He. 
Almost simultaneously calculations were carried out for the inclusive $(e,e')$ response function
$R_L(q,\omega)$~\cite{EfL97b} and for the $^4$He total photoabsorption cross section~\cite{EfL97a} 
(see also Ref.~\cite{BaE01}). 
The latter was calculated in unretarded dipole approximation, which is very reliable in the low-energy region. 
In fact in~\cite{GoS02} besides  benchmarking   FY and LIT calculations (shown in Fig.~\ref{figure_LIT_T}), 
a check of the validity of the unretarded dipole approximation 
was also made. It was found that retardation and higher multipole contributions are quite small (below 2\%)
up to moderate photon energies ($\omega\le 50$~MeV).
 
The potential models used in both the virtual and real photon calculations were the rather simple 
MTI-III and similar central potential models. For $R_L$ a fair 
agreement with experimental data was obtained in the considered momentum transfer range $300 \le q \le 500$ MeV/c.
It was found that FSI are quite important and that they are not negligible in the quasi-elastic peak, even at 
$q=500$ MeV/c~\cite{EfL97b,EfL98} (see Fig.~\ref{figure10_R}, where in addition a result of a more realistic calculation
with AV18+UIX is shown).

The situation was quite different for the $^4$He total photoabsorption cross section. As already mentioned above the experimental
results did not lead to a clear picture. In fact, the cross sections were rather different in
the giant dipole resonance region. The LIT result agreed quite well with those data sets exhibiting a high cross section
peak, while, on the other hand, data sets with a lower peak height were generally viewed more favorably
in the 1980s and 1990s, with the exception, however, of results deduced from elastic photon scattering off $^4$He~\cite{WeD92}. 
Also due to the findings of the LIT ab initio calculation new experimental interest arose and 
various new data became available with time~\cite{RaT12,NiA05,ShN05}. 
But before coming to a comparison of theory with the new experimental results we would first like to add some further information.
In the meantime LIT calculations for the $^4$He total photon absorption cross section were also performed with 
modern realistic nuclear forces (AV18+UIX in Ref.~\cite{GaB06}, chiral 2NF and 3NF in Ref.~\cite{QuN07}).
EIHH~\cite{GaB06} and NCSM~\cite{QuN07} techniques were used. The results of both calculations were rather similar and
confirmed the quite pronounced low-energy cross section peak observed in the first LIT calculation~\cite{EfL97a}.
We should mention two further recent calculations of the $^4$He total photoabsorption cross section with similar results
as in Refs.~\cite{GaB06,QuN07}, namely a LIT/EIHH
calculation with a UCOM version of AV18 by Bacca~\cite{Ba07} and a calculation by Horiuchi~et~al.~\cite{HoZ12} in a complex scaling
approach with the SVM method employing the AV8$'$ potential and an additional phenomenological 3NF.

As pointed out in Section~\ref{sec:LIT} the LIT method can also be applied to exclusive reactions.
In fact total cross sections for the $^4$He$(\gamma,n)^3$He and $^4$He$(\gamma,p)^3$H~\cite{QuL04} reactions were calculated with
the MTI-III potential. A nice agreement was obtained for the sum of both cross
sections (the only open channels at low energy) and the total low-energy photoabsorption cross section calculated with 
the same potential model and by using the LIT technique for inclusive reactions. At higher energies the difference of both
cross sections, i.e. the cross section for three- and four-body break-up reactions, was in fair agreement with corresponding
experimental data. 

Now we come back to the comparison of theory and experiment for the low-energy $^4$He photodisintegration.
In Fig.~\ref{figure8_R} we show the cross section of the reaction $^4$He$(\gamma,p)^3$H from the LIT 
calculation by Quaglioni~et~al.~\cite{QuL04} in comparison to the most recent experimental data. 
One sees very good agreement of the LIT results with the energy dependent cross section of Ref.~\cite{RaT12}, although
the theoretical results overestimate the data by about 10\%. However, there is a 
strong disagreement with the data of Ref.~\cite{ShN05}. It should be mentioned that recently Ref.~\cite{RaT12}
has cautioned against the use of the majority of radiative-capture data, like those of Ref.~\cite{ShN05} for $^4$He photodisintegration.
We want to remind the reader that the LIT calculation of Ref.~\cite{QuL04} was made with the simple MTI-III potential.
>From a comparison of the $^4$He total photo absorption cross section calculated with the MTI-III 
potential~\cite{EfL97a,BaE01} and with modern realistic nuclear Hamiltonians~\cite{GaB06,QuN07} one can estimate 
that the MTI-III result should be about 10\% too high. Thus one may expect to find a rather good agreement of a modern realistic
ab initio result with the cross section of Ref.~\cite{RaT12}.
\begin{figure}[ht]
\begin{minipage}[b]{0.56\linewidth}
\centering
\includegraphics[scale=0.3,angle=-90]{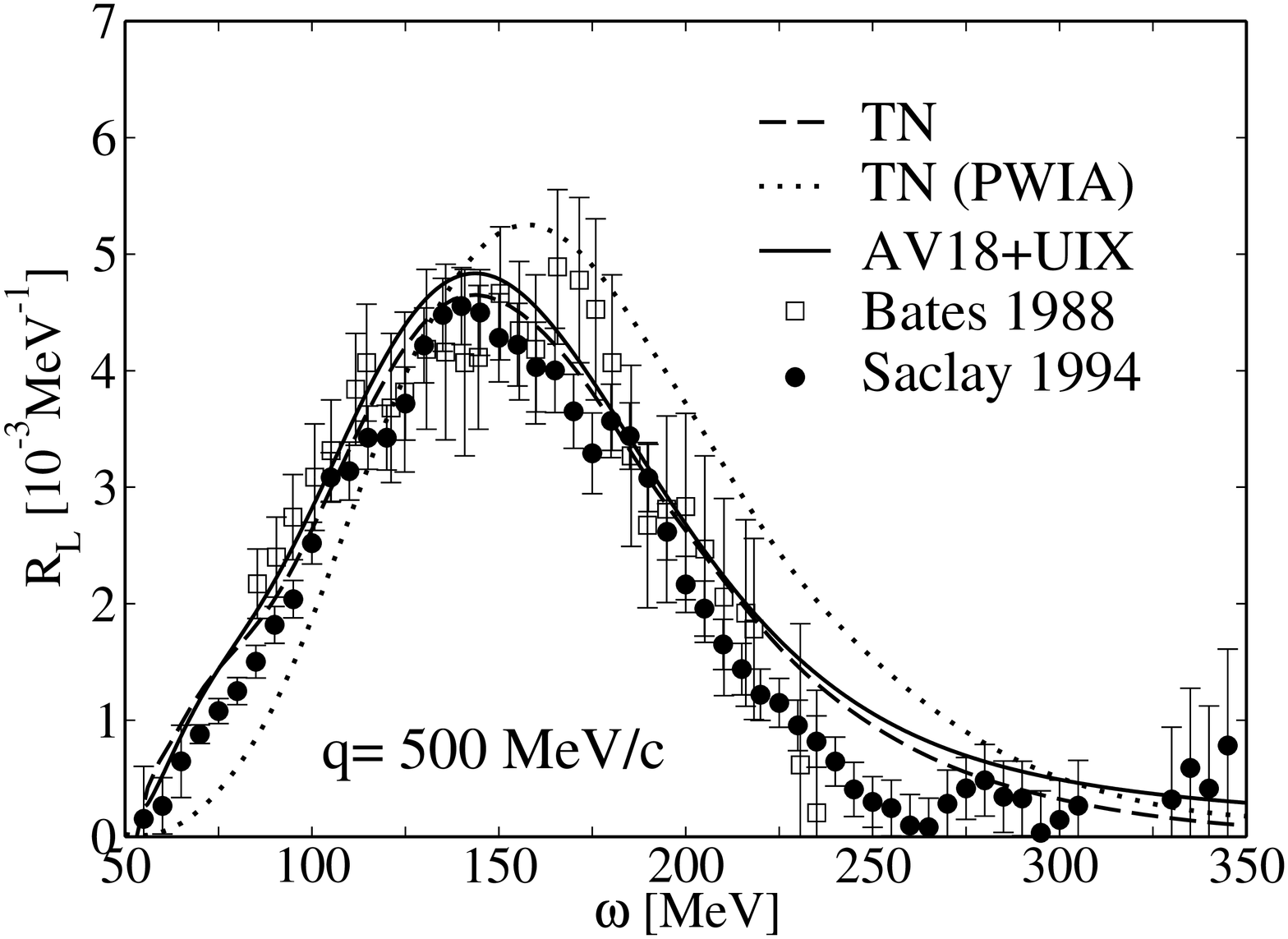}
\caption{$R_L(q,\omega)$ of $^4$He at $q$=500 MeV/c: LIT/CHH result~\cite{EfL97b,EfL98} with semi-realistic 
TN potential (dashed), LIT/EIHH result~\cite{BaB09a} with AV18+UIX (full), 
PWIA result (dotted) with TN potential in spectral function approach (see Ref.~\cite{EfL98}).
Experimental data from~\cite{DyB88} (open squares) and~\cite{Zetal94} (full dots).}
\label{figure10_R}
\end{minipage}
\hspace{0.5cm}
\begin{minipage}[b]{0.4\linewidth}
\centering
\includegraphics[scale=0.29,angle=-90]{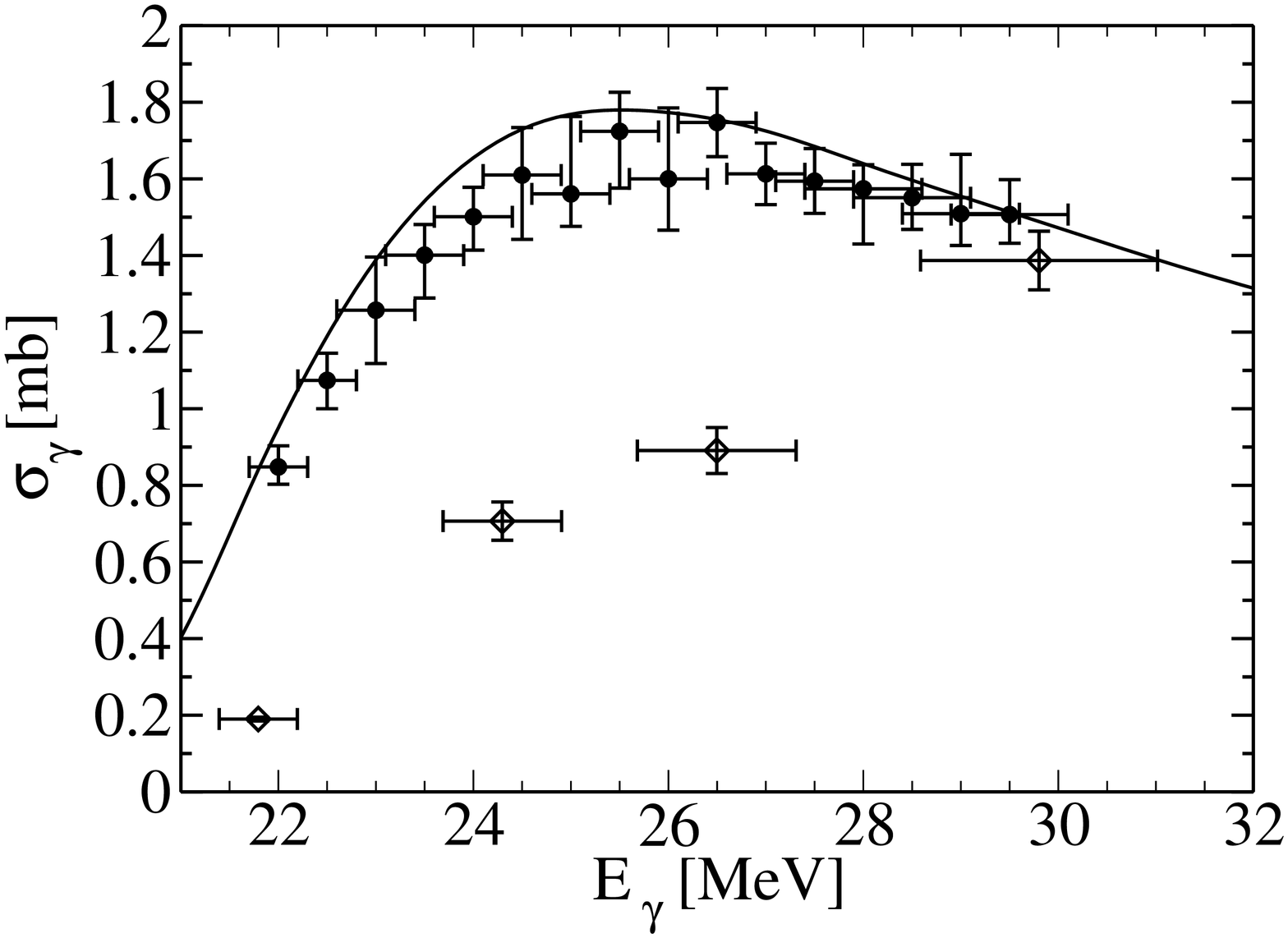}
\caption{$^4$He$(\gamma,p)^3$H cross section: Exclusive LIT/CHH calculation~\cite{QuL04} with MTI-III potential (full).
Experimental data from~\cite{RaT12} (full dots) and~\cite{ShN05} (open diamonds).}
\label{figure8_R}
\end{minipage}
\end{figure}

Recently LIT calculations with the EIHH method were carried out~\cite{BaB09a,BaB09b} for the inclusive response $R_L(q,\omega)$ of $^4$He
with modern realistic nuclear Hamiltonians (AV18 + UIX-3NF, AV18 + TM$'$-3NF~\cite{CoH01}). Note both Hamiltonians
lead essentially to the same $^4$He binding energy. 
A momentum transfer range between 50 and 500 MeV/c was considered. 
In particular, the 3NF effects on $R_L$ were investigated. For 300 MeV/c $\le q \le 500$ 
MeV/c 3NFs lead typically to effects not greater than 5$\,$-$\,$10$\,$\%. For lower momentum transfers however much larger effects 
were found, e.g. at $q$=50 MeV/c one observes reductions of more than 30\% due to the 3NF. The size of the effect depends even quite 
significantly on the choice of the 3NF~\cite{BaB09b} (see Fig.~\ref{figure9_R}). 

A first LIT calculation for the $^4$He $(e,e')$ transverse response function $R_T(q,\omega)$ 
was performed in Ref.~\cite{BaA07}. The MTI-III potential was employed and  a consistent meson exchange current (MEC)
was constructed. In contrast to  calculations of
Carlson and Schiavilla who used the Laplace transform approach with more realistic nuclear forces~\cite{CaS94,CaJ02},
no strong MEC 
effect could be found  with the semi-realistic MEC of the MTI-III potential at $q=300$ MeV/c.
 
In addition to the inclusive $(e,e')$ reactions off $^4$He the
following two exclusive reactions were calculated using the MTI-III potential and CHH expansions: 
$^4$He$(e,e'p)^3$H~\cite{QuE05} and $^4$He$(e,e'd)^2$H~\cite{AnQ06}. 
Last but not least we would like to mention a LIT benchmark calculation performed
with NCSM and EIHH techniques~\cite{StQ07}, where the test was made for various 
responses employing a simple NN potential.

Before concluding this section we have to mention that ab initio calculations were also performed for
inelastic neutrino scattering off $^4$He. These reactions play a crucial role in supernova explosion scenarios.
In the LIT calculation of Gazit and Barnea~\cite{GaB07} the interaction models 
AV8$'$, AV18, and AV18+UIX were used and HH expansions were employed with the EIHH technique. 
For the electroweak part one-body operators were taken into account, but such that vector MEC were included via 
the so-called Siegert operators. As a final result the temperature averaged cross section was given as function of the 
neutrino temperature. 
\begin{figure}[tb]
\begin{center}
\epsfig{file=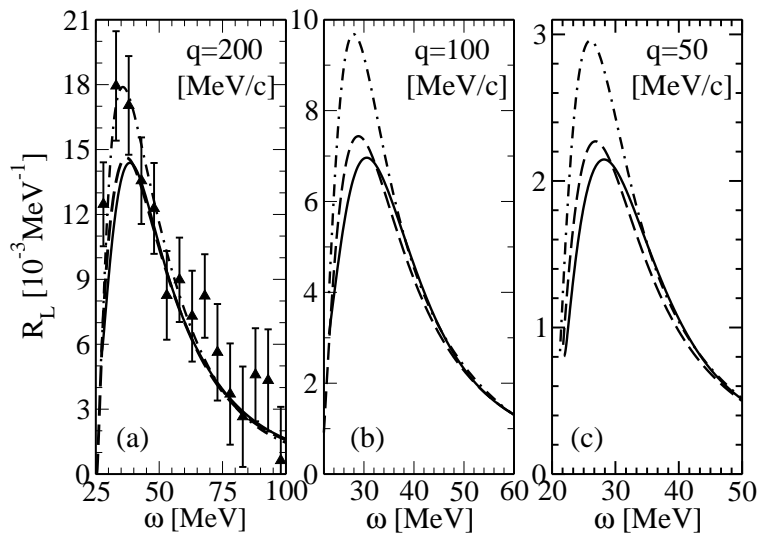,scale=0.4,angle=-90}
\caption{$R_L(q,\omega)$ of $^4$He at various values of $q$: LIT/EIHH results~\cite{BaB09a,BaB09b} 
with AV18 (dot-dashed), AV18+TM$'$ (dashed), and AV18+UIX (full) potentials;
experimental data from Ref.~\cite{BuT06}.}
\label{figure9_R}
\end{center}
\end{figure}

\subsection{Energy Spectra of $A>4$ Nuclei}
\label{sec:AG4_SPECTRA}

Another milestone in the development of ab initio calculations in few-body physics 
was set in a series of articles 
by the Urbana-Argonne-(Los Alamos) collaboration who by using the GFMC approach
paved the way for bound-state calculations of $A>4$ nuclear systems.
The first study was performed by Pudliner, Pandharipande, Carlson, and 
Wiringa in 1995~\cite{PuP95}. As interactions they
basically chose the AV18 NN potential and the UIX-3NF. However the V8 part of AV18 
together with the 3NF were calculated with GFMC while 
the remaining V9$\,$-$\,$V14 and Coulomb parts were considered in first order perturbation theory.
Further electromagnetic corrections were taken from  VMC. 

\begin{table}
\begin{center}
\caption{{\label{GFMC_A7}} Experimental and GFMC (AV18+UIX) energies of $A=6, \,7$ nuclei in MeV; also
given is $\Delta$, the deviation in MeV of the GFMC energy from the experimental one  (from Ref.~\cite{PuP97}).  }
\begin{tabular}{c||c|c|c||c||c|c|c}
\hline \hline
$^A$X($J^\pi;T$)  & GFMC  & Expt. & $\Delta$ & $^A$X($J^\pi;T$) & GFMC  & Expt. & $\Delta$ \\
\hline
$^6$He$(0^+;1)$ & --27.64(14) & --29.27 & 1.63 & $^7$He$({\frac{3}{2}}^-;{\frac{3}{2}})$ & --25.16(16) & --28.82 & 3.66 \\
$^6$He$(2^+;1)$ & --25.84(11) & --27.47 & 1.63 & $^7$Li$({\frac{3}{2}}^-;{\frac{1}{2}})$ & --37.44(28) & --39.24 & 1.80 \\
$^6$Li$(1^+;0)$ & --31.25(11) & --31.99 & 0.74 & $^7$Li$({\frac{1}{2}}^-;{\frac{1}{2}})$ & --36.68(30) & --38.76 & 2.08 \\
$^6$Li$(3^+;0)$ & --28.53(32) & --29.80 & 1.27 & $^7$Li$({\frac{7}{2}}^-;{\frac{1}{2}})$ & --31.72(30) & --34.61 & 2.89 \\
$^6$Li$(0^+;1)$ & --27.31(15) & --28.43 & 1.12 & $^7$Li$({\frac{5}{2}}^-;{\frac{1}{2}})$ & --30.88(35) & --32.56 & 1.68 \\
$^6$Li$(2^+;0)$ & --26.82(35) & --27.68 & 0.86 & $^7$Li$({\frac{3}{2}}^-;{\frac{3}{2}})$ & --24.79(18) & --28.00 & 3.21 \\
$^6$Be$(0^+;1)$ & --25.52(11) & --26.92 & 1.40 & - & - & -  & -\\
\hline \hline
\end{tabular}

\end{center}
\end{table}
Rather good results for the binding energies of the $A=6$ nuclei were obtained, namely 28.2(8) MeV $(^6$He)
and 32.4(9) MeV ($^6$Li), to be compared with the experimental values of 29.3 and 32.0 MeV, respectively.
Also the calculated $^6$Li proton point radius of 2.41 fm agreed well with the experimental value of 2.43 fm.
In addition, the $1/2^-$ and $3/2^-$ resonances of $^5$He were investigated, however, the calculated splitting
of 0.8(3) MeV was much smaller than the observed 1.4 MeV.

The GFMC study was continued in Ref.~\cite{PuP97}. The calculation
was extended to $A=7$. The same interaction model (AV18+UIX) was used and quite a number of 
results were presented. Here we illustrate only the excitation spectra  of the $A=6, \,7$ nuclei (see Table~\ref{GFMC_A7}).
It is evident that the energies are close, but somewhat above the experimental numbers. The authors believed that the discrepancy was
most likely due to the phenomenological short-range part of the 3NF. In the next GFMC
study also the $A=8$ nuclei were included~\cite{WiP00}, although the nuclear force was still described by the model AV18+UIX.
However now due to a constrained-path algorithm for the complex spatial and spin-isospin wave functions the precision of the calculation was improved.
The authors estimated that, for a given Hamiltonian, binding energies could be obtained with an accuracy of 1 to 2~\%.
In Table~\ref{GFMC_A8} we list the calculated ground-state energies of the various nuclei considered. Comparing the results
for the $A=6, \,7$ nuclei with those of Table~\ref{GFMC_A7} one finds for three of the four cases that the theoretical
 results moved closer to the experimental
values, however discrepancies still remained. In Ref.~\cite{WiP00} a large number of additional results are given, among them being the 
spectra of various nuclei. A particularly interesting finding was the strong 2$\alpha$ clustering in the $^8$Be ground state.
In the conclusion the authors stated that the general features of the light $p$-shell nuclei were described
fairly well, including the binding energies, and the correct ordering and approximate spacing of
the spectra. They further noted specific failures of the AV18+UIX Hamiltonian, namely a few percent underbinding in 
$N=Z$ nuclei (which increases as the $n$-$p$ asymmetry increases) and insufficient spin-orbit splittings.   
\begin{table}
\begin{center}
\caption{{\label{GFMC_A8}} Experimental and GFMC (AV18+UIX) energies of nuclear ground states in MeV; also
given is $\Delta$, defined as in Table~\ref{GFMC_A7} (from Ref.~\cite{WiP00}).  }
\begin{tabular}{c||c|c|c|c|c|c|c}
\hline \hline
$^A$X($J^\pi;T$)  & $^6$He$(0^+;1)$  & $^6$Li$(1^+;0)$ &  $^7$He$({\frac{3}{2}}^-;{\frac{3}{2}})$ & $^7$Li$({\frac{3}{2}}^-;{\frac{1}{2}})$ & 
  $^8$He$(0^+;2)$  & $^8$Li$(2^+;1)$ &  $^8$Be$(0^+;0)$ \\
\hline
GFMC & --28.11(9) & --31.15(11) & --25.79(16) & --37.78(14) & --27.16(16) & --38.01 & --54.44 \\
Expt. & --29.27 & --31.99 & --28.82 & --39.24 & --31.41 & --41.28 & --56.50 \\
$\Delta$ & 1.16 & 0.84 & 3.03 & 1.46 & 4.25 & 3.27 & 2.06 \\
\hline \hline
\end{tabular}
\end{center}
\end{table} 

In order to improve the comparison with experiment new 3NFs were developed, namely the five Illinois models (IL1-IL5)~\cite{PiP01}.
Compared to the UIX-3NF, which contains two terms, the Illinois-3NFs include two additional terms, a two-pion exchange $s$-wave 
term and three-pion ring term. The overall strengths of the four terms together with a cut-off factor in the radial dependence, were
adjusted to obtain best fits to the energies of 17 narrow states in $3 \le A \le 8$ nuclei (at most three parameters at 
a time could be uniquely determined and thus various models were constructed).
The effect of the Illinois 3NFs was then studied in a GFMC calculation for the $A=9, \,10$ nuclei~\cite{PiV02}.
In Table~\ref{GFMC_A910} we show some of the results obtained. Without a 3NF one observes a relatively large underbinding of
about 8 to 16 MeV which reduces to an underbinding of about 2 to 6 MeV
when the UIX-3NF is introduced. The Illinois 3NFs, however,
lead to much better results. The overall best result, including the 17 narrow states, was obtained with the IL2-3NF.
\begin{table}
\begin{center}
\caption{{\label{GFMC_A910}} Experimental energies of $A=9, \,10$ nuclei in MeV; also
given is $\Delta$(3NF-model), the deviation in MeV of the GFMC energy calculated with AV18 and a specific 3NF-model from 
the experimental energy (from Ref.~\cite{PiV02}).  }
\begin{tabular}{c||c|c|c|c|c|c}
\hline \hline
$^A$X($J^\pi$)  & Expt.  & $\Delta$(No-3NF) & $\Delta$(UIX) & $\Delta$(IL2) & $\Delta$(IL3) & $\Delta$(IL4) \\
\hline
$^9$Li$({\frac{3}{2}}^-)$ & --45.34 & 11.64(30)  & 4.44(30) & --0.66(40) & --1.36(50) & --2.26(40) \\
$^9$Li$({\frac{1}{2}}^-)$ & --42.65 & 8.65(30)   & 3.25(30) & --1.15(40) & --1.55(40) & --1.85(50)  \\
$^9$Li$({\frac{5}{2}}^-)$ & --39.96 & 7.86(30)   & 2.06(40) & --1.14(40) & --1.04(40) & --0.64(40) \\
$^9$Li$({\frac{7}{2}}^-)$ & --38.91 & 9.21(30)   & 3.71(30) & --0.09(40) &  0.21(40) & --1.89(40) \\
$^9$Be$({\frac{3}{2}}^-)$ & --58.16 & 11.76(40)  & 3.06(30) & --0.04(50) &  0.36(50) &  0.16(60) \\
$^9$Be$({\frac{5}{2}}^-)$ & --55.73 & 12.23(30)  & 4.43(50) & --0.07(50) &  1.63(40) &  0.63(50) \\
$^9$Be$({\frac{1}{2}}^-)$ & --55.36 & 10.36(40)  & 4.46(60) &  1.06(40) & --0.34(50) & --0.14(40) \\
$^{10}$Be$(0^+)$          & --64.98 & 12.98(50)  & 5.78(60) & --1.82(70) & --0.22(70) & --2.42(60) \\
$^{10}$Be$(2^+)$          & --61.61 & 13.91(50)  & 4.51(60) & --0.19(50) &  1.71(60) &  0.51(60) \\
$^{10}$B$(3^+)$           & --64.75 & 16.15(60)  & 5.75(40) & --0.85(50) &  0.65(50) & --0.85(60) \\
$^{10}$B$(1^+)$           & --64.03 & 12.43(60)  & 3.73(50) & --0.67(40) &  1.23(50) & --1.07(50) \\
\hline \hline
\end{tabular}

\end{center}
\end{table}
In fact the following root mean square  deviations from the experimental energies were found: 0.6 MeV (AV18+IL2), 0.8 MeV (AV18+IL3),
and 1.0 (AV18+IL4). This contrasts with rms deviations of 3.5 MeV for AV18+UIX and 9.9 MeV for AV18 alone. 
A further GFMC calculation using the AV18+IL2 Hamiltonian studied states with higher excitations
in $A\,$=$\,$6$\,$-$\,$8 nuclei~\cite{PiW04}.
This calculation benefited from a significant improvement of the VMC trial functions.
A good representation of the experimental spectra was found i.e. the rms deviations from 36 experimental energies with 
$6 \le A \le 8$ was only 0.6 MeV. In a later calculation even the $^{12}$C ground-state energy was calculated with the
AV18+IL2 Hamiltonian~\cite{Pi08}. Again, use of only the AV18 resulted in an underbinding of about 20 MeV
which was reduced to about 3 MeV with the addition of the IL2-3NF. 

\begin{table}
\begin{center}
\caption{{\label{6Li}} $^6$Li binding energy for Volkov and Minnesota~\cite{ThL77} (MN) potentials with various ab 
initio methods (note MN results are taken from Ref.~\cite{NaC04})  }
\begin{tabular}{c||c|c|c|c|c}
\hline \hline
Potential  & SVM~~\cite{VaS95} & HH~\cite{BaL99} & HH~\cite{GaK11} & NCSM~\cite{NaC04} & EIHH~\cite{BaL00} \\
\hline
Volkov & 66.25 & 66.57  & 66.49 & - & -  \\
MN & 36.51 & -   & - & 36.50(4) & 36.64(7) \\
\hline \hline
\end{tabular}
\end{center}
\end{table}
Other ab initio methods were not on a comparable level with the GFMC calculations for quite some time.
The SVM technique was applied to $A\,$=$\,$2$\,$-$\,$7 nuclei~\cite{VaS95} using simple NN potentials, like the
MTV and Volkov models.
Similar simple potential models were taken in a HH calculation~\cite{BaL99} based on an improvement of the IDEA approach 
(see Section~\ref{sec:HH}).
In Table~\ref{6Li} we list the $^6$Li binding energy for two of these simple potential models
as well as other available results from later ab initio studies.
Comparing the various results of Table~\ref{6Li} one finds a rather good agreement.
\begin{figure}[ht]
\begin{minipage}[b]{0.46\linewidth}
\centering
\includegraphics[scale=0.3,angle=-90]{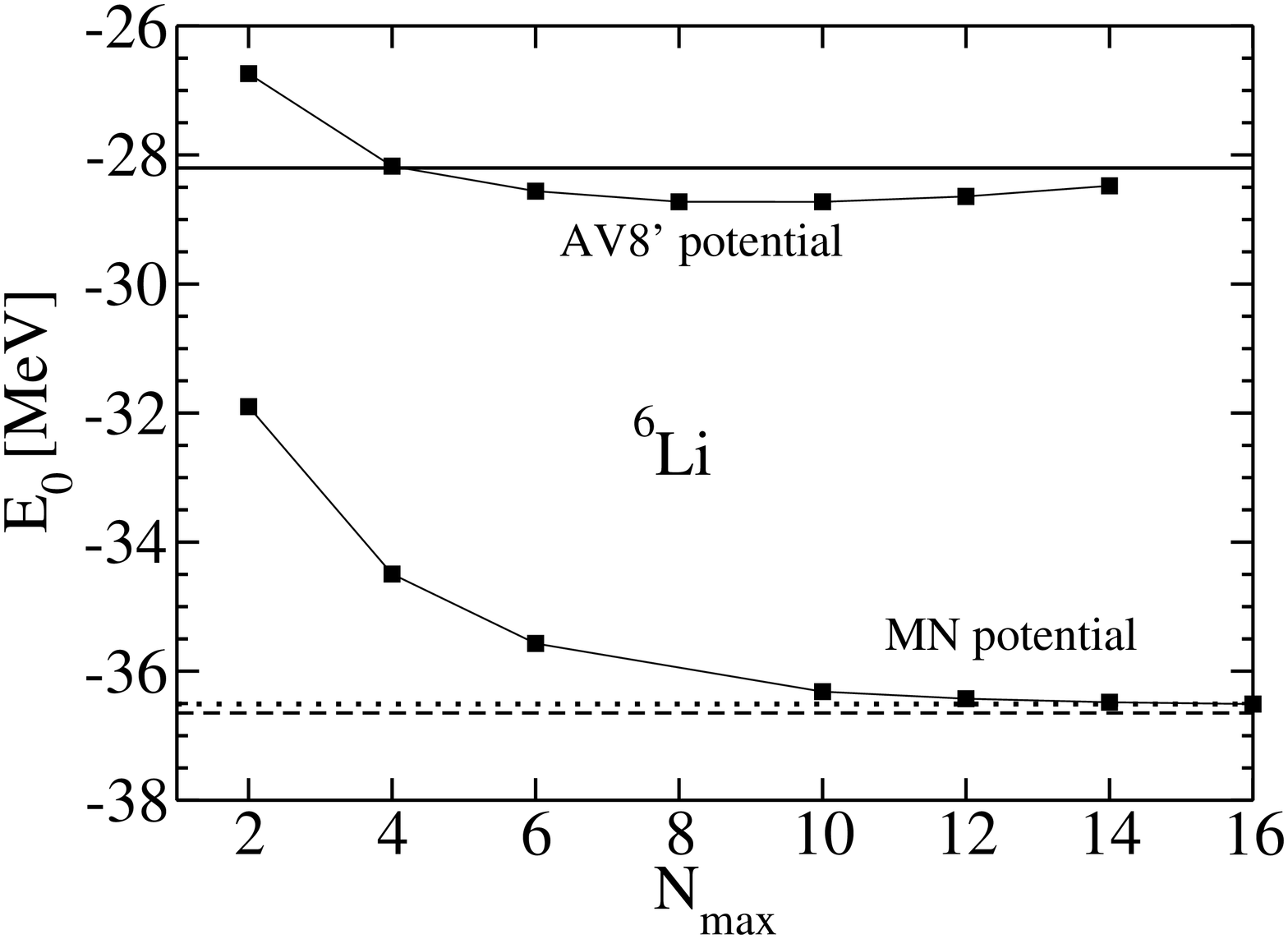}
\caption{Convergence pattern of $^6$Li-NCSM ground state energy (squares) with AV8$'$ and
Minnesota~\cite{ThL77} (MN) potentials (from~\cite{NaC04}), also given results for GFMC/AV8$'$ (full), 
SVM/MN (dotted), EIHH/MN (dashed).}
\label{figure11_R}
\end{minipage}
\hspace{0.5cm}
\begin{minipage}[b]{0.50\linewidth}
\centering
\includegraphics[scale=0.3,angle=-90]{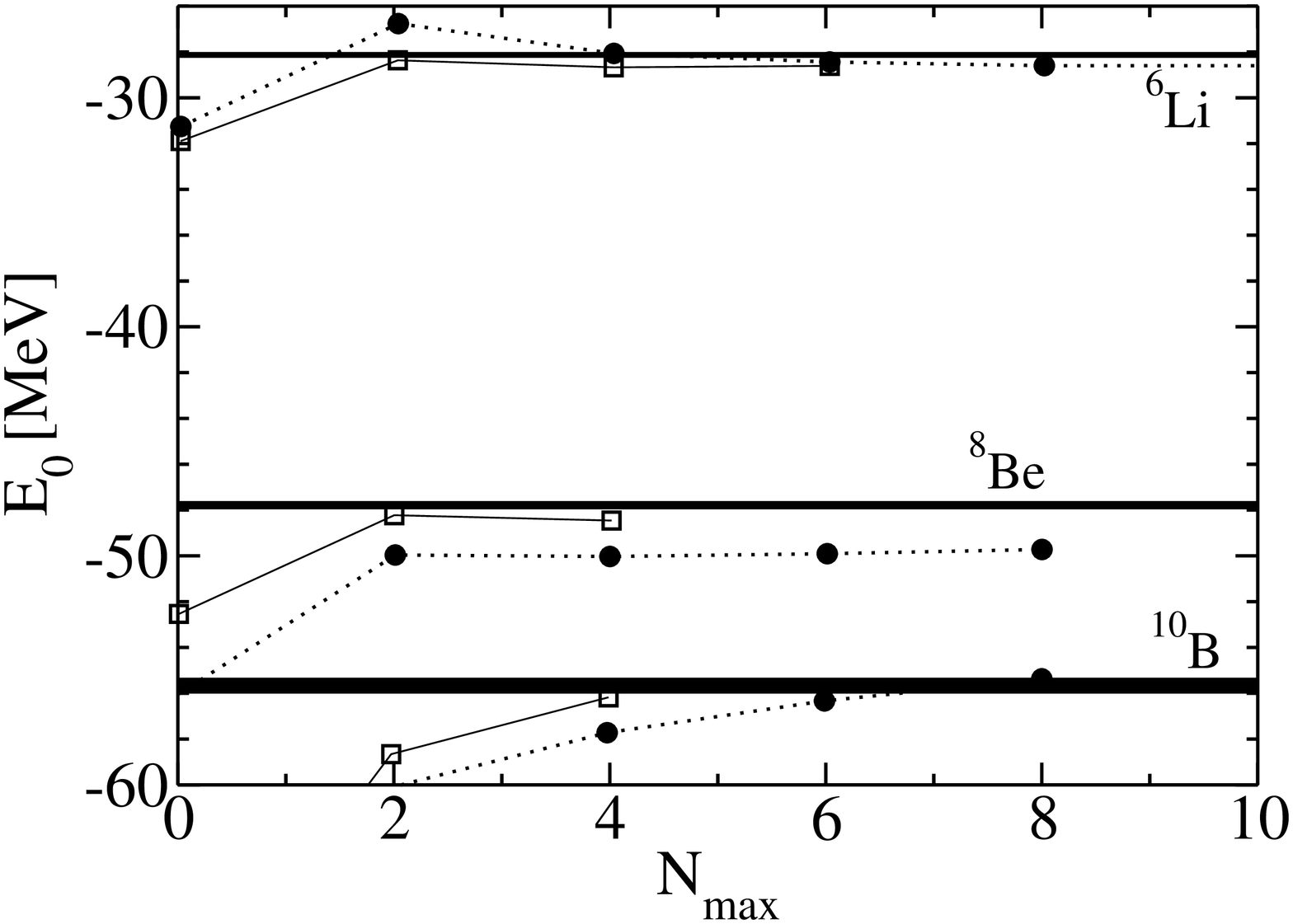}
\caption{As in Fig.~\ref{figure11_R}, but for $^6$Li, $^8$Be, $^{10}$B. NCSM results (AV8$'$)
with 2- (full circles) and 3-body (open squares) effective interaction (from~\cite{NaO02});
also given GFMC results (full lines) with statistical error. }
\label{figure12_R}
\end{minipage}
\end{figure}

It is interesting to investigate further progress 
in $A>4$ calculations with the various ab initio methods of Table~\ref{6Li}. The SVM technique and 
the HH approach of Ref.~\cite{BaL99} were not developed further to treat the more realistic potential models. The HH method 
of Ref.~\cite{GaK11} is quite recent and one has to await further developments. The EIHH approach was applied in several
studies with central potential models in $A\,$=$\,6,\,7$ nuclei (see Section~\ref{sec:AG4_FURTHER}).
In addition more realistic potentials were considered, e.g. $^6$Li with AV8$'$~\cite{BaL03},
although the results were not fully converged, 
particularly in the presence of a 3NF. The approach which has seen
the greatest progress is undoubtedly the NCSM, first presented
by Navr\'atil, Vary and Barrett~\cite{NaV00} and recently reviewed in~\cite{NaQ09}.

In various NCSM studies~\cite{NaO02,NaV01,CaN02,NaO03,NaC04,FoN05} spectra of $A>4$ nuclei were calculated for various 
potential models, among them also the AV8$'$ potential enabling comparisons with GFMC results of the 
Urbana-Argonne-(Los Alamos) group (see Table~\ref{NCSM_GFMC}). Inspection of Table~\ref{NCSM_GFMC} shows that
the NCSM results are rather close to those of the GFMC calculation. Thus, one may consider the NCSM results as
almost, but not yet fully, converged. To investigate this better,
we show in Fig.~\ref{figure11_R} 
the NCSM convergence pattern of the $^6$Li ground-state energy  with a two-body effective interaction. One sees that,
in order to have a convergent result, the number of considered HO quanta, $N_{\rm max}$, has to be quite large 
and certainly larger than the values given in Table~\ref{NCSM_GFMC}. As shown in Fig.~\ref{figure12_R} the convergence can 
be accelerated if a three-body effective interaction is used instead of a two-body one. In fact,
by inspecting the results of Table~\ref{NCSM_GFMC} one finds that to achieve the same quality
as given by the pure two-body effective interaction result in general a lower $N_{\rm max}$ is needed when
the three-body effective interaction is included.

\begin{table}
\begin{center}
\caption{{\label{NCSM_GFMC}} Comparison of GFMC and NCSM energies of various nuclei calculated with the AV8$'$ potential;
listed in MeV are the GFMC energies and $\Delta=E$(NCSM)$-E$(GFMC). NCSM results are
calculated with two-body and partially also with three-body effective interactions, $V_{\rm 2-eff}$ and $V_{\rm 3-eff}$,
respectively; number of considered HO quanta $N_{\rm max}$ in square brackets (from Refs.~\cite{NaO02,CaN02,FoN05}).  }
\begin{tabular}{c||c|c|c||c||c|c|c}
\hline \hline
$^A$X($J^\pi;T$)  & GFMC  & $\Delta(V_{\rm 3-eff})$ &  $\Delta(V_{\rm 2-eff})$ &  $^A$X($J^\pi$)  & GFMC  & 
   $\Delta(V_{\rm 3-eff})$ &  $\Delta(V_{\rm 2-eff})$ \\
\hline
$^6$Li$(1^+;0)$ & --28.19(5) & --0.42 [6] & --0.41 [10] & 
   $^8$Be$(1^+;0)$ & --32.77(15) & 1.64 [4]  & --0.03 [8]  \\
$^6$Li$(3^+;0)$ & --24.98(5) & --0.56 [6] & --0.60 [10] & 
   $^8$Be$(3^+;0)$ & --31.23(15) & 1.65 [4]  & 0.24 [8]   \\
$^6$Li$(0^+;1)$ & --24.15(4) & --0.61 [6]  & --1.05 [10] & 
   $^8$Be$(2^+;1)$ & --32.7(1) & 0.86 [4] & --0.70 [8] \\
$^6$Li$(2^+;0)$ & --24.12(4) &  0.15 [6] & --0.11 [10] & 
   $^9$Be$({\frac{3}{2}}^-;{\frac{1}{2}})$ & --49.9(2) & - & --0.30 [8]  \\
$^8$Be$(0^+;0)$ & --47.89(11) & --0.57 [4] & --1.83 [8] &  
   $^{10}$B$(1^+;0)$ & --55.67(26) & --0.52 [4] & 0.30 [8]  \\
$^8$Be$(2^+;0)$ & --45.62(11) & 0.82 [4] & --0.47 [8] & 
   $^{10}$B$(3^+;0)$ & --53.23(26) & --1.60 [4] & --0.60 [8]  \\
$^8$Be$(4^+;0)$ & --38.69(11) & 2.64 [4] & 1.44 [8] & - & - & - & - \\
\hline \hline
\end{tabular}
\end{center}
\end{table}

As evident from Fig.~\ref{figure11_R}
the NCSM calculations were also performed with NN potentials other than the AV8$'$.
The initially preferred model was the CD-Bonn potential, since 3NF effects are less pronounced in this case i.e.
neglect of the 3NF has less important consequences. Later on, the chiral I-N3LO potential~\cite{EnM03}
was used in addition~\cite{NaC04}. The first consideration of a 3NF in the NCSM calculations for $A>4$ nuclei
was made in Ref.~\cite{NaO03}, where the interaction model AV8$'$+TM$'$ was used and accordingly a three-body effective
interaction was employed. Taking $N_{\rm max}=6$ for $A=6, \,7$ nuclei and $N_{\rm max}=4$ 
for $A>7$ nuclei, NCSM calculations were carried out up to $^{13}$C. Though such a relatively low value of $N_{\rm max}$ did not lead to 
convergent results (see also comparison of GFMC and NCSM results with AV8$'$+TM$'$ in Ref.~\cite{Pi05}), 
interesting 3NF effects like e.g. an enhancement of spin-orbit effects were found. 
In the large basis NSCM calculation ($N_{\rm max}=8, \,9$) of $^{9,11}$Be~\cite{FoN05} the non-local NN potential INOY, 
which makes a 3NF obsolete, was added to the list of potentials. The INOY results suggested that a realistic 3NF
could have an important influence on the parity inversion of the $^{11}$Be ground state ($^{11}$Be has an unnatural
1/2$^+$ ground state).

In conjunction with the I-N3LO NN potential, two chiral 3NFs, models 3NF-A and 3NF-B, were used in the NCSM calculation of Ref.~\cite{NoN06}.  
The excitation spectrum of $^7$Li was calculated by taking $N_{\rm max}=6$, 
which did not allow fully convergent results to be reached. For the 
$^7$Li binding energy the following results were obtained, 38.0 MeV (3NF-A) and 36.7 MeV (3NF-B), whereas one has
a GFMC result (model AV18+IL2) of 38.9 MeV and an experimental value of 39.2 MeV. Thus the chiral forces, though having a
soft core, did not lead to overbinding. As the authors pointed out, it was previously believed that only 
the addition of a repulsive
core in the 3NF, as in the Urbana and Illinois 3NF models, could prevent overbinding. In addition, 
it was confirmed that the spectra depend on the choice of the specific 3NF. The study of NCSM calculations with chiral forces was 
continued in Ref.~\cite{NaG07}. The chiral 3NF was somewhat different from Ref.~\cite{NoN06},
in that the 3NF was chosen to be local
and the constraint to fit the $^4$He binding energy was loosened somewhat by considering additional observables of
$A>4$ nuclei in order to determine one of the 3NF parameters. Besides these other observables, the spectra 
of $A\,$=$\,$10$\,$-$\,$13 nuclei were calculated using 
$N_{\rm max}=6$. Overall, the 3NF contributed significantly to improve the agreement
between theory and experiment. This was
especially well demonstrated in the odd mass nuclei for the lowest few excited states.

Next we report on a recent NCSM study~\cite{RoL11} which employed an importance truncation scheme, where 
a criterion based on perturbation theory permitted certain states to be discarded. In this way a NCSM calculation 
with $N_{\rm max}=12$ became feasible 
even for $^{12}$C and $^{16}$O. In addition a SRG-transformed chiral Hamiltonian was employed (I-N3LO and a N2LO-3NF
version, where the 3NF parameters are constrained by both the triton binding energy and the triton 
$\beta$-decay~\cite{GaQ09}).
As described in Section~\ref{sec:SIM} the SRG introduces a prediagonalization of the Hamiltonian. Since it is important
to understand the SRG transformation correctly, we reiterate here the essential features. In order to have
a unitary transformation at the $A$-body level one should consider $A$-body operators in the SRG transformation.
However, in practice the SRG transformation is performed only at the two- or at maximum the three-body level. 
A SRG transformation at the two-body level leads to a phase shift equivalent NN interaction. 
Nonetheless, depending on the bare 
interaction and on the particle system under consideration, one may obtain a unitary transformation at the two-body (three-body) level, 
which can lead to results different from those of the untransformed Hamiltonian for $A>2$ ($A>3$) systems. The unitarity of the transformation  
on a specific N-body level can be checked by the independence of the results 
on the flow parameter $\alpha$. In other words, if one applies a SRG transformation at the two-body (three-body) level and obtains 
an $\alpha$ dependent binding energy in a calculation of an $A$-body system, then the transformation is not unitary, but induces
$\alpha$ dependent three-body (four-body) and/or higher order many-body forces. Note that the case of $\alpha$ independence does
not necessarily mean one is working with the original Hamiltonian, but that the induced many-body forces are $\alpha$ independent. 
On the other hand one is certain to be working with a NN phase shift equivalent interaction. It should be obvious that in presence of a bare
3NF the SRG transformation should be made at least at the three-body level.

We come back now to the calculation of Ref.~\cite{RoL11} where in fact the $\alpha$ dependence of the results was checked. 
By employing only the NN potential and the SRG transformation at the two-body level, it was shown that
the ground-state energies of the nuclei considered ($^4$He, $^6$Li, $^{12}$C, $^{16}$O) had a strong $\alpha$ dependence
thereby indicating that the transformation was not unitary on the $A$-body level.
By bringing the SRG transformation to the three-body level the results became approximately $\alpha$ independent
at $N_{\rm max}=12$. It is interesting to note that
an $^{16}$O binding energy of 119.7(64) MeV was obtained which is close to recent results of 121.0 and 119.5 MeV with  CC~\cite{HaP10} 
and UMOA~\cite{FuO09} techniques, respectively (for the latter see the end of Section~\ref{sec:NCSM}).
With the additional inclusion of the bare 3NF it was found that only the $^4$He binding energy is $\alpha$ independent.
However by extrapolating to larger $N_{\rm max}$
it was also possible to determine the binding energy of $^6$Li. On the contrary, even with
extrapolations, there still remained a pronounced $\alpha$ dependence 
for $^{12}$C and $^{16}$O showing that the induced $\alpha$ dependent 4N and higher-order forces are quite large.

The use of soft interactions makes full configuration NCSM calculations feasible. An example is found in Ref.~\cite{MaV09}, where
$A=6,8,12$ and 16 binding energies (and a few resonant states) were obtained. This was possible using the JISP potential~\cite{ShW07} 
and an extrapolation procedure
that was shown to be reliable in $A=2,3,4$ test cases, where fully converged results can be obtained.
Here one should also mention an interesting and promising exploration presented in  Ref.~\cite{AbM11} regarding the applicability
of the MCSMD (see Section~\ref{sec:MCSMD}) to $A>4$ systems. There a benchmark of the binding energies, point-particle 
rms matter radii, and electromagnetic moments for light nuclei ranging from $^4$He to $^{12}$C showed that   
results from full configuration interaction and MCSMD agree  within a few percent at most.

In recent years great progress was also made in the CC approach.
In fact just mentioned above was the CC result for $^{16}$O with the I-N3LO potential.
We already reported in Section~\ref{sec:CC} that CC binding energy results of $^4$He were successfully
benchmarked with FY results. The CC method is better suited for treating closed shell and neighboring nuclei
and can be applied to nuclei with considerably larger $A$ than with other ab initio methods. 
In Ref.~\cite{HaP10}, besides the above mentioned $^{16}$O case, even $^{40}$Ca and $^{48}$Ca were studied
with the I-N3LO NN potential in a $\Lambda$-CCSD(T) calculation. The model space comprising of 18 major HO shells
was sufficiently large. For $^{48}$Ca a similar underbinding per nucleon (0.4 MeV) as for $^{16}$O was found, whereas for
$^{40}$Ca a small overbinding of 0.08 MeV per nucleon was obtained. Thus a rather different effect of the 3NF would be 
needed for these nuclei. Further CC applications were performed mainly for $A > 16$ nuclei
(see e.g. \cite{JeH11}), which is certainly out of the scope of the present review.
First attempts of CC calculations in open shell nuclei with two nucleons (two holes) in the last shell were pursued
by considering the $^6$He nucleus~\cite{{JaH11}} and working in small model spaces. 
The results were quite encouraging and
prospects are good for future calculations in larger model spaces.

Also results from lattice simulations became available lately (see Section~\ref{sec:LCEFT}). A LCEFT calculation of
the spectra of the three lightest $\alpha$-nuclei was carried out in Ref.~\cite{EpK11}. Particularly interesting is the
result for the Hoyle state, where an energy of -85(3) MeV was found, which is close to the experimental value of
-84.51 MeV. 

Another very promising approach for the ab initio treatment of a larger number of particles is the AFDMC technique
mentioned in Section~\ref{sec:GFMC}. For finite nuclear systems, however, applications have only be made with simpler
interaction models like the AV6$'$ NN potential~\cite{GaP07}. 

\subsection{\it Further Observables for $A>4$ Nuclei}
\label{sec:AG4_FURTHER}

In addition to excitation spectra further observables exist for nuclear bound states. One has
charge and matter radii, magnetic and quadrupole moments, and electroweak transition
strengths. Here we do not discuss these observables in detail but just comment on a few selected points.
Since in most of the ab initio calculations wave functions of the various nuclear states are calculated, it is
in principle straight forward to evaluate the above mentioned observables. In fact in most of the NCSM calculations 
of the previous section such observables were calculated. In addition we mention here two further NCSM studies:
Ref.~\cite{CaN06} (charge radii of $^{4,6,8}$He) and Ref.~\cite{FoC09} (charge radii and electromagnetic moments
of Li and Be isotopes). In GFMC calculations
no wave functions are obtained, but nuclear radii and electromagnetic moments can be calculated nonetheless
(see e.g. Ref.~\cite{WiP00}). The GFMC calculation of electroweak transition matrix elements is
described in Ref.~\cite{PeP07}, where quite a number of transitions were calculated with the AV18+IL2 interaction model. In most cases 
a good agreement with experiment could be obtained. The calculation was further improved by including MEC~\cite{MaP08} 
when 
magnetic moments of $A\,$=$\,$2$\,$-$\,$7 ground states and M1 transitions in $A=6$,$ \,7$ nuclei were calculated.
The contributions from MEC were considerable and led to an overall good agreement with experiment.
Relevant MEC effects were also found in a HH hybrid calculation~\cite{VaB09} for the $^6$He $\beta$-decay using the JISP potential
and constructing a chiral weak current. The obtained decay rate was close to the experimental value (3\% difference).

Recently the radii of He isotopes were determined 
experimentally~\cite{MuS07,BrB12} and compared to results from ab initio methods. In a correlation plot of the $^6$He charge radius
versus the two-neutron separation energy only the GFMC calculation (the only one with an explicit 3NF) agreed 
well with experiment. In a further rather recent experiment excitation energies of two $2^+$ excited states of $^{10}$Be
as well as the corresponding $B(E2)$ values to the ground state were determined~\cite{McL09}.
These data were compared with the results of a GFMC calculation that employed the AV18+IL7 interaction model 
(the IL7-3NF has to be understood as an improved IL2-3NF).
A rather good agreement was found for three of the four observables, 
while the $B(E2)$ value from the higher of the two excited states turned out to be considerably smaller than experiment.

Now we turn our attention to reactions of nuclear systems with $A>4$. First we consider $n\,$-$\,\alpha$ scattering which
was calculated in Ref.~\cite{NoP07} in a GFMC calculation with the interaction models AV18, AV18+UIX, and AV18+IL2.
We depict those results in Fig.~\ref{figure13_R}, where the calculated phase shifts of the 
$^2s_{1/2}$ and $^2p_{3/2}$ $n\,$-$\,\alpha$ partial
waves together with a $R$-matrix fit are shown. For the $s$-wave phase shifts one already has a good description 
with AV18 alone, whereas this is not the case for the $p$-wave. For the latter a very interesting 3NF effect was found.
An inclusion of the UIX-3NF did not change the AV18 results much, whereas, at least for low energies, the phase shift moved 
very close to the $R$-matrix fit when the IL2-3NF was taken into account. The nice success of a GFMC calculation
for a low-energy scattering reaction in principle opened the door for further low-energy scattering calculations of other nuclear systems
and of corresponding calculations of electroweak cross sections.
\begin{figure}[ht]
\begin{minipage}[b]{0.51\linewidth}
\centering
\includegraphics[scale=0.3,angle=-90]{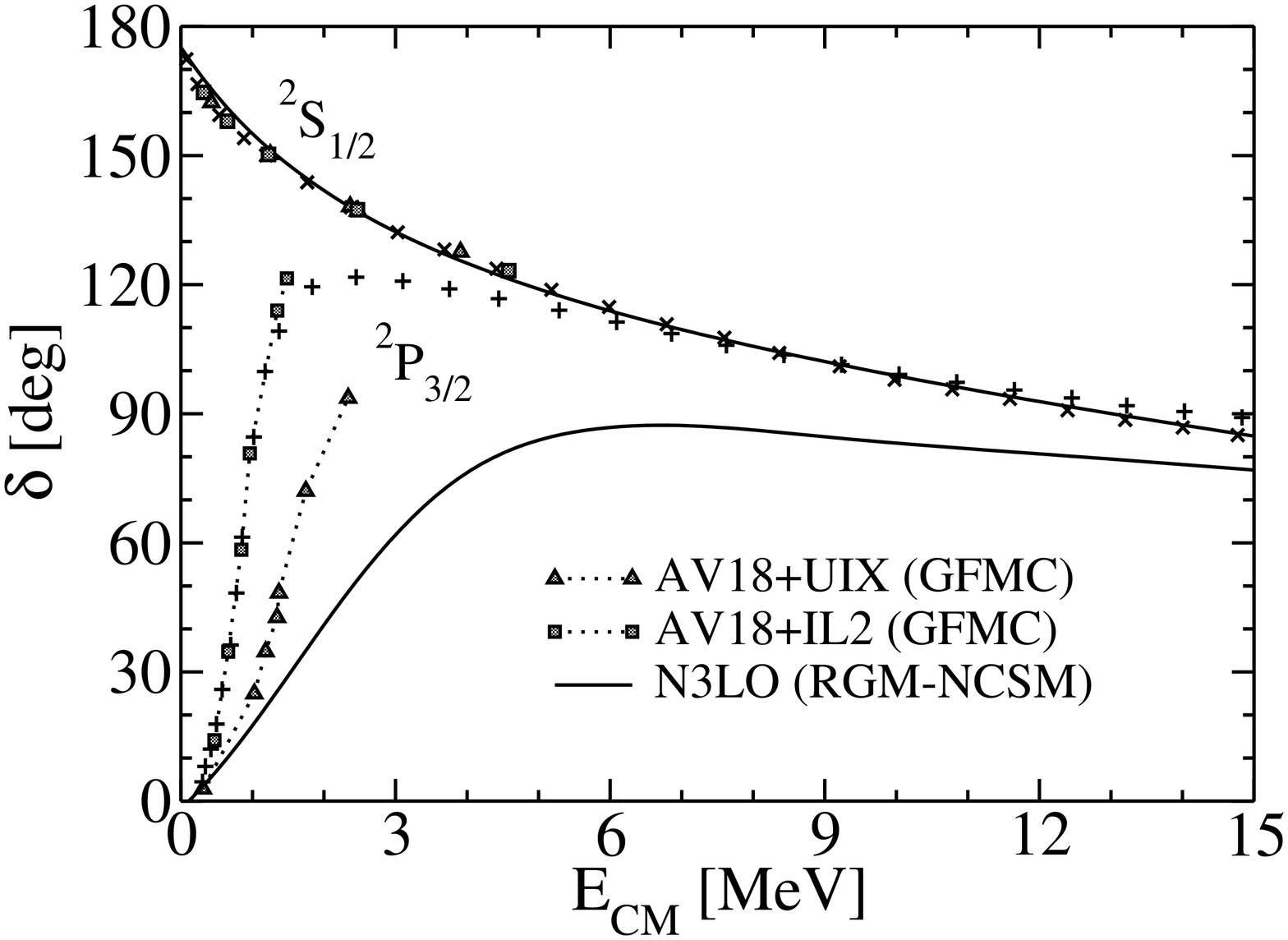}
\caption{$n\,$-$^4$He $^2s_{1/2}$ and $^2p_{3/2}$ phase shifts: GFMC results with AV18+UIX (triangles)
and AV18+IL2 (squares) from~\cite{NoP07}; RGM/NCSM results with I-N3LO (full curves) from~\cite{QuN09}.
Plus signs: R-matrix analysis results from~\cite{HaD-unp}. }
\label{figure13_R}
\end{minipage}
\hspace{0.5cm}
\begin{minipage}[b]{0.45\linewidth}
\centering
\includegraphics[scale=0.3,angle=-90]{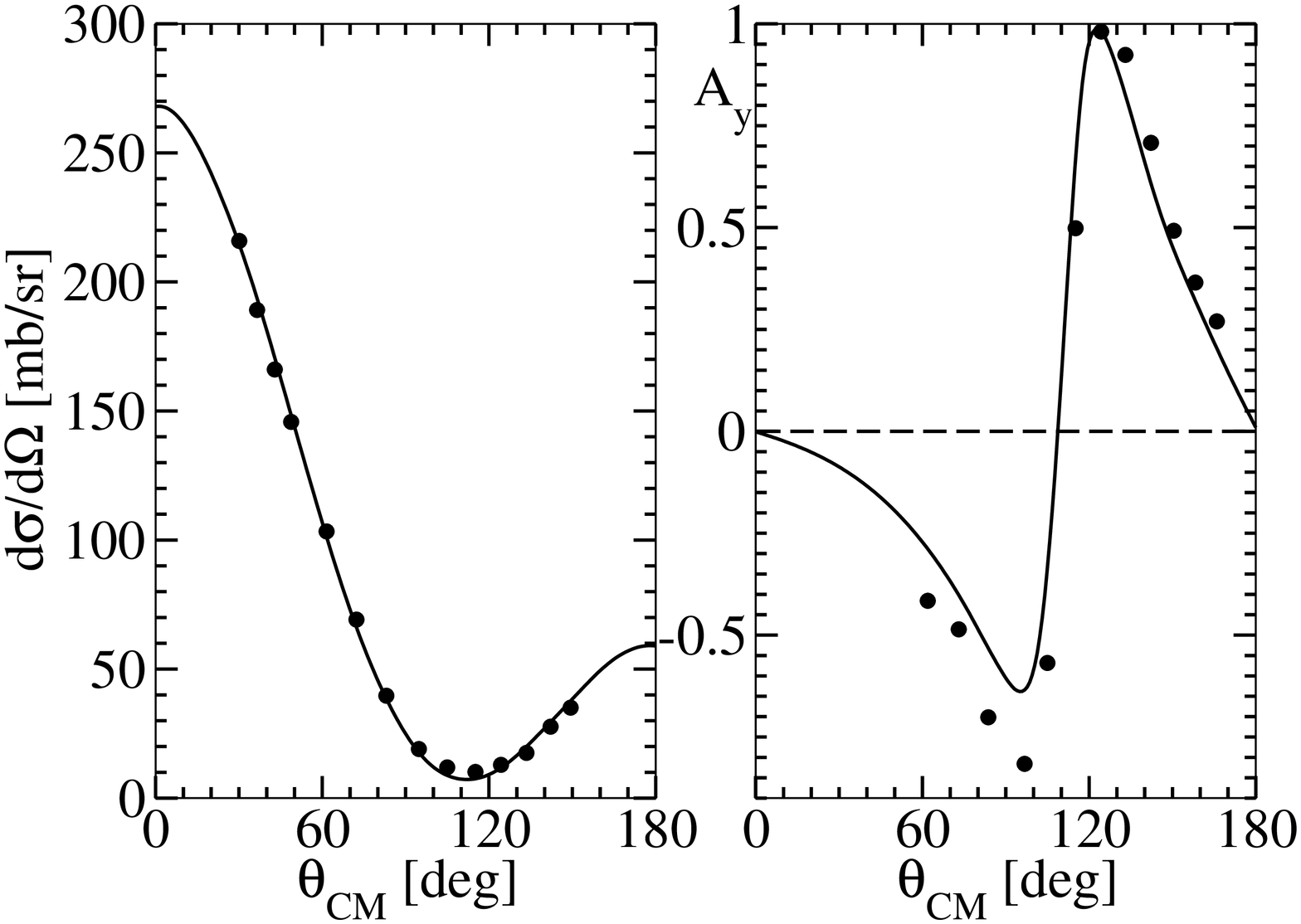}
\caption{$n\,$-$^4$He differential cross section (left panel) and analyzing power (right panel) 
at $E_n^{\rm lab}$=17 MeV. 
RGM/NCSM results (full curves) from~\cite{NaR10}; experimental data from~\cite{KrH84}. }
\label{figure14_R}
\end{minipage}
\end{figure}

In Fig.~\ref{figure13_R} we also show the I-N3LO results from a RGM calculation with the NCSM approach by 
Quaglioni and Navr\'atil~\cite{QuN09} who also used other NN potentials (CD-Bonn, $\vlowk$).
The potential model dependence was found to be rather mild. Only the $2p_{3/2}$ phase shift computed
with $\vlowk$ deviated a bit
from the results arising from the two other potentials. Fig.~\ref{figure13_R} shows good agreement with
the $R$-matrix fit for the $s$-wave,
whereas the $2p_{3/2}$ phase shifts are too small, which as a comparison with the GFMC results show, is because a 
proper 3NF is missing. The RGM calculations were also carried out for $p\,$-$\,\alpha$ scattering. Similar comparisons with the $R$-matrix fits
were obtained i.e. agreement for the $s$-wave and disagreement for the $p$-wave. We should mention that the
NCSM expansion included states up to $N_{\rm max}=17$, which provided nice convergence.
Besides the $^4$He ground-state, excitations of the $\alpha$
particle were also considered. No visible effect was found for the $s$-wave, while larger effects were present for the $p$-wave.
In Ref.~\cite{QuN09} further calculations were performed for the $s$-wave phase shift of $n\,$-$^{10}$B scattering. Due
to lack of data no comparison to experiment was possible but on the other hand the model space was
not sufficiently large ($N_{\rm max}=7$) to reach convergence.  

The convergence was improved in a following NCSM study by Navr\'atil, Roth, and Quaglioni~\cite{NaR10}. 
They used the importance truncated NCSM mentioned in Section~\ref{sec:AG4_SPECTRA} and
a SRG evolved (on the two-body level) Hamiltonian (I-N3LO). In this way the interaction was soft enough to reach convergence  
with $N_{\rm max}\,$=$\,14\,$-$\,$16 and thus
nucleon scattering on various nuclei ($^4$He, $^7$Li, $^7$Be, $^{12}$C, $^{16}$O) could be investigated. The $n\,$-$\,\alpha$ scattering 
was used as a test case.  In brief they found that {\bf (i)} very good convergence was obtained for the phase shifts (in the $N_{\rm max}$ range from 11 to 17
the results looked almost identical) and {\bf (ii)} the calculation was benchmarked with the full NCSM and a nice agreement was obtained.
The comparison to the $R$-matrix fit did not change much, with agreement obtained for the $^2s_{1/2}$, $^2p_{1/2}$, $^2d_{3/2}$
cases and, because of a missing 3NF, disagreement for the $^2p_{3/2}$ case. Comparison to experiment was made at 
$E_n=17\,$-$\,$19 MeV,
where 3NFs become less important. 
In Fig.~\ref{figure14_R}
we show the results at $E_n=17$ MeV where it is seen that the differential cross section 
agrees very well with data. An overall good agreement was obtained for $A_y$, but some differences were also present.
Comparisons to data with very similar outcomes could be made for $p\,$-$\,\alpha$ scattering as well. In the case of $n\,$-$^7$Li scattering
a successful benchmark with the full NCSM was made again. Besides the $^7$Li ($^7$Be) ground state two excited states (1/2$^-$, 7/2$^+$)
were also considered in the RGM calculation. Various phase shifts for the (7+1) scattering were calculated, and for the reaction
$^7$Li$(n,n')^7$Li$(1/2^-$) even a comparison with total cross section data could be made at very low energy and a fair agreement
was obtained (see Fig.~\ref{figure15_R}). With an additional 3NF the comparison to data could probably be improved further.  
In the case of N-$^{12}$C and N-$^{16}$O scattering, benchmarks with the full NCSM were no longer possible, since only the importance truncated 
NCSM could reach sufficiently high $N_{\rm max}$ values. The SRG parameter $\lambda \equiv \alpha^{-1/4}$ was increased
from 2.02 fm$^{-1}$ to 2.66 fm$^{-1}$, the higher value leading to a shorter evolution and less soft potentials. 
The use of a small $\lambda$ results in large overbinding of heavier nuclei
and a significant underestimation of their radii. We remind the reader that the transformed potential remains phase shift equivalent with
the bare interaction, but that $\alpha$ dependent three- and higher many-body forces are induced (see discussion of Ref.~\cite{RoL11} in
Section~\ref{sec:AG4_SPECTRA}). It is interesting to list the differences of the binding energies obtained with such a SRG evolved Hamiltonian
and the corresponding experimental values. 
Underbinding of 0.3, 1.04, and 8.7(8) MeV was found for $^3$H, $^4$He, and $^{12}$C respectively while
overbinding of 11.4(8) MeV  was found for $^{16}$O. 
Here we discuss neither further calculational details nor the results for the obtained phase shifts of 
$n\,$-$^{12}$C, $p\,$-$^{12}$C, and $p\,$-$^{16}$O, but refer the interested reader directly to Ref.~\cite{NaR10}.
We only mention that more difficulties
were encountered in the case of $^{16}$O and that the authors conclude that the inclusion of additional
excited states of the target nucleus would be beneficial in all studied systems particularly as A increases.
\begin{figure}[ht]
\begin{minipage}[b]{0.40\linewidth}
\centering
\includegraphics[scale=0.28,angle=-90]{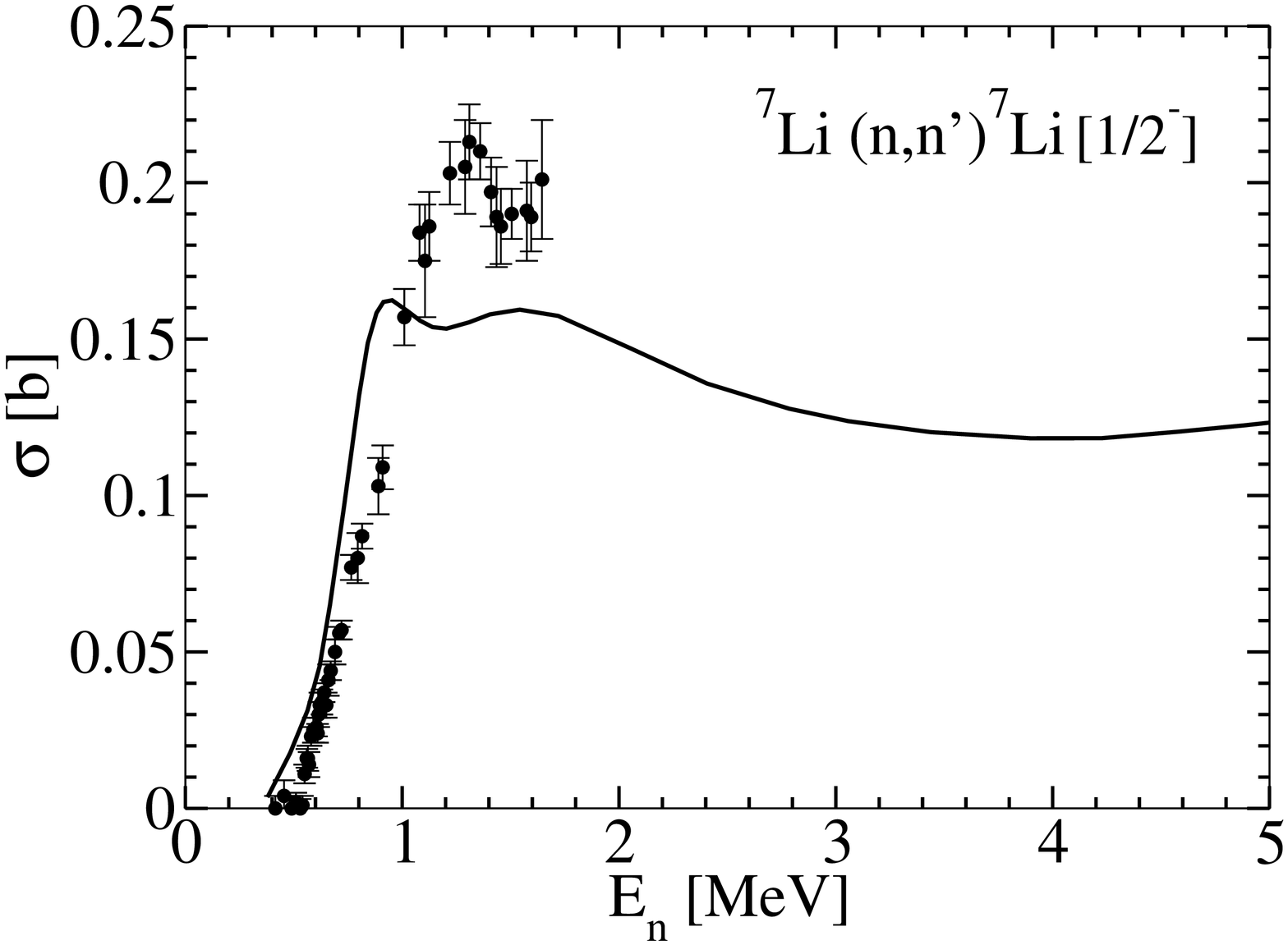}
\caption{Inelastic $^7$Li($n$,$n'$)$^7$Li(1/2$^-$) 
cross section from RGM/NCSM calculation of Ref.~\cite{NaR10} with SRG/I-N3LO NN potential; 
SRG parameter $\lambda$=2.02 fm$^{-1}$. Experimental data from Ref.~\cite{FlR55}.}
\label{figure15_R}
\end{minipage}
\hspace{0.5cm}
\begin{minipage}[b]{0.56\linewidth}
\centering
\includegraphics[scale=0.3,angle=-90]{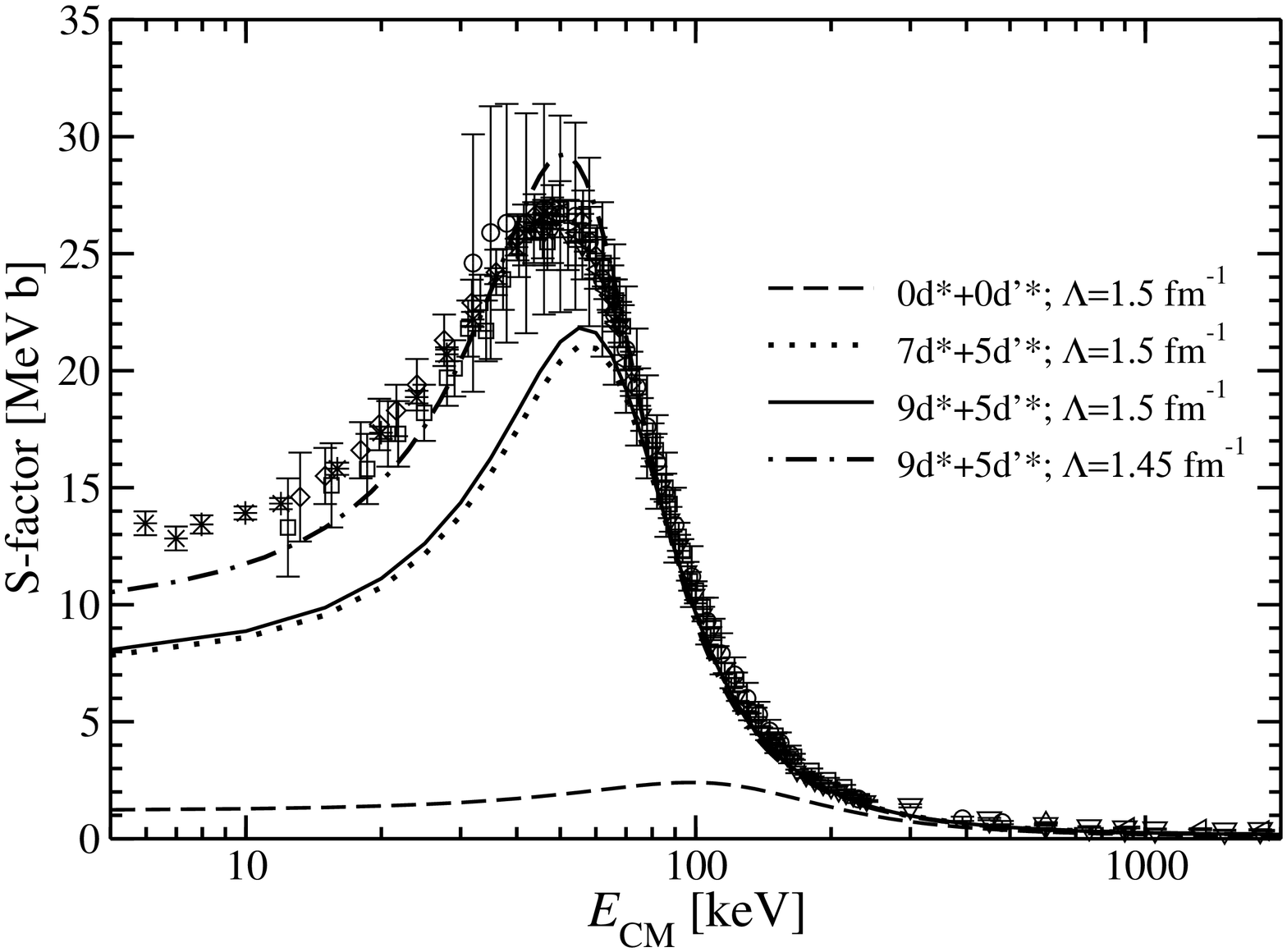}
\caption{$^3$H$(d,n)^4$He $S$-factor from NCSM/RGM calculation of Ref.~\cite{NaQ12} with SRG/I-N3LO NN potential 
including various numbers of deuteron virtual excitations in 
$^3$s$_1$-$^3$d$_1$ (d*) and $^3$d$_2$ (d$'$*) channels; SRG parameter $\lambda$=1.45 and 1.5 fm$^{-1}$.
Experimental data (see Ref.~\cite{DeA04}. }
\label{figure16_R}
\end{minipage}
\end{figure}

Another RGM calculation with the untruncated NCSM approach, but with a SRG transformed ($\lambda$=1.5 fm$^{-1}$) chiral 
Hamiltonian (I-N3LO) at the two-body level, 
was carried out for elastic $d\,$-$^4$He scattering~\cite{NaQ11}. It is interesting that also a $^6$Li ground-state calculation was performed 
with the RGM approach, where a fragmentation into $d\,$-$^4$He and $^4$He-($np$, $T$=0) clusters was considered. 
The virtual isoscalar ($np$)-states were 
taken into account up to an excitation energy of about 90 MeV, however, the RGM result obtained for the $^6$Li ground-state energy had not yet converged. 
In fact it lay 600 keV above the result of a six-body NCSM calculation for $^6$Li with the same Hamiltonian and where an equivalent HO model space was used. 
As discussed in
Ref.~\cite{NaQ11} the calculation should be further improved by taking into account excited states of $^4$He and other 
deuteron excitations not yet considered. Perhaps also a fragmentation into $^3$H-$^3$He could lead to non-negligible effects.
For the scattering calculation, i.e. for $s$- and $d$-wave phase shifts, it was checked that the HO model space ($N_{\rm max}=12$) and the number of virtual 
$np$ states  was sufficiently large. However, as already mentioned, unincluded $^4$He excitations and/or different fragmentation channels
could have non-negligible effects. The calculated phase shifts for $^3s_1$ and $^3d_3$ agree quite well with experimental data, whereas
differences were found for $^3d_1$ and $^3d_2$. In addition, differential cross sections were compared to data at various 
energies in a range of $E_d\,$=$\,3\,$-$\,$12 MeV. A fair agreement of theory and experiment was found. 
The results obtained are certainly very impressive, but
it should not be forgotten that in a rigorous ab initio
calculation the three-body break-up into $^4$He$\,$-$\,p\,$-$\,n\,$ should be included.

Recently the RGM/NCSM approach was applied to three reactions of astrophysical importance, namely to $^7$Be$(p,\gamma)^8$B~\cite{NaR11}
and to the fusion reactions $^3$H$(d,n)^4$He (see Fig.~\ref{figure16_R}) and $^3$He$(d,p)^4$He~\cite{NaQ12}. The latter two are also of interest for future energy 
generation on earth. In both cases a SRG evolved chiral Hamiltonian (I-N3LO), at the two-body level, was used, hence SRG evolved 
3NF as well as bare 3NF were not considered. In Ref.~\cite{NaR11} the SRG parameter $\lambda$ was chosen within what the authors
called a natural range with values from 1.73 to 1.95 fm$^{-1}$. With this choice the $^8$B separation energies could be well reproduced. On the contrary
for the fusion reactions $\lambda=1.5$ fm$^{-1}$ was taken in order to better describe the Q values of both reactions. It is evident that such reaction 
dependent choices of $\lambda$ are against the spirit of a true ab initio calculation. This is also admitted by the authors, who state that
it would be desirable to include a SRG transformation at the three-body level and a consideration of a bare 3NF. Then, in addition,
it should be either shown that the results do not depend on $\lambda$ or alternatively one may define a new potential model
choosing a fixed value for $\lambda$. In spite of all that
we consider these calculations to represent a considerable progress.

For the $^7$Be$(p,\gamma)^8B$ reaction it was necessary to use the NCSM truncation scheme in order to have sufficiently convergent results.
The $^8$B ground and continuum states were described as (1+7) fragmentations, where, obviously, the proton is the single particle, while the $^7$Be compound system
was described by the ground state and four additional excited states. 
For the radiative capture cross section only the dominant E1 transitions were
considered and were calculated by using the unretarded dipole approximation (see Section~\ref{sec:EXT_PR}). The resulting $S$-factor, 
$S_{17}(0)$, was found to be approximately 19.4 eV b, which was consistent with the latest evaluation of
20.8$\,\pm\,0.7$(expt)$\,\pm\,1.4$(theory) eV b~\cite{AdG11}. 

For the calculation of the fusion reactions, first the phase shifts for elastic scattering of $n\,$-$^4$He, $d\,$-$^3$H, 
and $d\,$-$^3$He were determined by including channel coupling, something which was not done in the previous NCSM studies 
of $n\,$-$^4$He scattering. The effect
of the channel coupling on the phase shifts remained quite small, but in some cases was not negligible. Also virtual $(np,\,T=0)$ states were
taken into account, which led to sizable effects in the $^4s_{3/2}$ phase shifts of the $d\,$-$^3$H and $d\,$-$^3$He channels. 
The results for the $S$-factors of the fusion reactions showed that the virtual $np$ states led to important  effects.
A rather good agreement with experimental data was obtained in the case of $^3$He$(d,p)^4$H (approximately correct peak position, 
slight overestimation of peak height), whereas a slightly wrong peak position was found for $^3$H$(d,n)^4$H. 
Interesting is the fact, that only a slight change of the 
SRG parameter $\lambda$ from 1.5 to 1.45 fm$^{-1}$ leads to quite a good description of the  $S$-factor for the 
$^3$H$(d,n)^4$H reaction, but presumably would lead to a stronger disagreement in the  $^3$He$(d,p)^4$H channel.
This demonstrates again that it would be desirable to perform a more complete SRG transformation and to include 
a bare 3NF, although one has to consider that such calculations are quite demanding.  

Another reaction of astrophysical relevance, $^3$He$(\alpha,\gamma)^7$Be, was calculated by Neff in unretarded dipole
approximation using the FMD approach~\cite{Ne11}. As a NN interaction a UCOM version of the AV18 was taken. 
Quite a good agreement with experimental data was obtained, both for $^3$He$\,$-$^4$He scattering phase shifts and  the
$^3$He$(\alpha,\gamma)^7$Be cross section. The reaction with the mirror nuclei, $^3$H$(\alpha,\gamma)^7$Li, was calculated as well.
The shape of the experimental cross section was reproduced, but the absolute normalization was about 15\% above the most 
recent experimental data. However, as already pointed out in Section~\ref{sec:FMD} the FMD technique still needs to be put on safer grounds
since benchmark tests with other ab initio methods are missing. It would be desirable to have such tests.  
\begin{figure}[tb]
\begin{center}
\epsfig{file=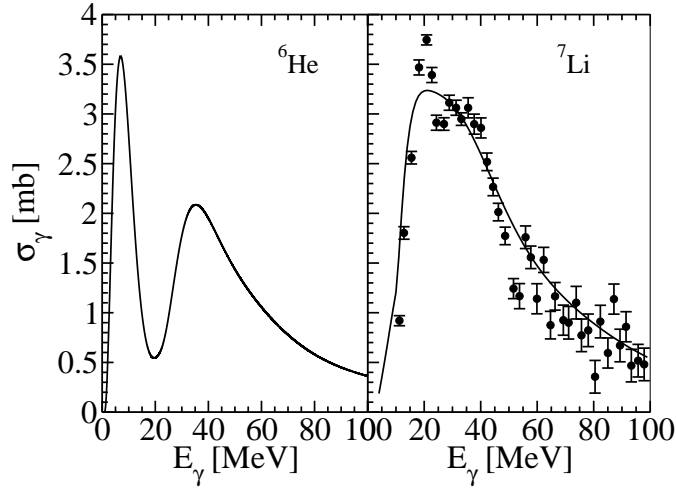,scale=0.35,angle=-90}
\caption{Total photoabsorption cross sections of $^6$He and $^7$Li calculated in Refs.~\cite{BaB04} and ~\cite{BaA04},
respectively, with the LIT/EIHH method and the AV4$'$~\cite{WiP02} potential. 
Experimental data from Ref.~\cite{Ahetal75}.}
\label{figure17_R}
\end{center}
\end{figure}

Coming to the end of this section we want to devote some space to ab initio calculations which reach further into the 
many-body continuum.
As already pointed out, such calculations can be carried out with the LIT method. Unfortunately,
LIT calculations have not yet been performed for $A>4$ nuclei with realistic nuclear Hamiltonians although
we hope that this will soon be possible at least for $A=6$,$\,7$.
Therefore we want to discuss some examples with simpler forces. In Fig.~\ref{figure17_R} we show the total photoabsorption
cross sections of $^6$He~\cite{BaM02,BaB04} and $^7$Li~\cite{BaA04} calculated by Bacca~et~al. with the central NN potential AV4$'$
and using the EIHH method. 
One notes that 
the $^6$He cross section exhibits a very interesting structure. In fact two separate peaks were found. One may interpret the low-energy peak
as a response due to the relative motion of the two halo neutrons with respect to the $\alpha$ core. Similarly one may interpret
the peak at higher energies in terms of the classical giant resonance picture, i.e. a collective response of protons with respect to the neutrons. 
For the latter case a break-up of the $\alpha$ core is necessary and thus the peak is located at higher energies. A similar structure of the 
cross section was not found for the $^7$Li total photoabsorption cross section where only a giant resonance peak is present. However,
it should be noted that whereas good quality experimental data exist for the $^7$Li case the same is not
true for the $^6$He case.
The theoretical LIT cross section describes the $^7$Li data fairly well far up into the many-body continuum, where many different break-up channels should
contribute to the total cross section. In fact with the LIT method the effect of all different channels is rigorously included 
within an ab initio
approach. 

\section{Summary and Outlook}\label{sec:SUM}

We have presented an overview of modern ab initio methods in few-nucleon systems with $A \ge 4$.
Our review shows that great progress has been made in the past, and in particular 
in the last two decades. As discussed in Section~\ref{sec:FB},  a number
of strategies have been followed for solving  the nuclear many-body problem.
Some of the methods have already been introduced
rather long ago (see Section~\ref{sec:4_hist}), while other approaches have appeared
on the scene only rather recently (see Section~\ref{sec:4_mod}).

As pointed out in the introduction we have qualified a method to be an ab initio method if it allows
a solution of the nuclear many-body problem without any uncontrolled approximation. Controlled expansions,
however, can be increasingly improved such that eventually a converged result is obtained.  Such results
are especially relevant for extracting reliable information about one of the fundamentals of nuclear
physics, i.e. the nuclear force. 

Precise ab initio results from different ab initio methods have to agree within a reasonable error range 
and hence they become benchmark results. In our report we have endeavored to show them whenever possible.
Benchmark results are important and often 
represent milestones in the development of a field. 

We have divided the discussion of results in four parts. In the first part (Section~\ref{sec:4_hist}) 
we have described calculations for the $A\,$=$\,4$ system from an earlier period (1970$\,$-$\,$1990). It is 
remarkable that already in 1982 a first milestone was set with benchmark results for the 
$^4$He binding energy with a semi-realistic NN potential (see Table~\ref{MTV}). The second part of our 
discussion (Section~\ref{sec:4_mod}) illustrates the more modern period of $A\,$=$\,4$ calculations. We place
their initiation at the end of the 1980s when a GFMC calculation with precise $^4$He ab initio results for a 
realistic NN potential has been  published. 
Somewhat later a similar success was achieved by the FY approach. In the following years a great innovation phase
had its beginnings. Various groups had begun to tackle the $^4$He ground-state problem with different methods. This
culminated in another milestone ten years later - a benchmark calculation for the $\alpha$ particle with a
realistic NN potential, where seven different groups using seven different methods participated
(see Tables~\ref{Bench_1} and \ref{Bench_2}). 

In Section~\ref{sec:4_mod} we reported on calculations for 
the 4N continuum, discussing results for hadronic and perturbation induced reactions. These calculations
are more complicated than for the 4N ground state, thus  accounting for the much later
appearance of the benchmarks.
In fact, only very recently have three groups  succeeded in doing so,
by considering  elastic (3+1) scattering below the break-up threshold (see Section~\ref{sec:4_had}).
The comparison to experiment, however, is still problematic in some cases, even when modern realistic
nuclear forces are used (see for example the total $n\,$-$^3$H elastic scattering cross section in
Fig.~\ref{figure3_R} or the proton $A_y$ in elastic $\vec p\,$-$^3$He scattering in
Fig.~\ref{figure5_R}). These discrepancies call for an improvement of the Hamiltonian and need further investigation.

For higher  energies, where not only two- but also more-body channels
are open, calculations are more complicated and exact computations of such continuum
states do not yet exist (with exception of the very recent Ref.~\cite{DeF12}). However, ab initio calculations of 
various perturbation-induced reaction cross sections,
among them the very much discussed $^4$He photoabsorption (see Fig.~\ref{figure8_R} and corresponding
discussion), have been carried out in this energy range with the LIT method.
We think that the power of this method, which bypasses the problem of calculating continuum states and
aims at computing observables directly using bound-state techniques, has not yet been fully explored. In particular
the application to the LIT of bound-state methods other than the HH or NCSM could be investigated. Also the complex
scaling approach, which has some similarities to the LIT method, should be further examined.

The energy spectra and other observables of $A>4$ systems have been discussed in sections~\ref{sec:AG4_SPECTRA} and 
\ref{sec:AG4_FURTHER}, respectively. The calculation of the spectra, pioneered by GFMC  has led
to a more detailed and better description of the 3NF. Improved phenomenological 3NF have allowed  a rather high precision description 
of the spectra of  $6 \le A \le 10$ nuclei (see also Table~\ref{GFMC_A910}). 
We have  also discussed quite a number of  NCSM studies. 
Precise ab initio results have been mainly obtained for
$A \le 6$ nuclei, while for $A>6$ nuclei,  particularly in the presence of a 3NF, the model space has 
often not been sufficiently large to reach converged results. However, in recent years the NCSM convergence has been
significantly improved by a controlled truncation of basis states. Also the introduction of SRG evolved Hamiltonians 
has allowed considerable acceleration towards convergence, although still not enough to reach full convergence for the ground states
of $^{12}$C and $^{16}$O in presence of a 3NF. Without a 3NF the NCSM ground-state energy of $^{16}$O agrees 
fairly well with CC and UMOA results. Evidently, the CC and UMOA techniques are very
powerful as well, even if up to now they have been restricted mainly to closed shell nuclei. On the other hand, very 
recently successful CC attempts
have been made to work in open-shell nuclei.  For further observables of $A>4$ nuclei (radii, magnetic moments, 
electromagnetic transition strengths, scattering observables, $S$-factors, total photoabsorption 
cross sections) there are a wealth of calculations. In~\ref{sec:AG4_FURTHER} we have limited ourself to a few selected examples 
of which we want to mention here only the great progress of scattering calculations with the RGM/NCSM approach.

Considering the progress that has been made in
few-nucleon physics over the years, and in particular in the last two decades, we think that
the future prospects are very promising and that further important progress will be made in 
the years to come. In fact, there still remains a great deal of work to be done. For example, it would be desirable
to have benchmark tests with realistic Hamiltonians for $A\,$=$\,4$ continuum calculations beyond the 
three-body break-up threshold, as well as for ground- and continuum-state quantities of $A>4$ 
nuclear systems. Another issue to be addressed concerns SRG 
transformations. In the last decade it has become evident that they play a relevant role
in present-day few-nucleon physics. One important aspect in such transformations is that they are 
applied only at the two- or three-body level, but the resulting Hamiltonian is used in the $A$-body
calculation. Higher order terms may be negligible and then the transformation is unitary on
the $A$-body level (independence of results on flow parameter), otherwise one has induced many-body forces which do not correspond to those 
of the unitary transformed $A$-body Hamiltonian. However, a three-body level SRG Hamiltonian of a bare two- plus 
three-nucleon force can be as legitimate as the original one, provided 
that the  flow parameter is adjusted once for all and then kept fixed. Otherwise
the predictive power of the theory would be lost.

For the  advancement of the field important contributions
must also come from experiment.
While there already exists a wealth of  data
on the bound-state properties of stable nuclei what is needed
are precise measurements of observables in low-energy reactions.
Particularly needed are those observables arising from electromagnetic probes (e.g. electron scattering
observables at low energy and low momentum transfer) since, as pointed
out in Section~\ref{sec:4_scatter}, they offer alternative insights into the nuclear
dynamics.
In addition, experiments with beams of
unstable nuclei such as $^6$He or other light neutron rich
nuclei,  will
be an important source of complementary information.
Present and future facilities with radioactive ion beams like
GANIL/SPIRAL2, GSI/FAIR and  RIKEN/RIBF are expected to play a key role in the determination
of such observables.

Concluding this review we express our confidence that the field of ab initio calculations will
continue to grow in importance in future nuclear physics.

\vspace{0.11cm}
{\bf Acknowledgements:}
Here we would like to thank in particular  Ed Tomusiak  who took the time to carefully read the entire manuscript, 
as well Sonia Bacca, Victor Efros and Sofia Quaglioni who read large parts of it. We thank all of them for giving 
helpful advices and criticisms, and for suggesting corrections. A special thank goes to Francesco Pederiva for 
advices and explanations concerning the Monte Carlo techniques.
Moreover, in alphabetic order, we  mention all the colleagues who have
given valuable comments on various parts of the paper
expressing  our gratitude towards them. They are:
N. Barnea, A. Deltuva, M. Gattobigio, A. Kievsky, L. Marcucci, T. Papenbrock, and N. Timofeyuk.


\begin{thebibliography}{00}

\itemsep -2pt 

\bibitem{GlK93a} W. Gl\"ockle and H. Kamada, \Journal{\PRL} {71} {971} {1993}
\bibitem{CiC99} F. Ciesielski, J. Carbonell, and C. Gignoux, \Journal{\PLB} {447} {199} {1999}
\bibitem{DeF05} A. Deltuva, A.C. Fonseca, and P.U. Sauer, \Journal{\PRC} {71} {054005} {2005}, 
                \Journal{\em ibid} {72} {054004} {2005}
\bibitem{Pi08} S.C. Pieper, \Journal{\em Rivista Nuovo Cim.}{31}{709}{2008}, and references therein
\bibitem{NaV00} P. Navr\'atil, J. P. Vary, and B. R. Barrett, \Journal{\PRL} {84} {5728} {2000};
                    \Journal{\PRC} {62}{054311}{2000}
\bibitem{EfL94} V.D. Efros, W. Leidemann, and G. Orlandini, \Journal{\PLB} {338} {130} {1994}
\bibitem{EpM11} E. Epelbaum and U.-G. Mei{\ss}ner, arXiv:1201.2136 
\bibitem{EpH09} E. Epelbaum, H.-W. Hammer, and U.-G. Mei{\ss}ner, \Journal{\RMP} {81}{1773}{2009}
\bibitem{MaE11} R. Machleidt and D.R. Entem, \Journal{\PREP} {503} {1} {2011} 
\bibitem{BoK03} S.K. Bogner, T.T.S Kuo, and A. Schwenk, \Journal{\PREP} {386} {1} {2003}
\bibitem{BoF10} S.K. Bogner, R.J. Furnstahl, and A. Schwenk, \Journal{\em Prog. Part. Nucl. Phys.}{65}{94}{2010}
\bibitem{RoN10} R. Roth, T. Neff, and H. Feldmeier, \Journal{\em Prog. Part. Nucl. Phys.}{65}{50}{2010}
\bibitem{ShW07} A.M. Shirokov, J.P. Vary, A.I. Mazur, and T.A. Weber, \Journal{\PLB}{644}{33}{2007}
\bibitem{KaE12} N. Kalantar-Nayestanaki, E. Epelbaum, J.G. Messchendorp, and A. Nogga, 
                \Journal{\em Rept. Prog. Phys.} {75}{016301} {2012}
\bibitem{BrH06} E. Braaten and H.-W. Hammer, \Journal{\PREP}{428}{259}{2006};
                E. Braaten and H.-W. Hammer, \Journal{\em Annals of Phys.}{322}{120}{2007};
                H.-W. Hammer and L. Platter, \Journal{\em Ann. Rev. Nucl. Part.Sci.}{60}{207}{2010}
\bibitem{Bl12}  D. Blume, \Journal{\em Rep. Prog. Phys.}{75}{046401}{2012}
\bibitem{GiP08} S. Giorgini, L.P. Pitaevskii, and S. Stringari, \Journal{\RMP}{80}{1215}{2008}
\bibitem{ReM02} S.M. Reimann and M. Manninen, \Journal{\RMP}{74}{1283}{2002}
\bibitem{PeH11} M. Pedersen Lohne et al., \Journal{\PRB}{84}{11530}{2011}
\bibitem{GoW04} M.L. Goldberger and K.W. Watson, {\it Collision Theory} (Dover, Mineola NY, 2004);
                R.G. Newton, {\it Scattering Theory Of Waves and Particles} (Dover, Mineola NY, 2003);
                J.R. Taylor, {\it Scattering Theory: The Quantum Theory of Nonrelativistic Collisions} (Dover, Mineola NY, 2006)
\bibitem{Fa61} L.D. Faddeev, \Journal{\em Sov. Phys. JETP} {12} {101} {1961}
\bibitem{Gl83} W. Gl\"ockle, {\it The Quantum Mechanical Few-Body Problem} (Springer Verlag Berlin 1983)
\bibitem{GlW96} W. Gl\"ockle et al., \Journal{\PREP} {274} {107} {1996)}
\bibitem{MeG76} S.P. Merkuriev, C. Gignoux, and A. Laverne, \Journal{\ANNP} {99} {30} {1976}
\bibitem{Me80} S.P. Merkuriev, \Journal{\ANNP} {130} {395} {1980}
\bibitem{Ya67} O.A. Yakubovsky, \Journal{\em Sov. J. Nucl. Phys.} {5} {93} {1967}
\bibitem{MeY83} S.P. Merkuriev and S.L. Yakovlev, \Journal{\it Theor. Math. Phys.} {56} {573} {1983}
\bibitem{MeY84} S.P. Merkuriev, S.L. Yakovlev, and C. Gignoux, \Journal{\NPA} {431} {125} {1984}
\bibitem{CiC98} F. Ciesielski and J. Carbonell, \Journal{\PRC} {58} {58} {1998}
\bibitem{La03} R. Lazauskas, Ph.D. thesis, Universit\'e Joseph Fourier, Grenoble, 1997,\\
               (http:/tel/ccsd.cnrs.fr/documents/archives0/0000/41/78/)
\bibitem{LaC04} R. Lazauskas and J. Carbonell, \Journal{\PRC} {70} {044002} {2004}
\bibitem{La09} R. Lazauskas, \Journal{\PRC} {79} {054007} {2009} 
\bibitem{KaG92a} H. Kamada and W. Gl\"ockle, \Journal{\NPA} {548} {205} {1992}
\bibitem{KaG92b} H. Kamada and W. Gl\"ockle, \Journal{\PLB} {292} {1} {1992}
\bibitem{GlK93b} W. Gl\"ockle and H. Kamada, \Journal{\NPA} {560} {541} {1993}
\bibitem{NaR78} M.M. Nagels, T.A. Rijken, and J.J. de Swart, \Journal{\PRD} {17} {768} {1978}
\bibitem{LaL80} M. Lacombe et al., \Journal{\PRC} {21} {861} {1980}
\bibitem{WiS84} R.B. Wiringa, R.A. Smith, and T.L. Ainsworth, \Journal{\PRC} {29} {1207} {1984}
\bibitem{NoK02} A. Nogga, H. Kamada, W. Gl\"ockle, and B.R. Barrett, \Journal{\PRC} {65} {054003} {2002}
\bibitem{ViD11} M. Viviani et al., \Journal{\PRC} {84} {054010} {2011}
\bibitem{GlW11} W. Gl\"ockle and H. Wita{\l}a, \Journal{\FBS}{51}{27}{2011} 
\bibitem{AlG67} E.O. Alt, P. Grassberger, and W. Sandhas, \Journal{\NPB} {2} {167} {1967}, 
                {\em JINR report No. E4-6688} (1972);
                P. Grassberger and W. Sandhas, \Journal{\NPB} {2} {181} {1967}
\bibitem{Fo87} A.C. Fonseca, \Journal{\em Lecture Notes in Physics} {273}{161}{1987}
\bibitem{Fo89} A.C. Fonseca, \Journal{\PRC} {40} {1390} {1989}
\bibitem{GiL76} B.F. Gibson and D.R. Lehman, \Journal{\PRC} {14} {685} {1976}; \Journal{\em ibid} {15} {2257} {1977}
\bibitem{GiL78} B.F. Gibson and D.R. Lehman, \Journal{\PRC} {18} {1042} {1978}
\bibitem{Ra870} J.M. Rayleigh, \Journal{\em Phil. Trans.} {161} {77} {1870}
\bibitem{Ri909} W. Ritz, 
   \Journal{\em Journal f\"ur die Reine und Angewandte Mathematik} {135} {1} {1909}
\bibitem{Ko48} W. Kohn, \Journal{\PREV} {74} {1763} {1948}
\bibitem{De63} Yu.N. Demkov, {\it Variational Principles in the Theory of Collisions} (Pergamon, New York, 1963)
\bibitem{Bo14} E. Borel, {\it Introduction g\'eometrique \`a quelques th\^eories physique}
              (Gauthier-Villars, Paris, 1914)
\bibitem{Gr37}  T.H. Gronwall, \Journal{\PREV} {51} {655} {1937}
\bibitem{Cl49}  R.E. Clapp, \Journal{\PREV} {76} {873} {1949}
\bibitem{HiD56} J.O. Hirschfelder and J. Dahler, \Journal{\em Proc. Natl. Acad. Sci. USA} {42} {363} {1956}
\bibitem{ZeB35} F. Zernike and H.C. Brinkman, \Journal{\em Proc. Kon. Ned. Acad. Wettensch.} {33} {3} {1935}
\bibitem{ViK65} N.Ya. Vilenkin, G.I. Kutznetsov, and Ya. A. Smorodinskii, 
              \Journal{\em Yad. Fiz.} {2} {906} {1965} [\Journal{\em Sov. J. Nucl. Phys.} {2} {645} {1966}]
\bibitem{Ba97} N. Barnea {\it Ph.D. Thesis, Hebrew University of Jerusalem}, 1997
\bibitem{Ef79} V.D. Efros, \Journal{\em Yad. Fiz.}{30}{84}{1979} [\Journal{\em Sov. J. Nucl. Phys.}{30}{43}{1979}]
\bibitem{GeZ50} I.M. Gel'fand and M.L. Zetlin, \Journal{\em Dokl. Akad. Nauk. USSR} {71}{825}{1950};
      I.M. Gel'fand, R.A. Minols, and Z. Ya. Shapiro, {\it Representations of the rotations and
      Lorentz groups and their applications}  (Pergamon, New York, 1963)  p.353
\bibitem{BaN97} N. Barnea and A. Novoselsky, \Journal{\ANNP} {256} {192} {1997}
\bibitem{BaL99} N. Barnea, W. Leidemann, and G. Orlandini, \Journal{\NPA}{650}{427}{1999}
\bibitem{Ti02} N.K. Timofeyuk, \Journal{\PRC} {65} {06430} {2002}, \Journal{\em ibid} {69} {034336} {2004}
\bibitem{Ti08} N.K. Timofeyuk, \Journal{\PRC} {78} {054314} {2008}
\bibitem{GaK11} M. Gattobigio, A. Kievsky, and  M. Viviani, \Journal{\PRC} {83}{024001}{2011}
\bibitem{FeE72} Y.I. Fenin  and V.D. Efros, \Journal{\em Yad. Fiz.} {15} {887} {1972} 
               [\Journal{\em Sov. J. Nucl. Phys.} {15} {497} {1972}]
\bibitem{EfL07} V.D. Efros, W. Leidemann, G. Orlandini, and N. Barnea, \Journal{\em J. Phys. G}
               {34} {R459} {2007}
\bibitem{RoK92} S. Rosati, A. Kievsky, and M. Viviani, \Journal{\em Few-Body Syst. Suppl.} {6} {563} {1992}
\bibitem{Ti07} N.K. Timofeyuk, \Journal{\PRC}{76}{044309}{2007}
\bibitem{Ef02} V.D. Efros, \Journal{\FBS} {32}{169}{2002} 
\bibitem{Fa84} M. Fabre de la Ripelle, \Journal{\em C.R. Acad. Sci Paris S\'erie II}{299}{839}{1984}; \Journal {\FBS}{1}{181}{1986}
\bibitem{ViK95} M. Viviani, A. Kievsky, and S. Rosati, \Journal{\FBS} {18} {25} {1995}
\bibitem{Vi98}  M. Viviani, \Journal{\NPA}{631}{111c}{1998}
\bibitem{KiR08} A. Kievsky et al., \Journal{\em J. Phys. G} {35}{063101}{2008}
\bibitem{ViM06} M. Viviani et al., \Journal{\FBS} {39}{159}{2006}
\bibitem{Ka88} M. Kamimura, \Journal{\PRA} {38} {621} {1988}
\bibitem{KaK90} H. Kameyama and M. Kamimura, \Journal{\NPA} {508} {17} {1990}
\bibitem{KuK77} V. I. Kukulin and V. M. Krasnopol'sk, \Journal{J. Phys.}{3} {795}{1977}
\bibitem{VaS95} K. Varga and Y. Suzuki, \Journal{\PRC} {52} {2885} {1995}
\bibitem{SuV98} Y. Suzuki and K. Varga, {\it Stochastic Variational Approach to Quantum Mechnical Few-Body Problems}
               (Springer-Verlag, Berlin, 1998)
\bibitem{Vo65} A. B. Volkov, \Journal{\NP} {74} {33} {1965}
\bibitem{BuA06} S. Bubin and L. Adamowicz, \Journal{\CHEM} {124}{224317}{2006}
\bibitem{BuA08} S. Bubin and L. Adamowicz, \Journal{\CHEM} {128}{114107}{2008}
\bibitem{VaU98} K. Varga, J. Usukura, and Y. Suzuki, \Journal{\PRL} {80} {1876} {1998};
                J. Usukura, K. Varga, and Y. Suzuki, \Journal{\PRA} {58} {1918} {1998}
\bibitem{WiT77} K. Wildermuth and Y.C. Tang, {\it A Unified Theory of the Nucleus} (Vieweg, Braunschweig, 1977)
\bibitem{TaL78} Y.C. Tang, M. LeMere, and D.R. Thompson, \Journal{\PREP} {47}{169}{1978}
\bibitem{Ho87} H. Hofmann, \Journal{\em Lecture Notes in Physics} {273}{243}{1987} 
\bibitem{HoH97} H.M. Hofmann and G. M. Hale, \Journal{\NPA} {613} {69} {1997}
\bibitem{HoH08} H.M. Hofmann and G. M. Hale, \Journal{\PRC} {77} {044002} {2008}
\bibitem{MaH87} R. Machleidt, K. Holinde, and Ch. Elster, \Journal{\PREP \,\,{\em C}} {149}{1}{1987}
\bibitem{Ca88}  J. Carlson, \Journal{\PRC} {38} {1879} {1988}
\bibitem{Fe90} H. Feldmeier, \Journal{\NPA} {515} {147} {1990}
\bibitem{FeS00} H. Feldmeier and J. Schnack, \Journal{\RMP} {72} {655} {2000}
\bibitem{BaF08} S. Bacca, H. Feldmeier, and T. Neff, \Journal{\PRC}{78}{044306}{2008}
\bibitem{NeF08} T. Neff and H. Feldmeier, \Journal{\em Eur. Phys. J. Special Topics}{156}{69}{2008}
\bibitem{HaP10} G. Hagen, T. Papenbrock, D.J. Dean, and M. Hjorth-Jensen, \Journal{\PRC}{82}{034330}{2010}
\bibitem{WiS95} R.B. Wiringa, V.G.J. Stoks, and R. Schiavilla, \Journal{\PRC} {51} {38} {1995}
\bibitem{Ok54} S. Okubo, \Journal{\PRO} {12} {603} {1954}
\bibitem{CoK60} F. Coester and H. K{\" u}mmel, \Journal{\em Nucl. Phys.}  {17} {477} {1960}
\bibitem{DaS64} J. da Providencia and C.M. Shakin, \Journal{\ANNP} {30} {95} {1964}
\bibitem{SuL80} K. Suzuki and S.Y. Lee, \Journal{\PRO} {64} {2091} {1980}
\bibitem{Su82} K. Suzuki, \Journal{\PRO} {68} {246} {1982}
\bibitem{SuO83} K. Suzuki and R. Okamoto, \Journal{\PRO} {70} {439} {1983}
\bibitem{We01} F.J. Wegner, \Journal{\PREP } {348} {77} {2001}
\bibitem{ZhB95} D.C. Zheng et al., \Journal{\PRC} {52} {2488} {1995} 
\bibitem{NaB96} P. Navr\'atil and B. R. Barrett, \Journal{\PRC} {54} {2986} {1996}
\bibitem{ShF74} A. de Shalit and H. Feshbach, {\em Theoretical Nuclear Physics: Nuclear Structure} 
               (John Wiley \& Sons Inc, 1974)
\bibitem{NaO02} P. Navr\'atil and V.E. Ormand, \Journal{\PRL} {88} {152502} {2002}
\bibitem{FuO09} S. Fujii, R. Okamoto, and K. Suzuki, \Journal{\PRL} {103} {182501} {2009}
\bibitem{Ma01} R. Machleidt, \Journal{\PRC} {63} {024001} {2001}
\bibitem{NaK00} P. Navr\'atil, G.P. Kamuntavi\^cius, and B.R. Barrett, \Journal{\PRC} {61} {044001} {2000}
\bibitem{NoN06} A. Nogga, P. Navr\'atil, B. R. Barrett, and J. P. Vary, \Journal{\PRC} {73} {064002} {2006}
\bibitem{NaQ09} P. Navr\'atil, S.Quaglioni, I.Stetcu and B. R. Barrett, \Journal{\em J.Phys. G} {36} {083101} {2009}
\bibitem{BaL00} N. Barnea, W. Leidemann, and G. Orlandini, \Journal{\PRC} {61} {054001} {2000}
\bibitem{BaL01} N. Barnea, W. Leidemann, and G. Orlandini, \Journal{\NPA} {693} {565} {2001}
\bibitem{BaL03} N. Barnea, W. Leidemann, and G. Orlandini, \Journal{\PRC} {67} {054003} {2003}
\bibitem{BaL10} N. Barnea, W. Leidemann, and G. Orlandini, \Journal{\PRC} {81} {064001} {2010}
\bibitem{BaL04} N. Barnea, V.D. Efros, W. Leidemann, and G. Orlandini, \Journal{\FBS}{35}{155}{2004}
\bibitem{Co58} F. Coester, \Journal{\NP}  {7} {421} {1958}
\bibitem{KuL78} H. K\"ummel, K.H. L\"uhrmann, and J.G. Zabolitzky, \Journal{\PREP} {36} {1} {1978}
\bibitem{BiF90b} R.F. Bishop et al., \Journal{\PRC} {42} {1341} {1990} 
\bibitem{HeM99} J.H. Heisenberg and B. Mihaila, \Journal{\PRC}  {59} {1440} {1999}
\bibitem{Ci66} J. Cizek, \Journal{\CHEM} {45} {425} {1966}; 
                          \Journal{\em Adv. Chem. Phys.} {14} {3} {1969}
\bibitem{BaM07} R.J. Bartlett and M. Musial, \Journal{\RMP} {79} {291} {2007}
\bibitem{DeH04} D.J. Dean and M. Hjorth-Jensen, \Journal{\PRC} {69} {054320} {2004}
\bibitem{HaD07} G. Hagen et al., \Journal{\PRC} {76}{044305} {2007}
\bibitem{NoB04} A. Nogga, S.K. Bogner, and A.Schwenk, \Journal{\PRC} {70} {061002(R)} {2004}
\bibitem{TaB08} A.D Taube and R.J. Bartlett, \Journal{\CHEM}{128}{044110}{2008}
\bibitem{EnM03} D.R. Entem and R. Machleidt, \Journal{\PRC} {68} {041001} {2003}
\bibitem{Za81}  J.G. Zabolitzky, \Journal{\PLB} {100} {5} {1981}
\bibitem{HaP09} G. Hagen, T. Papenbrock, and D.J. Dean, \Journal{\PRL} {103}{062503}{2009} 
\bibitem{Ro68}  D.J. Rowe,  \Journal{\RMP}  {40} {1531} {1968}
\bibitem{Ka62} M.H. Kalos, \Journal{\PREV} {128} {1791} {1962}
\bibitem{Ca87} J. Carlson, \Journal{\PRC} {36} {2026} {1987}
\bibitem{Hu59} J. Hubbard, \Journal{\PRL} {3} {77} {1959}
\bibitem{St57} R.L. Stratonovich, \Journal{\em Dokl. Akad. Nauk. SSSR} {115} {1097} {1957}
     [\Journal{\em Sov. Phys. Dokl.} {2} {416} {1957}]
\bibitem{ScF99} K. E. Schmidt and S. Fantoni, \Journal{\PLB} {446} {99} {1999} 
\bibitem{An74} J.B. Anderson, \Journal{\em J. Chem. Phys.} {63} {1499} {1974}
\bibitem{FoM01} W.M.C. Foulkes, L. Mitas, R.S. Needs, G. Rajagopal, \Journal{\RMP}{73} {34} {2001} 
\bibitem{PuP97} B.S. Pudliner et al., \Journal{\PRC} {56} {1720} {1997}
\bibitem{PiW04} S.C. Pieper, R.B. Wiringa, J. Carlson, \Journal{\PRC} {70} {054325} {2004}
\bibitem{BlL97} D. Blume, M. Lewerenz, P. Niyaz, K. B. Whaley, \Journal{\PRE} {55} {3664} {1997}  
\bibitem{NoP07} K. Nollett et al., \Journal{\PRL} {99} {022501} {2007}
\bibitem{Le09} D. Lee, \Journal{\em Prog. Part. Nucl. Phys.}{64}{117}{2009}
\bibitem{KoD97} S.E. Koonin, D.J. Dean, and K. Langanke, \Journal{\PREP}{278} {1} {1997}
\bibitem{HoM95} M. Honma, T. Mizusaki, and T. Otsuka, \Journal{\PRL} {75} {1284} {1995}
\bibitem{OtH01} T. Otsuka et al., \Journal{\em Prog. Part. Nucl. Phys.} {47} {319} {2001}
\bibitem{PiW01} S. C. Pieper and R. B. Wiringa, \Journal {\em Annu. Rev. Nucl. Part. Sci.} {51}{53}{2001}
\bibitem{ScP86} R. Schiavilla, V.R. Pandharipande, and R.B. Wiringa, \Journal{\NPA}{449}{219}{1986}
\bibitem{Wi91} R.B. Wiringa, \Journal{\PRC} {43} {1585} {1991}
\bibitem{NuC69} J. Nuttal and H. L. Cohen, \Journal{\PREV} {188} {1542} {1969}
\bibitem{Ef85} V.D. Efros, \Journal{\em Sov. J. Nucl. Phys.} {41}{949} {1985}
\bibitem{UzK03} E. Uzu, H. Kamada, and Y. Koike, \Journal{\PRC}{68}{061001}{2003}
\bibitem{DeF12} A. Deltuva, A. C. Fonseca, \Journal{\PRC}{86}{011001}{2012}
\bibitem{SuH10} Y.Suzuki, W.Horiuchi, D.Baye, \Journal{\PRO}  {123} {547} {2010}
\bibitem{HoZ12} W. Horiuchi, Y. Suzuki, and K. Arai, \Journal{\PRC} {85} {054003} {2012}
\bibitem{CaS92} J. Carlson and R. Schiavilla, \Journal{\PRL} {68} {3682} {1992}
\bibitem{EfL93} V.D. Efros, W. Leidemann, and G. Orlandini, \Journal{\FBS} {14} {151} {1993}
\bibitem{GoS02} J. Golak et al., \Journal{\NPA} {707} {365} {2002}
\bibitem{QuB04} S. Quaglioni et al.,  \Journal{\PRC}{69}{044002}{2004}
\bibitem{QuE05} S. Quaglioni, V.D. Efros, W. Leidemann, and G. Orlandini, \Journal{\PRC}{72}{064002}{2005}
\bibitem{AnQ06} D. Andreasi et al.,  \Journal{\em Eur. Phys. J. A}{27}{47}{2006}
\bibitem{BaL11} G. Bampa, W. Leidemann, and H. Arenh\"ovel, \Journal{\PRC}{84} {034005} {2011} 
\bibitem{BaE09} N. Barnea, V.D. Efros, W. Leidemann, and G. Orlandini, \Journal{\FBS}{47}{201}{2010}
\bibitem{MaB03} M.A. Marchisio, N. Barnea, W. Leidemann, and G. Orlandini, \Journal{\FBS} {33} {259} {2003}
\bibitem{CoS83} J.N.L. Connor, \Journal{\CHEM} {78} {6161} {1983}

\bibitem{BaC71} E. Balslev and J. M. Combes, \Journal{\em Commun. Math. Phys.} {22} {280} {1971}
\bibitem{Si73}  B. Simon, \Journal{\em Ann. Math.} {97} {247} {1973}; \Journal{\em Commun. Math. Phys.} {27} {1} {1992}
\bibitem{Mo98}  N. Moiseyev, \Journal{\PREP}{302} {211} {1998}
\bibitem{LaC11} R. Lazauskas and J. Carbonell, \Journal{\PRC}{84}{034002}{2011}
\bibitem{LaC05a} R. Lazauskas and J. Carbonell, \Journal{\PRC} {72} {034003} {2005} 
\bibitem{CaS98} J. Carlson and R. Schiavilla, \Journal{\RMP} {70}{743}{1998}
\bibitem{GoS05} J. Golak et al., \Journal{\PREP} {415}{89} {2005}
\bibitem{TaH67} Y.C. Tang and R.C Herndon, \Journal{\NPA} {93} {692} {1967}
\bibitem{AfT68} I.R. Afnan and Y.C. Tang, \Journal{\PREV} {175} {1337} {1968}; \Journal{\PRC} {1} {750} {1970} 
\bibitem{FaP70} S. Fantoni, L. Panottoni, and S. Rosati, \Journal{\em Nuovo Cim.} {69} {80} {1970}
\bibitem{YaY54} Y. Yamaguchi, \Journal{\PREV} {95} {1628} {1954}; 
                Y. Yamaguchi and Y. Yamaguchi, \Journal{\em ibid} {95} {1635} {1954}
\bibitem{Mi62} A.N. Mitra, \Journal{\em Nucl. Phys.} {32} {529} {1962}
\bibitem{AlG70} E.O. Alt, P. Grassberger, and W. Sandhas, \Journal{\PRC} {1} {85} {1970}
\bibitem{KhK72} V.F. Kharchenko and Y.E. Kuzmichev, \Journal{\NPA} {183} {606} {1972}
\bibitem{NaG73} I.M. Narodetsky, E.S. Galpern, and V.N. Lyakhovitsky, \Journal{\PLB} {46} {51} {1973}
\bibitem{Tj75} J.A. Tjon, \Journal{\PLB} {56} {217} {1976}
\bibitem{MaT69} R.A. Malfliet and A. Tjon, \Journal{\NPA} {127} {161} {1969}
\bibitem{DeP73} V.F. Demin, Yu.E. Pokrovsky, and V.D. Efros, \Journal{\PLB} {44} {227} {1973}
\bibitem{EiH71} H. Eikemeyer and H.H. Hackenbroich, \Journal{\NPA} {169} {407} {1971}
\bibitem{GoP70} D. Gogny, P. Pires, and R. de Tourreil, \Journal{\PLB} {32} {591} {1970} 
\bibitem{Za74} J.G. Zabolitzky, \Journal{\NPA} {228} {285} {1974}
\bibitem{Re68} R.V. Reid, \Journal{\ANNP} {50} {411} {1968}
\bibitem{Tj78} J.A. Tjon, \Journal{\PRL} {40} {1239} {1978}
\bibitem{SoF77} S. Sofianos, H. Fiedeldey, and N.J. McGurk, \Journal{\PLB} {68} {117} {1977}
\bibitem{PeS77} R. Perne and W. Sandhas, \Journal{\PRL} {39} {788} {1977}
\bibitem{SoM79} S. Sofianos, N.J. McGurk, and H. Fiedeldey, \Journal{\NPA} {318} {295} {1979}
\bibitem{SoF80} S. Sofianos, H. Fiedeldey, and H. Haberzettl, \Journal{\PRC} {22} {1772} {1980}
\bibitem{CaH82} A. Casel, H. Haberzettl, and W. Sandhas, \Journal{\PRC} {25} {1738} {1982}
\bibitem{FoH83} A.C. Fonseca, H. Haberzettl, and E. Cravo, \Journal{\PRC} {27} {939} {1983}
\bibitem{FoS76} A.C. Fonseca and P.E. Shanley, \Journal{\PRC} {14} {1343} {1976}
\bibitem{HaS81} H. Haberzettl and W. Sandhas, \Journal{\PRC} {24} {359} {1981}
\bibitem{Ba81} J.L. Ballot, \Journal{\ZPA} {302} {347} {1981}
\bibitem{ZaK81} J.G. Zabolitzky and M.H. Kalos, \Journal{\NPA} {356} {114} {1981}
\bibitem{ZaS82} J.G. Zabolitzky, K.E. Schmidt, and M.H. Kalos, \Journal{\PRC} {25} {1111} {1982}
\bibitem{CaP81} J. Carlson and V.R. Pandharipande,  \Journal{\NPA} {371} {301} {1981}
\bibitem{LoP81} J. Lomnitz-Adler, V.R. Pandharipande, and R.A. Smith, \Journal{\NPA} {361} {399} {1981}
\bibitem{CaP83} J. Carlson, V.R. Pandharipande, and R.B. Wiringa  \Journal{\NPA} {401} {59} {1983}
\bibitem{SchP86} R. Schiavilla, V.R. Pandharipande, and R.B. Wiringa, \Journal{\NPA} {449} {219} {1986}
\bibitem{NaA80} S. Nakaichi-Maeda, Y. Akaishi, and H. Tanaka, \Journal{\PRO} {64} {1315} {1980}
\bibitem{AkS74} Y. Akaishi, M. Sakai, J. Hiura, and H. Tanaka, \Journal{\em Suppl. Prog. Theor. Phys.} {56} {6} {1974}
\bibitem{Fo84} A.C. Fonseca, \Journal{\PRC} {30} {35} {1984}
\bibitem{Fe87} L.S. Ferreira, \Journal{\em Lecture Notes in Physics} {273}{100}{1987}
\bibitem{Or87} S. Oryu, \Journal{\em Lecture Notes in Physics} {273}{123}{1987} 
\bibitem{Pl87} W. Plessas, \Journal{\em Lecture Notes in Physics} {273}{137}{1987}
\bibitem{ReT71} I. Reichstein, D.R. Thompson, and Y.C. Tang \Journal{\PRC} {3} {2139} {1971}
\bibitem{PeP72} V.P. Permyakov, V.V. Pustolatov, Yu.I. Fenin, and V.D. Efros, \Journal{\em Sov. J. Nucl. Phys.}
        {14} {317} {1972}
\bibitem{KhL76} V.F. Kharchenko and V.P. Levashev, \Journal{\PLB} {60} {317} {1976}
\bibitem{Tj76} J.A. Tjon, Phys. \Journal{\PLB} {63} {391} {1976}
\bibitem{Fo79} A.C. Fonseca, \Journal{\PRC} {19} {1711} {1979}
\bibitem{HoZ81} H.M. Hofmann, W. Zahn, and H. St\"ove, \Journal{\NPA} {357} {139} {1981}
\bibitem{Fo86} A.C. Fonseca, \Journal{\FBS} {1} {69} {1986}
\bibitem{Fo89b} A.C. Fonseca, \Journal{\PRL} {63} {2036} {1989}
\bibitem{KaK89} H. Kameyama, M. Kamimura, and Y. Fukushima, \Journal{\PRC} {40} {974} {1989}
\bibitem{BiF90a} R.F. Bishop et al., in {\it The nuclear equation of
      state (Part A: Discovery of Nuclear Shock Waves and EOS)}, 
       edited by W. Greiner (Plenum, New York, 1990), p. 605 
\bibitem{OeS90} W. Oehm, S.A. Sofianos, H. Fiedeldey, and M. Fabre de la Ripelle, \Journal{\PRC} {42} {2322} {1990}
\bibitem{SchS92} N.W. Schellingerhout, J.J. Schut, and L.P. Kok, \Journal{\PRC} {46} {1192} {1992}
\bibitem{NoK00} A. Nogga, H. Kamada, and W. Gl\"ockle, \Journal{\PRL} {85} {944} {2000}
\bibitem{RoG06} D. Rozpedzik et al., \Journal{\em Acta Phys. Polon. B} {37}{2006}{2889}
\bibitem{HiK99} E. Hiyama, M. Kamimura, K. Miyazaki, and T. Motoba, \Journal{\PRC} {59} {2351} {1999}
\bibitem{NaB99} P. Navr\'atil and B.R. Barrett, \Journal{\PRC} {59} {1906} {1999}
\bibitem{KaN01} H. Kamada et al., \Journal{\PRC} {64} {044001} {2001}
\bibitem{WiP00} R.B. Wiringa, S.C. Pieper, J. Carlson, and V.R. Pandharipande, \Journal{\PRC} {62} {014001} {2000}
\bibitem{ViK05} M. Viviani, A. Kievsky, and R. Rosati, \Journal{\PRC} {71} {024006} {2005} 
\bibitem{GaB06} D. Gazit, S. Bacca, N. Barnea, W. Leidemann, and G. Orlandini, \Journal{\PRL} {96} {11230} {2006}
\bibitem{DeM03} A. Deltuva, R. Machleidt, P.U. Sauer, \Journal{\PRC} {68} {024005} {2003}
\bibitem{DeF08} A. Deltuva, A.C. Fonseca, and P.U. Sauer, \Journal{\PLB} {660} {471} {2008}
\bibitem{ViR98} M. Viviani, S. Rosati, and A. Kievsky, \Journal{\PRL} {81} {1580} {1998}
\bibitem{CiC98b} F. Ciesielski, J. Carbonell, and C. Gignoux, \Journal{\NPA} {631} {653c} {1998}
\bibitem{PhB80} T.~W. Phillips, B.~L. Berman, and J.~D. Seagrave,  \Journal{\PRC} {22}  {384} {1980}
\bibitem{DoB03} P. Doleschall, I. Borb\'ely, Z. Papp, and W. Plessas, \Journal{\PRC} {67} {064005} {2003}
\bibitem{Fo99} A.C. Fonseca, \Journal{\PRL} {83} {4021} {1999} 
\bibitem{PfH01} B. Pfitzinger, H.M. Hofmann, and G. M. Hale, \Journal{\PRC} {64} {044003} {2001}
\bibitem{ViK01} M. Viviani et al., \Journal{\PRL} {86} {3739} {2001}
\bibitem{LaC05b} R. Lazauskas et al., \Journal{\PRC} {71} {034004} {2005}
\bibitem{DeF07a} A. Deltuva and A.C. Fonseca, \Journal{\PRC} {75} {014005} {2007}
\bibitem{FiB06} B.M. Fisher et al., \Journal{\PRC}{74} {034001} {2006}
\bibitem{DeF07c} A. Deltuva and A.C. Fonseca, \Journal{\PRC} {76} {021001} {2007}
\bibitem{DeF07b} A. Deltuva and A.C. Fonseca, \Journal{\PRL} {98} {162502} {2007}
\bibitem{Do04} P. Doleschall, \Journal{\PRC} {69} {054001} {2004} 
\bibitem{DeF10} A. Deltuva and A.C. Fonseca, \Journal{\PRC} {81} {054002} {2010}
\bibitem{QuN08} S. Quaglioni and P. Navr\'atil, \Journal{\PRL} {101} {092501} {2008}
\bibitem{EpN02} E. Epelbaum et al., \Journal{\PRC} {66} {064001} {2002} 
\bibitem{Na07} P. Navr\'atil, \Journal{\FBS} {41} {117} {2007}
\bibitem{ViG10} M. Viviani et al., \Journal{\em EPJ Web of Conferences} {3} {05012} {2010} 
\bibitem{Ki11} A. Kievsky, \Journal{\em Few-Body Syst.} {50} {69} {2011}
\bibitem{AlK93} M.T. Alley and L.D. Knutson, \Journal{\PRC} {48} {1901} {1993}
\bibitem{FrL11} J. A. Frenje et al., \Journal{\PRL} {107} {122502} {2011}
\bibitem{KiV12} A. Kievsky, M. Viviani, and L.E. Marcucci, \Journal{\PRC} {85} {014001} {2012}
\bibitem{ViS07} M. Viviani et al., \Journal{\PRL} {99} {112002} {2007}
\bibitem{Fretal68} R.F. Frosch  et al., \Journal{\PRC} {160} {874} {1968}; 
                   R.G. Arnold et al., \Journal{\PRL} {40} {1429} {1978}
\bibitem{WeL88} H.R. Weller and D.R. Lehman, \Journal{\em Annu. Rev. Nucl. Part. Sci.} {38} {563} {1988}
\bibitem{SaU04} K. Sabourov et al., \Journal{\PRC} {70} {06460} {2004}
\bibitem{RaT12} R. Raut et al., \Journal{\PRL} {108} {042502} {2012} 
\bibitem{MaS00} L.E. Marcucci et al.,  \Journal{\PRL} {84} {5959} {2000}; \Journal{\PRC} {63} {015801} {2000}
\bibitem{PaM03} T.-S. Park et al., \Journal{\PRC} {67} {055206} {2003}
\bibitem{GiK10} L. Girlanda et al., \Journal{\PRL} {105} {232502} {2010}
\bibitem{AdG11}  E.G. Adelberger et al., \Journal{\RMP} {83} {195} {2011}
\bibitem{HiG04} E. Hiyama, B.F. Gibson, and M. Kamimura, \Journal{\PRC} {70} {031001} {2004}
\bibitem{FrR68} R. F. Frosch et al., \Journal{\NPA} {110} {657} {1968} 
\bibitem{Wa70} Th. Walcher,  \Journal{\PLB} {31} {442} {1970} 
\bibitem{KoO83} G. K\"obschall et al., \Journal{\NPA} {405} {648} {1983}
\bibitem{SpK89} M. Spahn et al., \Journal{\PRL} {63} {1574} {1989} 
\bibitem{ElS96} G. Ellerkmann, W. Sandhas, S.A. Sofianos, and H. Fiedeldey, \Journal{\PRC} {53} {2638} {1996}
\bibitem{UnH92} M. Unkelbach and H.M. Hofmann, \Journal{\NPA} {549} {550} {1992}
\bibitem{EfL97b} V.D. Efros, W. Leidemann, and G. Orlandini, \Journal{\PRL} {78} {432} {1997}
\bibitem{EfL97a} V.D. Efros, W. Leidemann, and G. Orlandini, \Journal{\PRL} {78} {4015} {1997}
\bibitem{BaE01} N. Barnea, V.D. Efros, W. Leidemann, and G. Orlandini, \Journal{\PRC} {63} {057002} {2001}
\bibitem{EfL98} V.D. Efros, W. Leidemann, and G. Orlandini, \Journal{\PRC} {58} {582} {1998} 
\bibitem{DyB88} S.A. Dytman et al., \Journal{\PRC} {38} {800} {1988} 
\bibitem{Zetal94}  A. Zghiche et al., \Journal{\NPA} {572} {513} {1994}  
\bibitem{WeD92} D.P. Wells et al., \Journal{\PRC} {46} {449} {1992}
\bibitem{NiA05} B. Nilsson et al., \Journal{\PLB}  {626} {65} {2005}
\bibitem{ShN05} T. Shima et al. \Journal{\PRC} {72} {044004} {2005}
\bibitem{QuN07} S. Quaglioni and P. Navr\'atil, \Journal{\PLB} {652} {370} {2007}
\bibitem{Ba07} S. Bacca, \Journal{\PRC}{75}{044001}{2007}
\bibitem{QuL04} S. Quaglioni et al., \Journal{\PRC} {69} {044002} {2004}
\bibitem{BaB09a} S. Bacca, N. Barnea, W. Leidemann, and G. Orlandini, \Journal{\PRL} {102} {16250} {2009}
\bibitem{BaB09b} S. Bacca, N. Barnea, W. Leidemann, and G. Orlandini, \Journal{\PRC} {80} {064001} {2009}
\bibitem{CoH01} S.~A. Coon and H.~K. Hahn, \Journal{\FBS}{30}{131}{2001}
\bibitem{BuT06}A. Yu. Buki, I. S. Timchenko, N. G. Shevchenko, and I. A. Nenko, \Journal{\PLB} {641} {156} {2006}
\bibitem{BaA07} S. Bacca et al., \Journal{\PRC} {76} {014003} {2007}
\bibitem{CaS94} J. Carlson and R. Schiavilla, \Journal{\PRC}{49}{R2880}{1994}
\bibitem{CaJ02} J. Carlson, J. Jourdan, R. Schiavilla, and I. Sick, \Journal{\PRC} {65} {024002} {2002}
\bibitem{StQ07} I. Stetcu et al., \Journal{\NPA} {785} {307} {2007}
\bibitem{GaB07} D. Gazit and N. Barnea, \Journal{\PRC} {70} {048801} {2004}; \Journal{\PRL} {98} {192501} {2007}
\bibitem{PuP95} B.S. Pudliner, V.R. Pandharipande, J. Carlson, and R.B. Wiringa, \Journal{\PRL}{74}{4396}{1995}
\bibitem{PiV02} S.C. Pieper, K.Varga, and R.B. Wiringa, \Journal{\PRC}{66}{044310}{2002}
\bibitem{PiP01} S.C. Pieper, V.R. Pandharipande, R.B. Wiringa, and J. Carlson, \Journal{\PRC}{64}{014001}{2001}
\bibitem{ThL77} D.R. Thomson, M.LeMere, and Y.C. Yang, \Journal{\NPA} {286} {53} {1977} 
\bibitem{NaV01} P. Navr\'atil, J.P. Vary, W.E. Ormand, and B.R. Barrett, \Journal{\PRL}{87}{172502}{2001}
\bibitem{CaN02} E. Caurier, P. Navr\'atil, W.E. Ormand, and J.P. Vary, \Journal{\PRC}{66}{024314}{2002}
\bibitem{NaO03} P. Navr\'atil and W.E. Ormand, \Journal{\PRC}{68}{034305}{2003}
\bibitem{NaC04} P. Navr\'atil and E. Caurier, \Journal{\PRC} {69} {014311} {2004}
\bibitem{FoN05} C. Forss\'en, P. Navr\'atil, W.E. Ormand, and E. Caurier, \Journal{\PRC}{71}{044312}{2005}
\bibitem{Pi05} S.C. Pieper, \Journal{\NPA}{751}{516}{2005}
\bibitem{NaG07} P. Navr\'atil et al., \Journal{\PRL}{99}{042501}{2007}
\bibitem{RoL11} R. Roth et al., \Journal{\PRL}{107}{072501}{2011}
\bibitem{GaQ09} D. Gazit, S. Quaglioni, and P. Navr\'atil, \Journal{\PRL}{103}{102502}{2009}
\bibitem{MaV09} P. Maris, J. P. Vary, and A. M. Shirokov, \Journal{\PRC} {79} {014308} {2009}
\bibitem{AbM11} T. Abe et al., \Journal{\em AIP Conf.Proc.} {1355} {173} {2011} 
\bibitem{JeH11} O. Jensen, G. Hagen, M. Hjorth-Jensen, and J.S. Vaagen, \Journal{\PRC}{83}{021305}{2011}
\bibitem{JaH11} G. R. Jansen, M. Hjorth-Jensen, G. Hagen, and T. Papenbrock, \Journal{\PRC}{83}{054306} {2011}
\bibitem{EpK11} E. Epelbaum, H. Krebs, D. Lee, and Ulf-G. Mei{\ss}ner, \Journal{\PRL} {106} {192501} {2011} 
\bibitem{GaP07} S. Gandolfi, F. Pederiva, S. Fantoni, and K.E. Schmidt, \Journal{\PRL} {99} {022507} {2007}
\bibitem{CaN06} E. Caurier and P. Navr\'atil, \Journal{\PRC}{73}{021302}{2006}
\bibitem{FoC09} C. Forss\'en, E. Caurier, and P. Navr\'atil, \Journal{\PRC}{79}{021303}{2009}
\bibitem{PeP07} M. Pervin, S.C. Pieper, and R. B. Wiringa, \Journal{\PRC}{76}{064319}{2007}
\bibitem{MaP08} L.E. Marcucci et al., \Journal{\PRC}{78}{065501}{2008}
\bibitem{VaB09} S. Vaintraub, N. Barnea, and D. Gazit, \Journal{\PRC}{79}{065501}{2009}
\bibitem{MuS07} P. Mueller et al., \Journal{\PRL}{99}{252501}{2007}
\bibitem{BrB12} M. Brodeur et al., \Journal{\PRL} {108} {052504} {2012}
\bibitem{McL09} E. A. McCutchan et al., \Journal{\PRL} {103} {192501} {2009}
\bibitem{QuN09} S. Quaglioni and P. Navr\'atil, \Journal{\PRC} {79} {044606} {2009}
\bibitem{HaD-unp} G.M. Hale, D.C. Dodder, and K.Witte, unpublished
\bibitem{NaR10} P. Navr\'atil, R. Roth, and S. Quaglioni, \Journal{\PRC} {82} {034609} {2010} 
\bibitem{KrH84} H. Krupp et al., \Journal{\PRC} {30} {1810} {1984} 
\bibitem{FlR55} J. M. Freeman, A. M. Lane, and B. Rose, \Journal{\em Phil. Mag.} {46} {17} {1955} 
\bibitem{NaQ11} P. Navr\'atil and S. Quaglioni, \Journal{\PRC}{83}{044609}{2011}
\bibitem{NaR11} P. Navr\'atil, R. Roth, and S. Quaglioni, \Journal{\PLB}{704}{379}{2011}
\bibitem{NaQ12} P. Navr\'atil and S. Quaglioni, \Journal{\PRL}{108}{042503}{2012}
\bibitem{DeA04} P. Descouvemont et al. \Journal{\em Data Nucl. Data Tables} {88} {2003} {2004}
\bibitem{Ne11} T. Neff, \Journal{\PRL}{106}{042502}{2011}
\bibitem{BaM02} S. Bacca et al., \Journal{\PRL}{89}{052502}{2002} 
\bibitem{BaB04} S. Bacca, N. Barnea, W. Leidemann, and G. Orlandini, \Journal{\PRC}{69} {057001} {2004}
\bibitem{BaA04} S. Bacca et al., \Journal{\PLB}{603}{159}{2004}
\bibitem{WiP02} R.B. Wiringa and S.C. Pieper, \Journal{\PRL} {89} {182501} {2002}
\bibitem{Ahetal75} J. Ahrens et al., \Journal{\NPA} {251} {479} {1975}

\end{thebibliography}
\end{document}